\documentclass[aps,prl,notitlepage,superscriptaddress,longbibliography,twocolumn,nofootinbib,floatfix]{revtex4-2} 

\usepackage[utf8]{inputenc}
\usepackage{graphicx}
\usepackage{hyperref}
\usepackage[dvipsnames]{xcolor}
\usepackage{graphicx}

\usepackage{tikz,tikz-3dplot}
\usetikzlibrary{decorations.pathreplacing}
\usetikzlibrary{decorations.pathmorphing}
\tikzset{snake it/.style={decorate, decoration={snake,amplitude=.6mm,segment length=1.5mm}}}
\usepackage{pgfplots}

\usepackage{lipsum}

\usepackage{amsmath,amsthm,amssymb,mathtools}
\usepackage{multirow}
\usepackage{braket}
\usepackage{bbm,bm}
\usepackage[bb=boondox]{mathalfa}

\newcommand{\ie}{{\it i.e.},\ }

\newcommand{\id}{\mathbb{1}}

\newcommand{\ii}{\operatorname{i}}
\newcommand{\ee}{\operatorname{e}}

\newcommand{\integral}[3]{\int_{#2}^{#3} \!\! \mathrm{d} #1 \,}

\theoremstyle{thmstyleone}

\theoremstyle{thmstyletwo}

\theoremstyle{thmstylethree}

\newcommand{\dyad}[1]{\vert #1 \rangle \langle #1 \vert}

\newcommand{\cor}[1]{#1}

\begin{document}

\title{Random pure Gaussian states and Hawking radiation}

\author{Erik Aurell}
\affiliation{KTH – Royal Institute of Technology, Alba Nova University Center, SE-106 91 Stockholm, Sweden}

\author{Lucas Hackl}
\affiliation{School of Mathematics and Statistics, The University of Melbourne, Parkville, VIC 3010, Australia}
\affiliation{School of Physics, The University of Melbourne, Parkville, VIC 3010, Australia}

\author{Pawe{\l} Horodecki}
\affiliation{International Centre for Theory of Quantum Technologies, University of Gdańsk, Wita Stwosza 63, 80-308 Gdańsk, Poland}
\affiliation{Faculty of Applied Physics and Mathematics, National Quantum Information Centre, Gdańsk, University of Technology, Gabriela Narutowicza 11/12, 80-233 Gdańsk, Poland}

\author{Robert H. Jonsson}
\affiliation{Nordita, Stockholm University and KTH Royal Institute of Technology, Hannes Alfvéns väg 12, SE-106 91 Stockholm, Sweden}

\author{Mario Kieburg}
\affiliation{School of Mathematics and Statistics, The University of Melbourne, Parkville, VIC 3010, Australia}

\begin{abstract}
A black hole evaporates by Hawking radiation. Each mode of that radiation is thermal. If the total state is nevertheless to be pure, modes must be entangled. Estimating the minimum size of this entanglement has been an important outstanding issue. We develop a new theory of constrained random symplectic transformations, based on that the total state is pure and Gaussian  with given marginals.
In the random constrained symplectic model we then compute the distribution of mode-mode correlations, from which we bound mode-mode entanglement. 
Modes of frequency much larger than $\frac{k_B T_{H}(t)}{\hbar}$ are not populated at time $t$ and drop out of the analysis.
Among the other modes we find that correlations and hence entanglement between  relatively thinly populated modes (early-time high-frequency modes and/or late modes of any frequency) to be strongly suppressed. 
Relatively 
highly populated modes (early-time low-frequency modes) can on the other hand be strongly correlated, but a detailed analysis reveals that they are nevertheless very unlikely to be entangled. 
Our analysis hence establishes that restoring unitarity after a complete evaporation of a black hole does not require any significant
quantum entanglement between any pair of Hawking modes. 
Our analysis further gives exact general expressions for the distribution of mode-mode correlations in random, pure, Gaussian states with given marginals, which may have applications beyond black hole physics.
\end{abstract}

\maketitle

Hawking's \textit{black hole information paradox} has for almost half a century held the place of one of the most important unsolved problem in fundamental physics~\cite{Hawking76}.
The paradox is based on two premises. The first is Hawking's own discovery that black holes radiate~\cite{Hawking74,Hawking75} in the presence of quantum fields. Hawking showed that the state of one mode of frequency about $\omega$ localized around some time slice is thermal at the Hawking temperature, which is determined by the black hole mass.

The second premise is that quantum mechanics holds everywhere. Accordingly, closed quantum systems must evolve unitarily if the probabilistic interpretation of quantum theory is to hold true.
This implies that an initially pure state will remain pure forever. One can imagine a pure state of matter collapsing into a black hole, and then radiating away in Hawking radiation. The total state of all the radiation, after the entire black hole has dissolved and given back its content to the remaining universe, should thus be pure once again. The question is hence how this can be possible if the states of all the single modes are thermal. A selected set of recent reviews discussing the paradox and ways to resolve it are~\cite{Harlow},~\cite{Marolf} and~\cite{UnruhWald17}. Page was the first to point out that if the second premise holds, then a resolution of the paradox requires that modes be entangled~\cite{Page1980,Page1993}.

We recall that  the \textit{marginal} over a subsystem $A$ of a large pure state $\dyad{\Psi}$ is the density matrix $\rho_A =\mathrm{Tr}_{B}\dyad{\Psi}$, where $B$ is the complement of $A$~\cite{Horodecki4}.
A resolution of Hawking's paradox following Page is therefore an instance of the quantum marginal problem~\cite{Klyachko2004,Klyachko2006}: \emph{Given a fixed  Hilbert space  and a set of disjoint sub-systems $\{A_i\}_{i=1}^L$
with complements  $\{B_i\}_{i=1}^L$  and
density matrices  $\{\rho_{i}\}_{i=1}^L$, does there exist a global pure state $\dyad{\Psi}$ in ${\cal H}$, such that $\rho_{i} =\mathrm{Tr}_{B_i}\dyad{\Psi}$
for all $i=1,\ldots,L$?} 
In general, this is a too involved problem, on the current level of quantum information science. However, if we additionally assume that the total state $\dyad{\Psi}$ is a pure Gaussian quantum state, the problem is greatly simplified. In~\cite{Aurell2022}, two of the present authors showed that the constraints of the Gaussian quantum marginal problem for bosons~\cite{Eisert2008} are easily satisfied for a macroscopic black hole.

In this work, we show that given any two modes, their expected entanglement is vanishingly small when averaging over all pure Gaussian states whose marginals match Hawking's calculation. 
To this end, we introduce a new method to the study of random Gaussian states: We consider the family of $N$-mode pure Gaussian states whose marginals are thermal states. This set of states can be shown to be compact and equipped with a natural measure induced from the Haar measure of the symplectic group. This enables us to find the probability distribution over correlations between two modes 
at asymptotic infinity $\mathcal{I}^+$.
Most of these correlations are very small, which implies that the corresponding mode-mode entanglement is also very small. Some specific pairs of modes (to be described below) can be strongly correlated in the constrained random Gaussian pure state ensemble, but are nevertheless also very unlikely to be entangled. 

\textbf{Relation to other approaches.}
The literature motivated by Hawking's information
paradox is large and still growing. After the
publication of the 
general reviews~\cite{Harlow,Marolf,UnruhWald17}, substantial advances have been made:  
\cor{
Based on ideas from holographic models and AdS~\cite{Almheiri2019,Penington2020}, they led to a partial understanding of microscopic origin of Bekenstein-Hawking entropy through the ideas of entanglement wedges and islands~\cite{Almheiri2020,Almeiri21,Penington2022}. And recently, a family of black hole microstates were  proposed as explanation for the microscopic origin of astrophyscial black hole entropy~\cite{Balasubramanian2024b,Balasubramanian2024a}.}
Our approach is complementary to these developments, as we focus on properties of random states, and consider them as \cor{as a model for} Hawking radiation which has escaped the black hole. 
This point of view was initiated by Page, who showed that the density matrix of a small subsystem is almost surely close to thermal, if the possible pure states of the entire system are sampled uniformly~\cite{Page1993b}.

Page's approach has been applied by substituting the unknown dynamics inside a black hole with a random unitary transformation, and then estimating entanglement between Hawking radiation emitted before and after this internal mixing \cite{Page1993b,Hayden2007,Braunstein2013}. Compared to Page's approach, and those derived from it, we consider not only one subsystem but a large number of subsystems consisting of all the modes of the system, and we assume that their marginals are all thermal, as they would be in Hawking's theory.
Furthermore, motivated by the observation that most modes of the Hawking radiation emanate in a region where gravity is not strong at the black hole horizon, we here consider Gaussian states as approximation to the physical state of the Hawking radiation.

More realistic microscopic models of black hole radiation, however, may yield non-Gaussian states exhibiting subleading corrections. 
If we had access to a zoo of self-consistent models (classical collapsing
black hole spacetime equipped with quantum field, such that Einstein’s equations are satisfied
with stress energy tensor of the quantum field) to compute the final state, we could analyze
how close the respective final states are to the ‘typical’ one, we studied. As the few existing
models come with their own  assumptions (lower dimensions, violations of Einstein’s
equation, truncation of quantum degrees of freedom etc.), such a comparison is not available
for now. Let us 
therefore emphasize that our claim is not that we found the correct final state of
an evaporating black hole, but that we demonstrated that potential resolutions of the black hole
information paradox may require relatively little entanglement between the individual modes.


\textbf{Pure Gaussian states with fixed marginals.} 
Let $\hat{x}^a=(\hat{q}_1,\hat{p}_1,\ldots,\hat{q}_N,\hat{p}_N)$ be quadratures (generalized canonical positions and momenta) of a system with $N$ bosonic degrees of 
freedom. 
\cor{We here focus on states with vanishing expectation values $\left<\hat x^a\right>=0$.}
The covariance matrix of \cor{such} a quantum state, contains the  (symmetrized) expectation values,
\begin{equation}
C^{ab}=\frac{1}{2}\left<\hat{x}^a\hat{x}^b+\hat{x}^b\hat{x}^a\right>\,.
\end{equation}
$C$ is a positive semi-definite real symmetric matrix. Its quantum origin is reflected in the Robertson-Schrödinger uncertainty relations that $C+i\Omega\geq0$ is  positive semi-definite as well, where $i\Omega^{ab}\hat\id=[\hat{x}^a,\hat{x}^b]=\hat{x}^a\hat{x}^b-\hat{x}^b\hat{x}^a$ is the symplectic form.

Gaussian bosonic states (mixed or pure) are fully characterized by their covariance matrix, which can be decomposed as
\begin{equation}
\label{eq:symplectic-C}
C= S D S^T
\end{equation}
where $S\in \mathrm{Sp}(2N)$ is a real symplectic transformation and $D=\mathrm{diag}(d_1,d_1,\ldots,d_N,d_N)$ gives the symplectic eigenvalues. For a pure state, we have $d_i=1$, \ie $D=\id$, which means every pure Gaussian state is the vacuum with respect to its normal modes.

The marginal state 
over one pair of quadrature variables $(q_i,p_i)$
is Gaussian. 
In general this state is mixed.
It is
characterized by the $2\times 2$ covariance matrix $C_{(i)}$ 
which by construction is real, positive and positive semi-definite, and also satisfies the Robertson-Schr\"odinger 
uncertainty relations.
It can therefore be 
\begin{equation}
C_{(i)} = S_2 \operatorname{diag}(\mu_i,\mu_i) S_2^T\,,
\end{equation}
where $\mu_i\geq1$ is its
symplectic eigenvalue and 
$S_2\in \mathrm{Sp}(2)$ is a two-dimensional real symplectic transformation.

Important examples of Gaussian states, the covariance matrices of which are proportional to the identity, are thermal equilibrium states of a single harmonic oscillator mode. For a mode of frequency $\omega_i$ and temperature $T_i$, their covariances are diagonal
\begin{equation}
\label{eq:Cn-thermal}
    C_{(i)}=\text{diag}[ \mu_i,\mu_i ]
\quad 
{\rm with}\quad \mu_i=\coth\left(\frac{\hbar \omega_i}{2k_B T_i} \right).
\end{equation}
Here, $\tfrac12(\mu_i-1)$
is the expected number of quanta in the mode, which
vanishes only in the vacuum state 
when the temperature $T_i$ is zero.

In the following, we consider the set $\mathcal{M}_{\hat C}$ of all pure Gaussian $N$-mode state covariance matrices $C$ whose marginals are identical to some prescribed single-mode marginals $C_{(i)}$, collectively referred to as $\hat C=\mathrm{diag}\left(C_{(1)},\dots,C_{(N)}\right)$.
This set can be equipped with a natural choice of measure $\rho(C|\hat C)$: 
As detailed in~\cite{supp}, we obtain it from the restriction of
the Haar measure on all symplectic transformations to the compact subset of transformations which (via~\eqref{eq:symplectic-C}) generate a covariance matrix $C$ with the prescribed marginals $\hat C$. 
The resulting distribution $\rho(C|\hat C)$ thus allows us to study the typical two-mode correlations in the ensemble $\mathcal{M}_{\hat C}$.

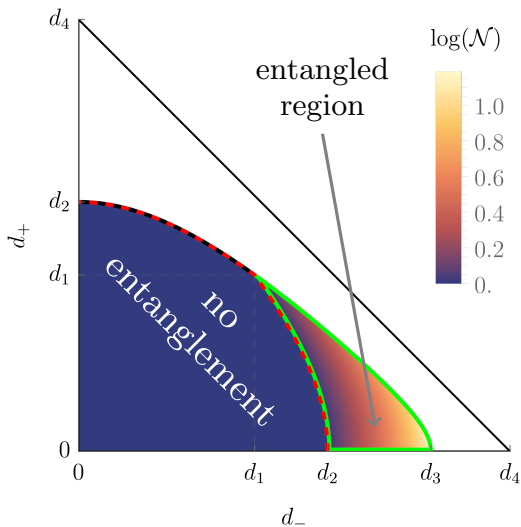
\begin{figure}[t]
\centering
\begin{tikzpicture}
\draw (0,0) rectangle (1,1);
\draw (0,0) node{\includegraphics[width=.8\linewidth]{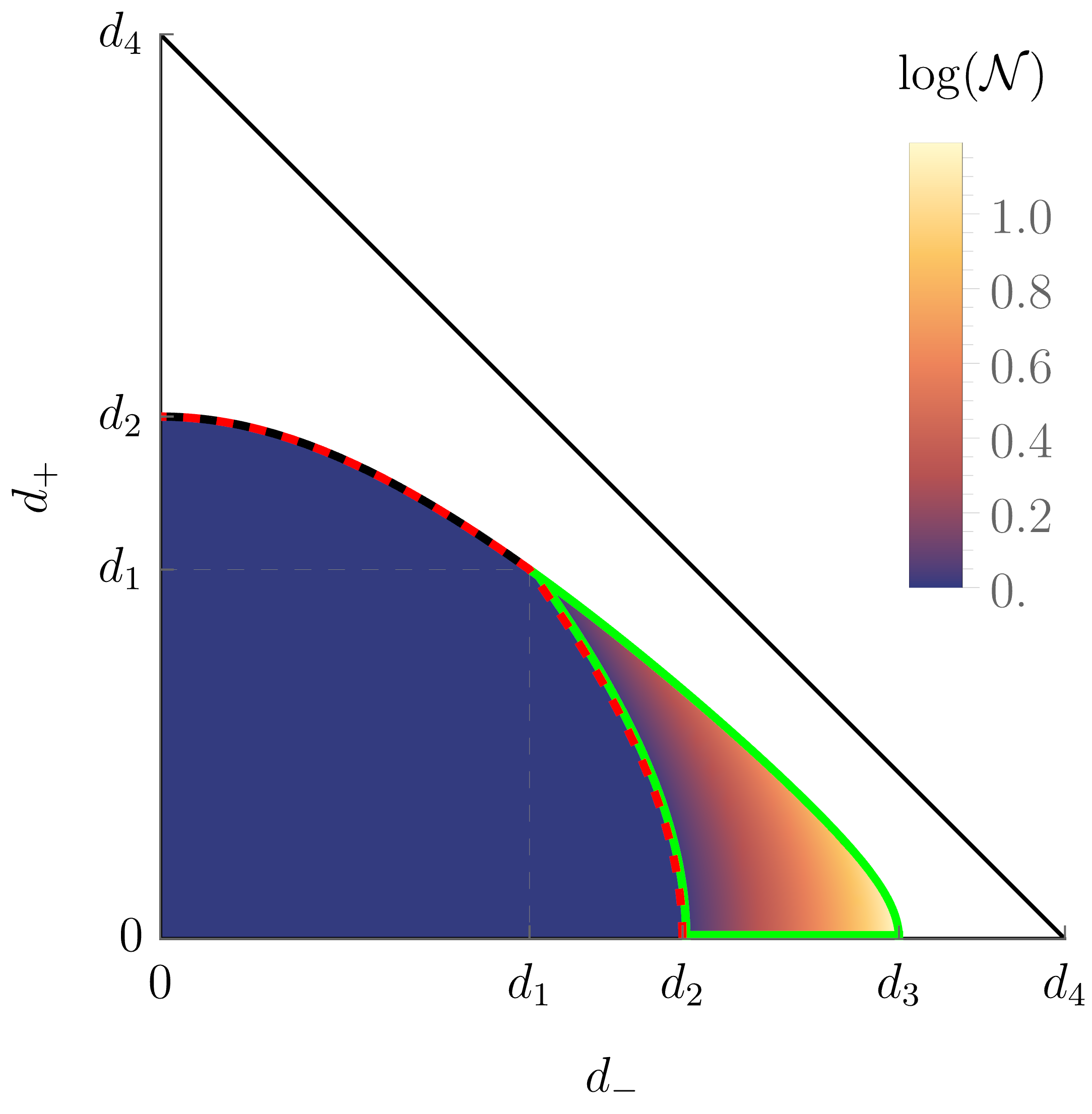}};
\draw (-.8,-.8) node[white,scale=1.6,text width=2cm,align=center,rotate=-45]{no\\ entanglement};
\draw[very thick,gray,->] (.8,1.8) node[scale=1.3,above,black,text width=2cm,align=center]{entangled\\region} -- (1.5,-2.1);
\end{tikzpicture}
\caption{
\emph{Correlations and entanglement between two jointly Gaussian modes.} 
Colored region shows the allowed region in the variables ($d_+,d_-$) parametrizing the mode-mode correlation matrix block $C_{(12)}$, where in this
example the two diagonal blocks 
have symplectic eigenvalues 
$\mu_1=3$ and $\mu_2=2$ 
(see main text).
Blue region indicates no entanglement (separable states). Region bounded by green curve indicates entangled states, 
colored from blue to yellow by increasing logarithmic negativity.
Maximal entanglement is found at the tip at $(d_+,d_-)=(0,d_3)$. 
Marked points in the figure are $d_1=\sqrt{(\mu_1^2-1)(\mu_2^2-1)/(2\mu_1\mu_2)}$, $d_2=\sqrt{2(\mu_1-1)(\mu_2-1)}$ and $d_3=\sqrt{2(\mu_1+1)(\mu_2-1)}$, see
\cite{supp}.
}
\label{fig.correlations}
\end{figure}

\textbf{Two-mode correlations and entanglement.} 
To study correlations between two modes of a given $N$-mode Gaussian state, we consider 
its reduction to two modes which we call $1$ and $2$.
All the other modes are hence numbered
$3, 4,\ldots,N$.
The correlation within and between modes $1$ and $2$
is described by their covariance matrix
\begin{equation}
    C_{1 \wedge 2}=\left(\begin{array}{c|c}C_{(1)}& C_{(12)}\\ \hline C_{(12)}^\intercal & C_{(2)}
    \end{array}\right).\label{eq:C1plus2}
\end{equation}
where $C_{(1)}$ and $C_{(2)}$ are the covariance matrices of
modes $1$ and $2$ as introduced
above, and $C_{(12)}$ captures the correlations across  variables in mode $1$ and mode $2$. In the total (large)  covariance matrix $C$, $C_{(1)}$ and $C_{(2)}$ are diagonal 
$2$-by-$2$ blocks,
and $C_{(12)}$ is an off-diagonal 
$2$-by-$2$ block.

It is well established~\cite{serafiniQuantumContinuousVariables2017}
that one may first separately diagonalize
$C_{(1)}$ and $C_{(2)}$ using two symplectic transformations, to get $C_{(i)}=\mu_i \id_{2}$ ($i=1,2$) and 
then use special orthogonal transformations (which leave 
the $\mu_i \id_{2}$-form of on $C_{(i)}$ unchanged) 
to get \begin{align}
    C_{(12)}=C_{(21)}= \text{diag}[c_+,c_- ]
\end{align}
with $c_+\geq |c_-|$.
The $4$-by-$4$ matrix $C_{1 \wedge 2}$
in \eqref{eq:C1plus2} can therefore
without restriction be taken to consist of four 
diagonal $2$-by-$2$ submatrices. For convenience 
we shall use the parametrisation 
$d_\pm=\frac{1}{\sqrt{2}}(c_+\pm c_-)$,
where both $d_\pm$ are non-negative.

The matrix $C_{1 \wedge 2}$ in \eqref{eq:C1plus2}
can be a  covariance  matrix of a quantum state if it satisfies the Robertson-Schr\"odinger uncertainty relations, which is the case if both its symplectic eigenvalues $\nu_{+}$, $\nu_-$ (being functions  of $\mu_1$, $\mu_2$, $d_+$
and $d_-$ ~\cite{supp}, sec.~II, eq.~8), are larger or equal to one.
As shown in Fig.~\ref{fig.correlations}, valid Gaussian states lie below a ($\mu_1$- and $\mu_2$-dependent)
curve in the $d_\pm$-plane. 

A two-mode Gaussian state is entangled
if and only if the Peres-Horodecki criterion is 
satisfied~\cite{serafiniQuantumContinuousVariables2017}.
This criterion is generally given in terms of 
positivity after a partial transpose, 
which for two-mode Gaussian states reduces to 
the statement that smallest symplectic 
eigenvalue ($\nu^{PT}_-$) of a partially transposed 
density matrix is less than 
one.
The dashed red line 
in Fig.~\ref{fig.correlations} 
is given by $\nu^{PT}_-=1$, and hence 
separates entangled from not entangled (separable) states. 
The symplectic eigenvalues $\nu^{PT}_\pm$, 
are the same algebraic functions as $\nu_\pm$, with only the roles
of $d_+$ and $d_-$ interchanged.
To illustrate the above, in Fig.~\ref{fig.correlations} we plot the logarithmic negativity $\mathcal{N}=\max\left[0,-\ln\nu^{PT}_-\right]$ of the state 
in color coding.
We see that entangled states only occur in a relatively small corner towards the maximum value of $d_-$. For large values of $\mu=\mu_1\approx\mu_2$ detailed estimates
show that the size of the area corresponding to entangled states relative to the size of all allowed states, falls off at least as $\log(\mu)/\mu^2$,
for illustration see Fig.~\ref{fig:fig3_tripple_fig_black_hole_distribution}[c].

\textbf{Induced probability distribution over two-mode correlations.}
The distribution over the ensemble of all pure Gaussian states with fixed marginals $\hat C=\mathrm{diag}\left(C_{(1)},\dots,C_{(N)}\right)$ induces---through the two-mode covariance matrix~\eqref{eq:C1plus2}---a distribution on its $2\!\times\! 2$-off-diagonal-block $C_{(12)}$
(as before the pair $(12)$ may stand for any fixed pair of modes).
This probability distribution,  denoted by $\rho(C_{(12)}|\hat C)$, is one of the central quantities in this paper since it gives the {\it typical behaviour of two-mode correlations in a random $N$-mode Gaussian state with fixed single-mode marginals.} 
We construct it explicitly (see Supp. material \cite{supp}), showing, in particular, that it may be reinterpreted just as a probability
measure  $\rho(d_+,d_-|\hat C)$ over the two real, non-negative  numbers $d_+,d_-\geq 0$, already known
from above to basis-independently characterise the state's two-mode correlations. 
In the limit of many marginals 
$\rho(d_+,d_-|\hat C)$
takes the form
\begin{equation}\label{eq:rho-exact}
    \rho(d_+,d_-|\hat C) \propto d_+d_-\frac{\chi (d_+,d_-) \prod_{i=3}^N \Phi(\mu_1,\mu_2,C_{(12)},\mu_i)}{\left[(\nu_+^2-1) (\nu_-^2-1)\right]^2}
\end{equation}
where $\chi(d_+,d_-)=\Theta(\nu_--1)\Theta(\sqrt{2\mu_1\mu_2}-d_+-d_-)$ denotes the characteristic function of the region of valid states.
Only states $i$ with $\mu_i$
strictly larger than one contribute to 
the product;
the explicit form of $\Phi$ together with its properties \footnote{In particular,
$\Phi(\mu_1,\mu_2,C_{(12)},\mu_i=1)=1$
while for $\mu_i>1$
$\Phi$ takes values in $[0,1]$, attains its global maximum over the region at the origin of $C_{(12)}$, i.e.,
$\Phi(\mu_1,\mu_2,C_{(12)}=0,\mu_i)=1$, is convex over the region, and vanishes on the boundary where $\nu_-=1$.} 
are derived and discussed  in~\cite{supp}. 

\begin{figure}[t!]
\centering
\begin{tikzpicture}

\begin{scope}[xshift=-5.3cm,yshift=1cm,scale=1.2]
    \draw[thick, snake it] (0,0) -- (1.5,0);
    \draw[thick] (1.5,0) -- (3,-1.5) -- (1.5,-3);
    \draw[thick, snake it] (0,-3) -- (1.5,-3);
    \draw[dashed] (0,-1.5) -- (1.5,0);
    \draw[dashed] (0,-1.5) -- (1.5,-3);
    \draw[thick] (0,0) -- (0,-3);
    \draw[very thick,red] (.5,0) arc (-180:-45:1);
    \draw (1.5,-3.7) node[text width=4cm]{(a) Stationary black hole (mass $M_1$)};
\end{scope}

\begin{scope}[scale=.6]
    \draw[thick] (0,0) -- (0,-4) -- (4,0) -- (1.5,2.5) -- (1.5,0);
    \draw[thick, snake it] (0,0) -- (1.5,0);
    \draw[dashed] (0,-1.5) -- (1.5,0);
    \draw (1.5,0) -- (2.75,1.25) node[right,yshift=2mm]{$M=0$};
    \draw (1,0) arc (-180:-45:.5) --(3.10355,0.896447);
    \draw[red,very thick] (.5,0) arc (-180:-45:1) --(3.45711,0.542893) node[right,yshift=2mm]{$M_1$};
    \draw[very thick,blue] (1.2,0) arc (-540:135:.3) -- (.7,1.2) node[above,text width=1.3cm,align=left,xshift=-.5cm,yshift=-.4cm]{strong \& \\
    quantum\\gravity};
    \draw (3,-5.72) node[text width=4cm]{(b) Evaporating\\black hole};
\end{scope} 
\end{tikzpicture}
\caption{Penrose diagrams representing stationary and 
evaporating black holes (following~\cite{wald1994quantum}). Dashed lines indicate horizons, wiggly lines  singularities, and $45^{\circ}$ lines 
future/past null infinity ($\mathcal{I}^{\pm}$). (a) Diagram of a stationary black hole with fixed mass. Red curve: a time slice of modes  escaping to $\mathcal{I}^{+}$. Lower singularity is in the past (white hole), upper singularity in the future (black hole). (b) Diagram of an evaporating black hole obtained by joining time slices of decreasing black hole mass. This spacetime only has  one singularity inside the black hole.
The red curve corresponds to the same time slice as in (a).
The blue ring indicates the region where the curvature near the horizon becomes large enough, such that back reaction and eventually quantum gravitational effects become relevant.
Most modes of the outgoing radiation emanate outside this region motivating our assumption of Gaussianity.}
\label{fig.conformal-diagrams}
\end{figure}
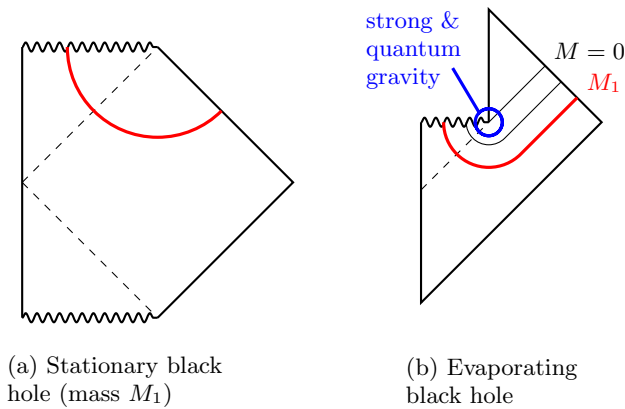

\begin{figure*}
    \centering
    \begin{tikzpicture}
        \draw (-5.8,0) node[draw=red, line width=.5mm,inner sep=0.5\pgflinewidth]{\includegraphics[height=5cm]{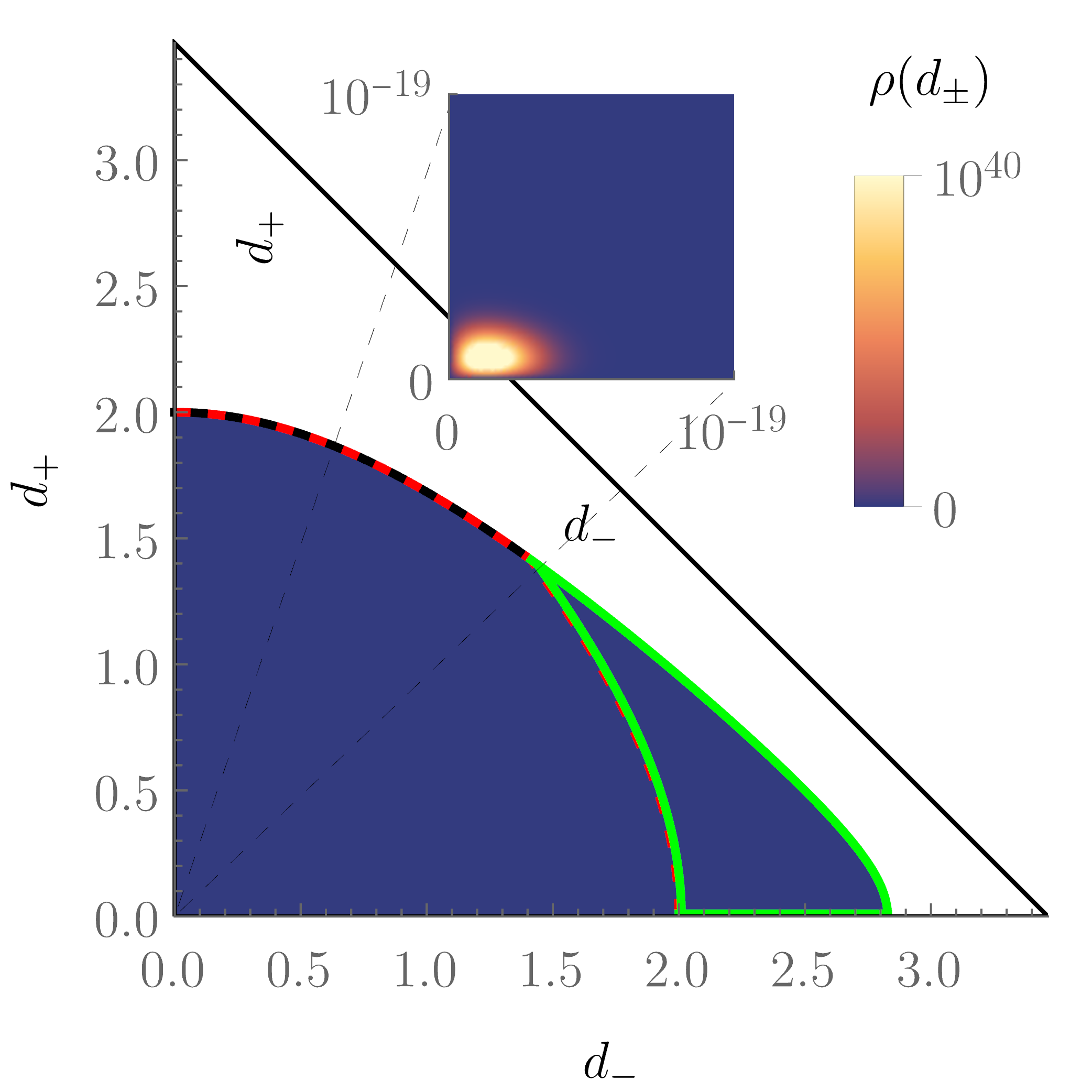}};
        \draw (0,0) node{\includegraphics[height=4.7cm]{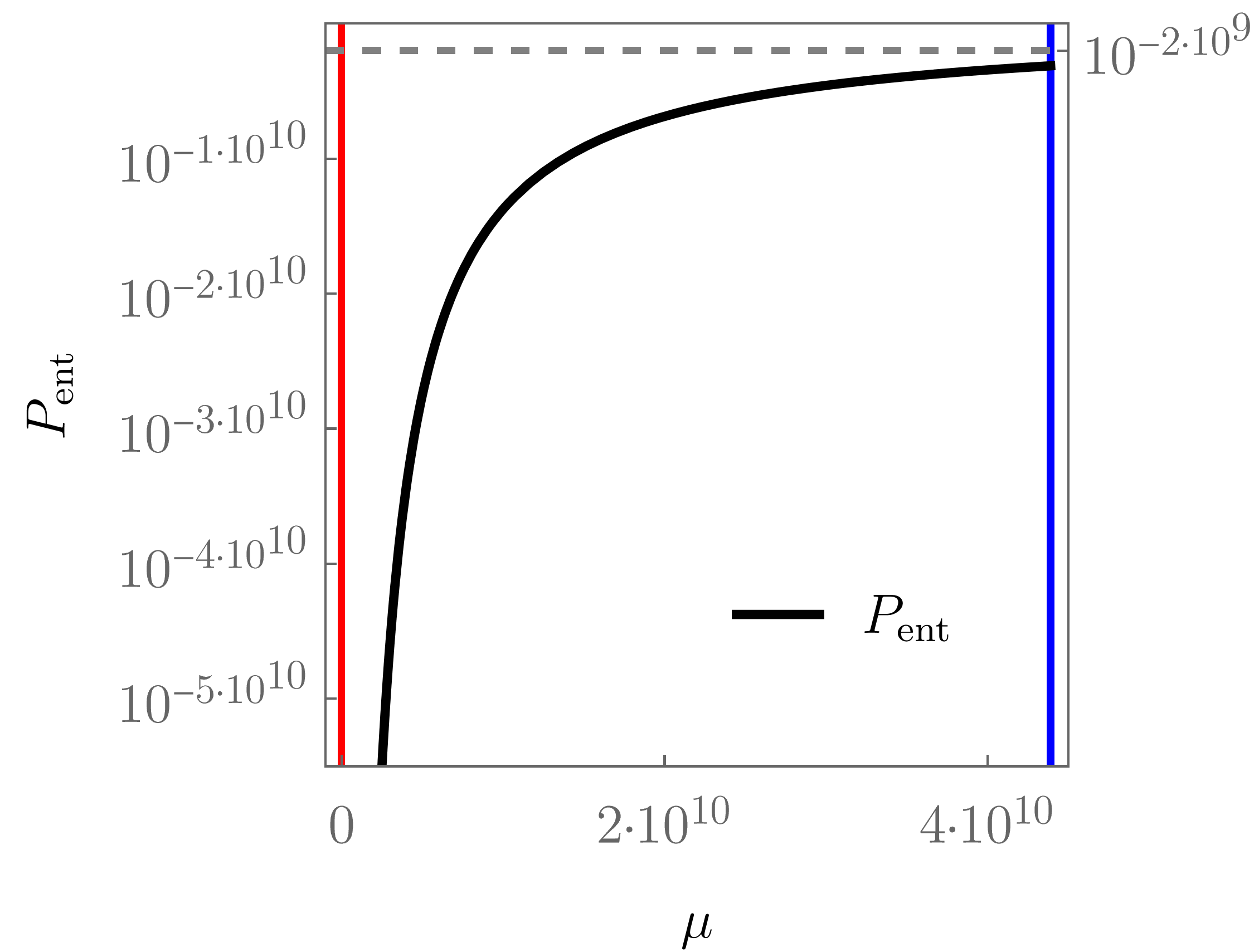}};
        \draw (6,0) node[draw=blue, line width=.5mm,inner sep=0.5\pgflinewidth]{\includegraphics[height=5cm]{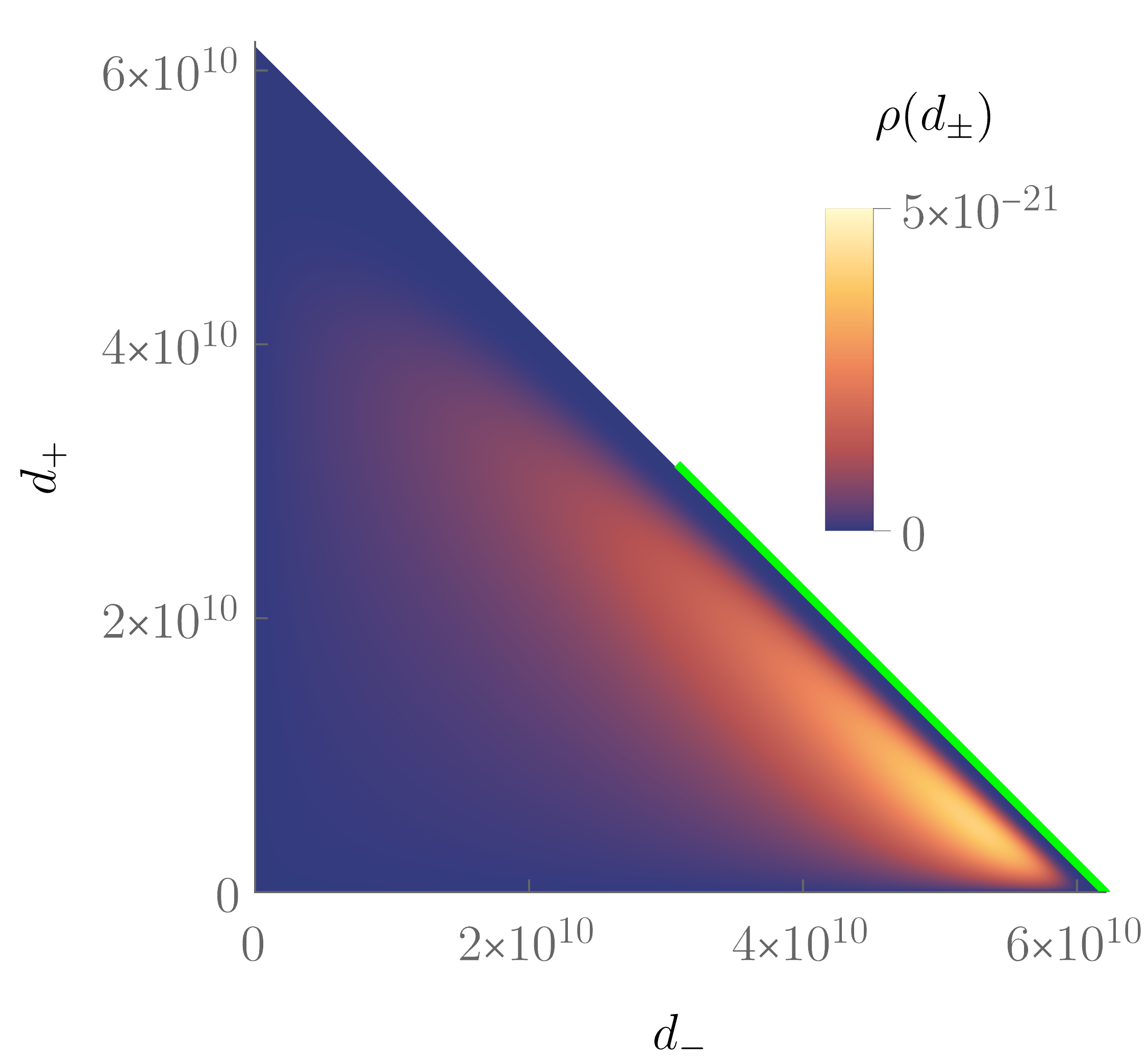}};
        \draw (-7.8,2.8) node{(a)};
        \draw (-7.5,2.8) node[right]{$\mu_1=3, \mu_2=2$};
        \draw (0.2,2.8) node{(b)};
        \draw (4,2.8) node{(c)};
        \draw (4.2,2.8) node[right]{$\mu_1\approx \mu_2=4.39\times10^{10}$};
        \draw[very thick,red,->] (-3.3,-2.535) -- (-1.7,-2);
        \draw[very thick,blue,->] (3.3,-2.535) -- (2.2,-2);
\end{tikzpicture}
\caption{
\emph{Allowed mode-mode correlations, entanglement and probability distributions for Hawking modes.} The probability distribution $\rho(d_+,d_-)$
for different values of $\mu_1$ and $\mu_2$ on the background of a black hole spectrum,
and its integrated weight over the entangled region.
$M_i/m_P$
is $10^{10}$ and 
$\kappa$
is $10^3$. 
[a] 
$\mu_1=3$ and $\mu_2=2$. 
$\rho(d_+,d_-)$ is strongly peaked 
near the origin [insert].
It is well described by the Gaussian 
approximation (Eq.~\protect\ref{eq:rho-gaussian}),
and has almost vanishing support in the entangled region (region bounded by green curve).
[b] 
$\mu_1 = \mu_2= \mu$.
$P_{\mathrm{ent}}$, the  
probability
$\rho(d_+,d_-)$
integrated over the entangled region, as function of $\mu$.
Dashed horizontal line
marks exponential
bound
(Eq.~
\protect\ref{eq:P_ent-asymptotics})
on 
$P_{\mathrm{ent}}$
for two modes of Hawking radiation from a black hole.
[c] 
$\mu_1$ and $\mu_2$ 
are for the two most excited
modes from this black hole 
(corresponding to $\mu_{10}$ and $\mu_{20}$ in main text), both approximately equal to $\mu=4.39\times10^{10}$.
$\rho(d_+,d_-)$,
(here obtained from Eq.~\protect\ref{eq:rho-exact}),
is peaked 
towards the lower right corner.
The probability mass
of the entangled region
(marked in green)
is nevertheless small,
since it
is here but a thin slice located beyond the bulk of the probability mass. 
}
\label{fig:fig3_tripple_fig_black_hole_distribution}
\end{figure*}
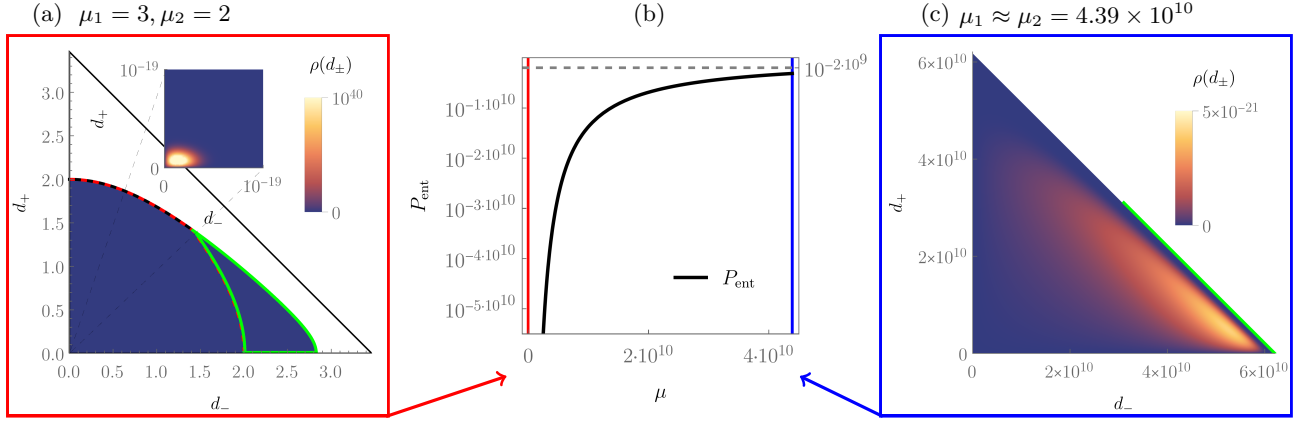

For a large number of modes, these properties typically
lead to a concentration of the probability distribution around $d_+=d_-=0$. 
In this case, the probability density can be approximated by a Gaussian matrix model which is obtained by applying $\exp(\log(\dots))$ to $\rho(d_+,d_-|\hat C)$, and performing a leading order expansion of $\log\Phi$ in $d_+$ and $d_-$. This results in
\begin{equation}\label{eq:rho-gaussian}
    \rho(d_+,d_-|\hat C)\approx4\Lambda_+\Lambda_- d_+d_-\ee^{-\Lambda_+ d_+^2-\Lambda_- d_-^2},
\end{equation}
where $\Lambda_+$ and $\Lambda_-$ are eigenvalues of the bilinear form describing the Gaussian distribution,
which can be written as (for other forms
of the sums, see~\cite{supp}) 
\begin{equation}
\label{eq:Gaussian-eigenvalues}
\Lambda_{\pm}
=
\sum_{j=3}^N\left[
\frac{\mu_1\mu_2 \pm 1}{(\mu_1^2-1)(\mu_2^2-1)}
- \frac{\mu_j^4\mu_1\mu_2\pm \mu_j^2}{(\mu_j^2\mu_1^2-1)(\mu_j^2\mu_2^2-1)}\right].
\end{equation}

\textbf{Correlations in evaporating black hole Hawking radiation.} 
Hawking in~\cite{Hawking74,Hawking75}, showed that in a Schwarzschild black hole spacetime formed from gravitational collapse,
observers at future null infinity $\mathcal{I}^+$ -- far outside the black hole and long after its formation -- observe a quantum field, which at past infinity was in its vacuum state, to be in a thermal state. That is, that the black hole emits thermal radiation at the Hawking temperature $T_H(M)=m_P^2c^2/(8\pi k_B M)$.

Hawking's derivation uses wavepacket modes, 
\begin{equation}
\label{eq:a-def}
a_{nj}=\frac1{\sqrt{\Delta \omega}}\integral\omega{j \Delta \omega}{(j+1) \Delta \omega} \ee^{\ii 2\pi n \omega/\Delta \omega} a_\omega,
\end{equation}
obtained from the continuum  $s$-wave modes $a_\omega$ with positive frequency.
These modes are characterized by their band width $\Delta \omega$
which is a free parameter in Hawking's theory.
The inverse band width $\Delta t=2\pi/ \Delta \omega$ is the width of the mode in time
and increasing $n\to n+1$ shifts the center of the mode by $\Delta t$.
This construction is given in the original papers \cite{Hawking75,Wald75}, 
and for convenience reviewed in~\cite{supp}.

Hawking showed that the expectation value $\mu_{nj}=2\braket{a^\dagger_{nj}a_{nj}}+1$ for late times (large $n$) and sufficiently large frequencies (large $j$) is thermal as in~\eqref{eq:Cn-thermal}. 
In~\cite{Wald75}, Wald showed that $\braket{a^\dagger_{nj}a^\dagger_{nj}}=\braket{a_{nj}a_{nj}}=0$
so that the covariance matrix of one mode has the form~\eqref{eq:Cn-thermal}.
Wald further also showed that for the same time slice (same value of $n$) two-mode correlations vanish \cite{Wald75}. 
Page and others considered non-$s$-wave terms. 
These have the form~\eqref{eq:Cn-thermal} with $(\mu_i-1)$ multiplied by greybody factors. 
These decrease quickly with $l$ such that the radiated power is dominated by  $s$-wave modes.

For the evaporating black hole we posit  that its radiation is as in Hawking's theory with suitable $\Delta t$. 
On the one hand $\Delta t$ must be chosen (much) shorter than the remaining life-time of the black hole, which scales as $\left(\frac{M}{m_P}\right)^3$. On the other hand $\Delta t$ must be longer than the internal time scale of the black hole which different authors have variously suggested to scale as $\left(\frac{M}{m_P}\right)$\cite{Maldacena} or $\left(\frac{M}{m_P}\right)^2$\cite{SekinoSusskind}.
We here opt for the larger $\Delta t$, which we call an \textit{epoch}.
Note that epochs cannot be of constant length, but must become shorter as the black hole loses mass.
Including numerical constants we define the length of epoch $n$ of a black hole
of mass $M_n$ to be the time it takes that black hole to radiate away 
one Planck mass in Hawking's theory, which means 
\begin{equation}
\Delta_n= \kappa  \,t_P(M_n/m_P)^2.
\end{equation}
One epoch corresponds to a slice of Schwarzschild spacetime as indicated in Fig.~\ref{fig.conformal-diagrams}.
The dimensionless constant $\kappa$ depends on the type and number of fields taken into account, and Page estimated it to be on the order of $10^4$ or below~\cite{Page2005}.

To each epoch, we now associate one infinite family 
of modes $a_{nj}$ as above, where $n$ indicates the epoch and $j=0,1,\dots$ for each epoch. 
In particular, we now assume the band widths of the modes to depend on $n$ as $\Delta \omega_n=2\pi/\Delta_n$ so that their  width in time corresponds to the epoch's length.
We assume that these modes have thermal marginals, that is they have the form~\eqref{eq:Cn-thermal} with
\begin{equation}
\begin{split}
	\mu_{nj}& = \coth((j+1)\eta_n),\quad \eta_n=\frac{\hbar \Delta\omega_n}{2T_n k_B}=\frac{8\pi^2 m_P}{\kappa M_n}.
\end{split}
\end{equation}

Using that $\eta_n\ll1$ remains small, the product in~\eqref{eq:rho-exact} and the sum in~\eqref{eq:rho-gaussian} can be approximated by integrals~\cite{supp}.
To this end, in~\eqref{eq:rho-exact} we write 
$\prod_{n,j}\Phi_{nj}=\exp(\sum_{n,j}\log\Phi_{nj})$
and approximate the summation over $j$ by an integral for each epoch $n$. This integral takes a value F which depends only on $\mu_1,\mu_2$ and $d_\pm$ but not on $n$, which is divided by $\eta_n$. Hence the summation over $n$ factorizes out and yields a factor of
\begin{equation}\begin{split}
        \sigma=&\sum_n\frac1{\eta_n} \approx  \frac{\kappa }{16\pi^2}  \frac{M_i^2-M_R^2}{m_P^2},
    \end{split}
\end{equation}
where $M_i$ is the initial mass of the black hole, and $M_R$ is some final mass
entering the problem definition, for instance beyond which Hawking's derivation is no longer valid.
For the following $M_R^2$ can be neglected compared to $M_i^2$.
With this integral approximation the probability distribution reads
\begin{equation}
    \rho(d_+,d_-|\hat C)\! =\!\frac{  d_+d_-\left[(\nu_+^2-1) (\nu_-^2-1)\right]^{-2}  \chi (d_+,d_-) \ee^{\sigma F} }{\prod_{i=1,2} \Phi(\mu_1,\mu_2,C_{(12)},\mu_i)},
\end{equation}
where 
\begin{equation}
    \begin{split}
        F\!=\!&\mathrm{arccoth}^2(\mu_1)+\mathrm{arccoth}^2(\mu_2)\\
        &
-\mathrm{arccoth}^2(\nu_+)-\mathrm{arccoth}^2(\nu_-).
    \end{split}
\end{equation}
Similarly, the integral approximation for the Gaussian approximation~\eqref{eq:rho-gaussian} reads~\cite{supp}
\begin{equation}
\begin{split}
&\Lambda_\pm \approx
 \sigma
 \frac{ \left((\mu_1^2-1 )  \mathrm{arccoth}(\mu_2)\mp\mathrm{arccoth}(\mu_1) (\mu_2^2-1)  \right)}{(\mu_1\mp\mu_2)  \left(\mu_1^2-1\right) \left(\mu_2^2-1\right) }.
\end{split}
\end{equation}  

For most modes in our ansatz, their symplectic eigenvalue $\mu_{nj}$ is not very large and the Gaussian approximation applies.
In particular due to the large numerical prefactor from $\sum_n\eta_n^{-1}$, being proportional to $\kappa\approx 10^4$ and to the square of the initial black hole mass, the probability distribution $\rho(d_+,d_-|\hat C)$ is narrowly concentrated near $d_\pm=0$, \ie at vanishing correlations.
In particular, as seen in Fig.~\ref{fig:fig3_tripple_fig_black_hole_distribution}(a), this means that although for modes with lower $\mu_{nj}$ the entangled area corresponds to a sizeable portion of the state space, the probability distribution is peaked away from entangled states.

For the most highly excited modes in our ansatz the Gaussian approximation does not apply but the probability distribution takes the shape seen in Fig.~\ref{fig:fig3_tripple_fig_black_hole_distribution}(c).
The two most highly excited modes are the modes with $j=0$ from the earliest two epochs, \ie the modes $a_{10}$ and $a_{20}$.
They have symplectic eigenvalues of
$
    \mu_{10}\approx  \kappa M_i/(8\pi^2 m_P),
$
and $
    \mu_{20}\approx  \kappa (M_i-m_P)/(8\pi^2 m_P)
$. 
Considering the probability distribution for these two most highly excited modes, it is useful to express the symplectic eigenvalues of their joint two-mode state relative to $\mu_{10}$, \ie by setting
\begin{equation}
    \nu_\pm=\frac{\kappa M_i}{8\pi^2 m_P} y_\pm.
\end{equation}
By an asymptotic analysis of the exact probability distribution
it can then be shown that the probability distribution for large black hole masses converges to
\begin{equation}
  \rho(d_+,d_-|\hat C)\propto \frac{4\pi^2}{\kappa} \frac{y_+^2-y_-^2}{ y_+^3 y_-^3} \ee^{-\frac{4\pi^2}\kappa \frac{y_+^2+y_-^2}{y_+^2y_-^2}}
\chi(d_+,d_-).
\end{equation}
The shape of the distribution relative to $\mu_{10}$ is
hence strongly dependent
on the non-dimensional constant $\kappa$. This effect can be traced back to our choice of the duration of each epoch, and on the assumption that the marginals of each wavepacket mode with these lengths in time are thermal.
As seen in Fig.~\ref{fig:fig3_tripple_fig_black_hole_distribution}(c), already at a value of $\kappa=10^3$ this distribution peaks at relatively large correlations, and the peak moves further towards the lower right corner when $\kappa$ is increased further.
However, only a negligibly small portion of the probability distribution reaches into the region of entangled states.
This is due to the size of this region relative to the region of all allowed states decreasing rapidly for large $\mu\approx\mu_1\approx\mu_2$, as 
stated above. In fact, one can show that the probability to find an entangled state between the two most excited modes in the radiation of the black hole falls of at least as fast as
\begin{equation}
\label{eq:P_ent-asymptotics}
    P_{\rm ent}= \mathcal{O}\left( \kappa^2 (M_i/m_P)^7 \ee^{-\frac{M_i}{4\pi^2 m_P}} \right).
\end{equation}
For a derivation of this result and a discussion of other asymptotics, see~\cite{supp}

\textbf{Discussion.} We have developed the general theory of constrained symplectic transformations,
resulting in a mathematically consistent picture of random  multimode pure Gaussian states with local modes being thermal. 
We applied the theory to the Hawking radiation. 
Considering any fixed two modes in this picture we have shown the vast majority of pure Gaussian states with thermal marginals have zero entanglement between them. They may still admit quantum correlations as characterized by quantum discord~\cite{ollivier2001quantum,adesso2010quantum}.
The partner mode construction~\cite{hotta2015partner,trevison2018pure,trevison2019spatially,hackl2019minimal} guarantees that in a pure Gaussian state to every mode with a mixed state marginal there exists a partner mode such that the two modes are entangled and in a product state with the remainder of the system. Here, our findings suggest that the partner mode for a Hawking mode must be a potentially complicated combination of all other Hawking modes.
\cor{(In the context of AdS/CFT, see also~\cite{papa,Papadodimas2014}.)}
Quite remarkably, the present result harmonises 
very well with a result of Page who calculated that two-mode correlations due to recoil effect in Hawking radiation, though non-zero, are exceedingly small~\cite{Page1980}.
The present picture may hence suggest that in the search for dynamical mechanisms of black hole  evaporation, the assumption of Gaussianity of the resulting radiation may be a good working hypothesis. 
Independently the theory of constrained  randomised symplectic transformations developed here may be applied in other fields, including in particular the research on thermalisation  of subsystems of a closed quantum or classical system,
see \cite{Popescu2006, Bianchi2022, Baldovin2023}, and references therein.

\textbf{Acknowledgements.}
LH and RHJ thank Eduardo Martin-Martinez for inspiring discussions on the topic.
EA acknowledges support of the Swedish Research Council through grant 2020-04980.
LH~acknowledges support by the Alexander von Humboldt Foundation and by the grant $\#$62312 from the John Templeton Foundation, as part of the \href{https://www.templeton.org/grant/the-quantuminformation-structure-ofspacetime-qiss-second-phase}{‘The Quantum Information Structure of Spacetime’ Project (QISS)}.
RHJ gratefully acknowledges support by the Wenner-Gren Foundations and, in part, by the Wallenberg Initiative on Networks and Quantum Information (WINQ).
Nordita is supported in part by NordForsk.
PH acknowledges support of the National Science Centre in Poland through grant Maestro (2021/42/A/ST2/0035). MK acknowledges support by the
Australian Research Council via Discovery Project grant DP210102887 and the hospitality of RIMS during the workshop ``Random Matrices and Applications'' (5--9 June 2023).
The opinions expressed in this publication are those of the authors and do not necessarily reflect the views of the respective funding organization.

\bibliography{prl}

\end{document}


\title{Supplementary Material: Hawking radiation and random pure Gaussian states}

\author{Erik Aurell}
\email{eaurell@kth.se}
\affiliation{KTH – Royal Institute of Technology, Alba Nova University Center, SE-106 91 Stockholm, Sweden}

\author{Lucas Hackl}
\email{lucas.hackl@unimelb.edu.au}
\affiliation{School of Mathematics and Statistics, The University of Melbourne, Parkville, VIC 3010, Australia}
\affiliation{School of Physics, The University of Melbourne, Parkville, VIC 3010, Australia}

\author{Pawe{\l} Horodecki}
\email{pawel.horodecki@pg.edu.pl}
\affiliation{International Centre for Theory of Quantum Technologies, University of Gdańsk, Wita Stwosza 63, 80-308 Gdańsk, Poland}
\affiliation{Faculty of Applied Physics and Mathematics, National Quantum Information Centre, Gdańsk, University of Technology, Gabriela Narutowicza 11/12, 80-233 Gdańsk, Poland}

\author{Robert H. Jonsson}
\email{robert.jonsson@su.se}
\affiliation{Nordita, Stockholm University and KTH Royal Institute of Technology, Hannes Alfvéns väg 12, SE-106 91 Stockholm, Sweden}

\author{Mario Kieburg}
\email{m.kieburg@unimelb.edu.au}
\affiliation{School of Mathematics and Statistics, The University of Melbourne, Parkville, VIC 3010, Australia}

\maketitle

\tableofcontents

\section{Introduction and summary}\label{sec:SI-introduction-summary}
This Supplementary Information document contains background
material and the details of some of the calculations 
underlying the results reported in the main body of the paper.

The document is organized as follows.
In Section \ref{sec:Gaussian-general} we state the
important result on the Gaussian marginal problem
obtained in \cite{Eisert2008} and discuss in detail
properties of 2-mode Gaussian states which we use in the
analysis. In Section \ref{sec:measure} we construct a measure
on random symplectic transformations constrained by their marginals, and in Section \ref{RMT} we use this compute the 
probability of mode-mode correlations. We also 
combine this with the discussion of the two-mode
case to estimate the probability that two modes are
entangled, given their respective symplectic eigenvalue,
their mode-mode correlation, and the symplectic 
eigenvalues of all the other modes.
In Section \ref{app:bh_calculation} we finally specify 
that the two modes, and all the other modes, are modes
of Hawking radiation. We recall some details of 
Hawking's construction. 

\section{Gaussian marginal problem, and properties of the two-mode Gaussian state}\label{sec:Gaussian-general}

\subsection{Gaussian marginal problem}
Given the set-up described in the main body of the
paper, we have a putative pure Gaussian state on $N$ degrees of freedom and its one-mode marginals. By Gaussianity
both the pure state and each marginal density matrix
are completely specified by the correlation matrix
and its block diagonal submatrices.
Using the symplectic representation of correlation matrices
the Gaussian quantum marginal problem is whether there exists a global Gaussian state with symplectic eigenvalues $\lambda_1,\lambda_2,\ldots,\lambda_N$  such that the symplectic eigenvalues of the diagonal blocks are $\mu_1,\mu_2,\ldots,\mu_N$. 

Eisert, Ty\v{c}, Rudolph and Sanders \cite{Eisert2008}
showed that this problem has the following
solution:  \emph{Given $N$ single mode covariance matrices $C_{(i)}$ with symplectic eigenvalues $\mu_1\leq\dots\leq\mu_N$, there exists an $N$-mode Gaussian state, with symplectic eigenvalues $\lambda_1\leq\dots\leq \lambda_N$, 
the covariance matrix of which has diagonal blocks $C_{(1)},\dots,C_{(n)}$ if and only if (a) $\sum_{i=1}^k \lambda_i \leq
\sum_{i=1}^k \mu_i$ for all $k\leq N$ and (b)
$\mu_N-\sum_{i=1}^{N-1} \mu_i\leq
\lambda_N-\sum_{i=1}^{N-1} \lambda_i$.} 
From this, one can derive an immediate consequence for  globally pure states, \ie when all $\lambda_i$ are equal to one,
as this requires $\mu_N-1\leq \sum_{i=1}^{N-1} (\mu_i-1)$.

\subsection{Two-Mode Gaussian States ($N=2$)}\label{sec:two-mode}

Recall from the main text that we first fix a mode basis labelled by $i=1,\dots, N$. With this in mind, $C_{(i)}$ then refers to the diagonal $2\!\times\!2$-block of the covariance matrix for mode $i$, while $C_{(ij)}$ refers to the respective $2\!\times\!2$-block describing the correlations between mode $i$ and mode $j$. We thus naturally have $C_{(ji)}=C_{(ij)}^\intercal$ due to the symmetry of the covariance matrix.

We here consider general properties of a two-mode bosonic Gaussian state, that may be mixed or pure.
The first question we need to address is what the symplectic eigenvalues are in terms of the matrices $C_{(1)}$, $C_{(2)}$ and $C_{(12)}$. For this purpose, we need to solve the symplectic eigenvalue equation
\begin{equation}\label{sev.eq}
\begin{split}
    0=&\det\left[\begin{array}{cc}
        C_{(1)}-\nu\tau_2 & C_{(12)} \\
        C_{(12)}^\intercal &  C_{(2)}-\nu\tau_2
    \end{array}\right]\\
    =&\det(C_{(1)}-\nu\tau_2)\det(C_{(2)}-\nu\tau_2)\det\left[\mathbf{1}_{2}-\frac{C_{(12)}(\tau_2C_{(1)}\tau_2+\nu\tau_2)C_{(12)}^\intercal(\tau_2C_{(2)}\tau_2+\nu\tau_2)}{\det(C_{(1)}-\nu\tau_2)\det(C_{(2)}-\nu\tau_2)}\right],
\end{split}
\end{equation}
where we have used the simple relation $H^{-1}=\tau_2 H^\intercal\tau_2/\det H$ for an arbitrary invertible $2\!\times\!2$ matrix $H$. For an arbitrary $2\!\times\!2$ matrix we have $\det(\mathbf{1}_2-H)=1-\Tr H+\det H$ which simplifies the expression further. Additionally, we point out the following relations for  the symplectic eigenvalues $\mu_j\geq 1$ of the marginals $C_{(j)}$:
\begin{equation}\label{sev.rel}
    \det C_{(j)}=\mu_j^2\qquad{\rm and}\qquad \det (C_{(j)}-\nu\mathbf{1}_2)=\mu_j^2-\nu^2.
\end{equation}
This simplifies~\eqref{sev.eq} to 
\begin{equation}
    0=\nu^4-(\mu_1^2+\mu_2^2+2\det C_{(12)})\nu^2-\Tr [C_{(12)}\tau_2C_{(2)}\tau_2C_{(12)}^\intercal\tau_2C_{(1)}\tau_2]+\mu_1^2\mu_2^2+\det C_{(12)}^2,
\end{equation}
which leads to the squares of the symplectic eigenvalues
\begin{equation}\label{sev.2mod}
    \nu_\pm^2=\frac{\mu_1^2+\mu_2^2}{2}+\det C_{(12)}\pm\sqrt{\frac{(\mu_1^2-\mu_2^2)^2}{4}+(\mu_1^2+\mu_2^2)\det C_{(12)}+\Tr [C_{(12)}\tau_2C_{(2)}\tau_2C_{(12)}^\intercal\tau_2C_{(1)}\tau_2]}.
\end{equation}
This equation for the symplectic eigenvalues holds true for both mixed and pure bosonic Gaussian states. Particularly, we apply it to the study of modes $1$ and $2$ when we consider a general number $N$ of modes.

As outlined in the main body of the paper
we can simplify the expression~\eqref{sev.2mod} by diagonalising
\begin{equation}
    C_{(j)}=\mu_jS_{(j)}S_{(j)}^\intercal\quad\text{with real symplectic matrices}\ S_{(j)}\in{\rm Sp}(2)
\end{equation}
and
\begin{equation}
    S_{(1)}^{-1}C_{(12)}(S_{(2)}^\intercal)^{-1}=O_1{\rm diag}(c_+,c_-)O_2^\intercal\quad{\rm with}\ O_1,O_2\in{\rm SO}(2),\ 0\leq c_+<\infty\ {\rm and}\ -c_+< c_-\leq c_+.
\end{equation}
The variables $c_+$ and $|c_-|$ are the singular values of $S_{(1)}^{-1}C_{(12)}(S_{(2)}^\intercal)^{-1}$. We allow a sign in $c_-$ to account for the sign of $\det C_{(12)}$ and then choose only special orthogonal matrices $O_1$ and $O_2$ which drop out in the analysis.

As stated in the main body of the paper it is convenient to change to the variables
\begin{equation}\label{def.dpm}
    d_\pm=\frac{1}{\sqrt{2}}(c_+\pm c_-)\in\left[0,\sqrt{2\mu_1\mu_2}\right],
\end{equation}
where the upper bound follows from the positivity of the two-mode correlation matrix minus $\tau_2\otimes\mathbf{1}_2$.
The symplectic eigenvalues $\nu_\pm$ then take the simpler form
\begin{equation}\label{sev.2mod.b}
    \nu_\pm^2=\frac{\mu_1^2+\mu_2^2}{2}+\frac{d_+^2-d_-^2}{2}\pm\sqrt{\frac{(\mu_1^2-\mu_2^2)^2}{4}+(\mu_1^2+\mu_2^2)\frac{d_+^2-d_-^2}{2}+\mu_1\mu_2(d_+^2+d_-^2)}.
\end{equation}

The covariance matrix originates from a bosonic Gaussian states when $\nu_-\geq1$, $\sqrt{2\mu_1\mu_2}\geq d_++d_-\geq |d_+-d_-|\geq 0$, which enforces the positive definiteness of the covariance matrix. Solving the extreme situation $\nu_-=1$ in $d_+$, we find that the eligible region is given by
\begin{equation}\label{region}
\begin{split}
    0\leq &d_+ \leq \mathcal{B}_+(d_-,\mu_1,\mu_2) \quad {\rm and} \quad 0\leq d_- \leq \min\{\sqrt{2(\mu_1-1)(\mu_2+1)},\sqrt{2(\mu_1+1)(\mu_2-1)}\},
\end{split}
\end{equation}
with the functions
\begin{equation}\label{boundary}
\mathcal{B}_\pm(x,\mu_1,\mu_2)=\left\{\begin{array}{cl} \sqrt{x^2+2(\mu_1\mu_2\pm 1)-2\sqrt{(\mu_1\pm\mu_2)^2+2\mu_1\mu_2 x^2}}, & x\leq d_2, \\ 0, & {\rm otherwise},\end{array}\right.
\end{equation}
The function $\mathcal{B}_+$ is plotted (for one numerical
example) as the curve separating valid states in Figure~1 in the main article.
The second function $\mathcal{B}_-$  will be needed below.

The crossing points of the curve $d_+=\mathcal{B}_+(d_-,\mu_1,\mu_2)$ with the axes are
\begin{equation}\label{axes.cross}
\begin{split}
    d_2=&\sqrt{2(\mu_1-1)(\mu_2-1)}\qquad\text{with the $d_+$--axis},\\
    d_3=&\min\{\sqrt{2(\mu_1-1)(\mu_2+1)},\sqrt{2(\mu_1+1)(\mu_2-1)}\}\qquad\text{with the $d_-$--axis},
\end{split}
\end{equation}
and this gives the intersection of the curve and the two axes
in Figure~1 in the main article.

The bounds for the eligible region also implies bounds for larger of the two symplectic eigenvalues $\nu_+$ as it is strictly increasing in $d_+$ and decreasing in $d_-$ so that
\begin{equation}\label{lower.bound.nup}
    \nu_+\geq \nu_+|_{d_+=0,d_-=d_3}=|\mu_1-\mu_2|+1
\end{equation}
and 
\begin{equation}\label{upper.bound.nup}
    \nu_+\leq \nu_+|_{d_+=\mathcal{B}_+(d_-,\mu_1,\mu_2)}\leq \nu_+|_{d_+=d_2,d_-=0}=\mu_1+\mu_2-1.
\end{equation}
For the latter we used the fact that $\mathcal{B}_+(x,\mu_1,\mu_2)$ is decreasing in $x$ in the limits of its argument $x=d_-$ given by the bounds on $d_-$ in \eqref{region}.
These bounds agree with those found by Eisert et al.~\cite{Eisert2008}
 when using the fact $\nu_-\geq 1$ which can be reached for $d_+=\mathcal{B}_+(d_-,\mu_1,\mu_2)$. The theoretical upper bound $\nu_-=\nu_+$  for the second symplectic eigenvalue can never be attained when $\mu_1\neq\mu_2$. It is replaced by~\cite{Eisert2008}
\begin{equation}\label{upper.bound.num.a}
    \nu_-\leq  \nu_+-|\mu_1-\mu_2|,
\end{equation}
where equality is reached along the line $d_+=0$. Therefore, it is
\begin{equation}\label{upper.bound.num.b}
    \nu_-\leq\nu_-|_{d_+=0}\leq\sqrt{\frac{\mu_1^2+\mu_2^2-d_-^2}{2}-\frac{|\mu_1-\mu_2|}{2}\sqrt{(\mu_1+\mu_2)^2-2d_-^2}}\leq\min\{\mu_1,\mu_2\}=\nu_-|_{d_+=d_-=0},
\end{equation}
once again agreeing with the analysis by Eisert et al.~\cite{Eisert2008}.

\subsection{Entangled Two-Mode Gaussian States}\label{sec:entangled-two-mode}
The remaining question we would like to address is when the states are entangled and when they are separable. To do so we anew start for the general case of mixed states as we need it also in the ensuing sections. 

The point is that we consider $1\oplus 1$ bipartition so that Peres-Horodecki criterion is applicable and separability is equivalent with the fact that the partial transposition of the density matrix must be positive definite, see~\cite{werner_wolf_bound_2001,weedbrookGaussianQuantumInformation2012,braunsteinQuantumInformationContinuous2005,serafiniQuantumContinuousVariables2017}. In terms of the quadrature operators, 
standard quantum time reversal of, say, subsystem $2$
is a partial transpose effectuated by  
reflecting the momentum operator $\hat{p}_2\to-\hat{p}_2$.
In the covariance matrix $C$ this means that the following blocks 
are transformed
\begin{equation}\label{part.trans}  
    C_{(2)}\to\tau_3C_{(2)}\tau_3 \qquad{\rm and}\qquad C_{(12)}\to C_{(12)}\tau_3
\end{equation}
while the other blocks remain the same. The diagonalisation of $\tau_3C_{(2)}\tau_3$ is then done by the real symplectic matrix $\tau_3S_{(2)}\tau_3$. Therefore, the singular values of $S_{(1)}^{-1}C_{(12)}\tau_3(\tau_3S_{(2)}^\intercal\tau_3)^{-1}=S_{(1)}^{-1}C_{(12)}(S_{(2)}^\intercal)^{-1}\tau_3$ remain the same namely $c_+$ and $|c_-|$. The only change is in the sign of $c_-\to-c_-$ or, in terms of $d_\pm$,  $d_+$ and $d_-$ are interchanged.

The positivity of the density operator is hence given by the fact that the new symplectic eigenvalues (after partial transpose) $\nu^{PT}_+\geq \nu^{PT}_-$ satisfy the condition $\nu^{PT}_-\geq1$ as the other inequalities are already satisfied. Thus, when the smaller of the two symplectic eigenvalues $\nu^{PT}_-$ drops below $1$ the two-mode state changes from being separable to being entangled. 

As $d_+$ has been interchanged with $d_-$, the symplectic eigenvalues of the transformed covariance matrix are
\begin{equation}\label{sev.2mod.part.trans}
    \left(\nu^{PT}_\pm\right)^2=\frac{\mu_1^2+\mu_2^2}{2}-\frac{d_+^2-d_-^2}{2}\pm\sqrt{\frac{(\mu_1^2-\mu_2^2)^2}{4}-(\mu_1^2+\mu_2^2)\frac{d_+^2-d_-^2}{2}+\mu_1\mu_2(d_+^2+d_-^2)}.
\end{equation}
For given $\mu_1$,
$\mu_2$ and $d_-$
the condition
$\nu^{PT}_-\geq 1$
can be written
$d_+ \leq 
\mathcal{B}_-(d_-\mu_1,\mu_2)$ where we identify
$\mathcal{B}_-$
from \eqref{boundary}.

Therefore, the region where the reduced density matrix of the two modes are separable is given by
\begin{equation}\label{region.separable}
\begin{split}
    \Sigma_{\rm sep}=\biggl\{(d_-,d_+)\biggl|0\leq &d_+ <  \min\left\{\mathcal{B}_+(d_-,\mu_1,\mu_2),\mathcal{B}_-(d_-,\mu_1,\mu_2)\right\}\quad {\rm and}\quad 0\leq d_- \leq \sqrt{\frac{(\mu_1^2-1)(\mu_2^2-1)}{2\mu_1\mu_2}}=d_1\biggl\},
\end{split}
\end{equation}
while when it is entangled it is given by
\begin{equation}\label{region.entangled}
\begin{split}
    \Sigma_{\rm ent}=\biggl\{(d_-,d_+)\biggl|&\mathcal{B}_-(d_-,\mu_1,\mu_2)\leq d_+ \leq \mathcal{B}_+(d_-,\mu_1,\mu_2)\qquad{\rm and}\\
    &d_1=\sqrt{\frac{(\mu_1^2-1)(\mu_2^2-1)}{2\mu_1\mu_2}}\leq d_- \leq \min\{\sqrt{2(\mu_1-1)(\mu_2+1)},\sqrt{2(\mu_1+1)(\mu_2-1)}\}=d_3\biggl\}.
\end{split}
\end{equation}
The two functions $\mathcal{B}_\pm(d_-,\mu_1,\mu_2)$ cross at the point
\begin{equation}\label{d1.def}
    d_+=d_-=d_1=\sqrt{\frac{(\mu_1^2-1)(\mu_2^2-1)}{2\mu_1\mu_2}}.
\end{equation}
The expression is obtained by first observing that
at the crossing-point $d_+=d_-$, and
then solving for $\nu_- = 1$ using $d=d_+=d_-$ in \eqref{sev.2mod.b} or equivalently \eqref{sev.2mod.part.trans}.
The region can be also expressed into the equivalent form
\begin{equation}\label{region.entangled.b}
\begin{split}
    0\leq &d_+\leq \sqrt{\frac{(\mu_1^2-1)(\mu_2^2-1)}{2\mu_1\mu_2}}=d_1\quad {\rm and}\quad \mathcal{B}_+(d_+,\mu_1,\mu_2)\leq d_-\leq \mathcal{B}_-(d_+,\mu_1,\mu_2).
\end{split}
\end{equation}

Two bounds of the symplectic eigenvalues $\nu_\pm$ for entangled two-mode states can be found as
\begin{equation}\label{entangled.bound.nup}
    |\mu_1-\mu_2|+1=\nu_+|_{d_+=0,d_-=d_3}\leq\nu_+\leq \nu_+|_{d_+=d_-=d_1}=\sqrt{\mu_1^2+\mu_2^2-1}
\end{equation}
and
\begin{equation}\label{entangled.bound.num}
    1\leq\nu_-\leq \nu_-|_{d_+=\mathcal{B}_-(d_-,\mu_1,\mu_2)}\leq \nu_-|_{d_+=0,d_-=d_2}=\sqrt{\frac{(\mu_1-\mu_2)^2}{2}+\mu_1+\mu_2-1-|\mu_1-\mu_2|\sqrt{\frac{(\mu_1-\mu_2)^2}{4}+\mu_1+\mu_2-1}}.
\end{equation}
The latter is derived by noticing that $\nu_-$ is decreasing in $d_+$ and that $\nu_-|_{d_+=\mathcal{B}_-(d_-,\mu_1,\mu_2)}$ is strictly increasing in  $d_-$. To show the later one may substitute $d_-=\sqrt{[y^2-(\mu_1-\mu_2)^2]/(2\mu_1\mu_2)}$ and then take the derivative in $y>|\mu_1-\mu_2|$ to see that there is no critical point in $y$ and hence not in $d_->0$. It 
can be checked that $\nu_-|_{d_+=0,d_-=d_2}\leq \min\{\mu_1,\mu_2\}$ as it should be.

\section{Symmetries of the covariance matrix and the Group Invariant Measure}\label{sec:measure}

We consider a bosonic Gaussian states which is described by the covariance matrix
\begin{equation}\label{def.C}
C=SS^\intercal
\end{equation}
with $S\in\mathrm{Sp}(2N)$ a real symplectic matrix, meaning it satisfies 
\begin{equation}\label{sym.S}
S=S^* \qquad {\rm and} \qquad S^{-1}=\hat{\tau}_2 S^\intercal \hat{\tau}_2,
\end{equation}
where $X^*$ is the complex conjugation, $X^\intercal$ is the transposition, and $X^\dagger$ the Hermitian adjoint of a matrix $X$.
We are employing the three Pauli matrices $\tau_j$ with $j=1,2,3$ and embed them into $2N\times 2N$ matrices as follows
\begin{equation}
\hat{\tau}_j=\mathbf{1}_N\otimes\tau_j.
\end{equation}
The symmetries of $S$ carry over to $C$ which satisfies
\begin{equation}\label{sym.C}
C=C^*=C^\intercal \qquad {\rm and} \qquad C^{-1}=\hat{\tau}_2 C \hat{\tau}_2.
\end{equation}
Moreover, $C$ is positive definite so that it lies in the image of the exponential map of the Lie-algebra coset ${\rm sp}(2N)/{\rm u}(N)$, meaning it is
\begin{equation}
C=\exp[\hat{B}]\qquad{\rm with}\ \hat{B}=\left[\begin{array}{cc} B_1 & B_2 \\ B_2 & -B_1 \end{array}\right]=B_1\otimes\tau_3+B_2\otimes\tau_1,
\end{equation}
where $B_1,B_2\in\mathrm{Sym}(N)$ are real symmetric $N\times N$ matrices.

The matrix $\hat{B}$ is actually chiral, meaning it has a $2\times2$ block struture with its diagoal blocks being zero, though it is not written in its chiral basis. 
Let $V\in{\rm U}(2)$ be the unitary matrix that corresponds to the following $2\pi/3$--rotation
\begin{equation}\label{def.V}
V\tau_1V^\dagger=\tau_3,\ V\tau_2V^\dagger=\tau_1,\ V\tau_3V^\dagger=\tau_2. 
\end{equation}
This matrix transforms the quadratures $(q_j,p_j)$ to the ladder opperators $(a_j^\dagger,a_j)$, i.e.,
\begin{equation}
    V^\dagger\left[\begin{array}{c} q_j \\ p_j \end{array}\right]=\left[\begin{array}{c} a^\dagger_j \\ a_j \end{array}\right].
\end{equation}
We embed $V$ in $\mathrm{U}(2N)$ as follows $\hat{V}=\mathbf{1}_N\otimes V$. Then, it becomes clear what the matrix $\hat{B}$ is, namely, \begin{equation}\label{def.B}
\hat{B}=\hat{V}(B_1\otimes\tau_1+B_2\otimes\tau_2)\hat{V}^\dagger=\hat{V}\left[\begin{array}{cc} 0 & B \\ B^\dagger & 0 \end{array}\right]\hat{V}^\dagger
\end{equation}
with $B=B_1-iB_2\in\mathrm{Sym}_{\mathbb{C}}(N)$ a complex symmetric $N\times N$ matrix. The corresponding symmetric matrix space from which $\hat{B}$ is drawn is the symmetric space  CI, \ie the quotient $\mathrm{Sp}(2N,\mathbb{R})/\mathrm{U}(N)$. Indeed, it is on the Lie-algebra level ${\rm sp}(2N)/{\rm u}(N)=i{\rm usp}(2N)/{\rm u}(N)$ where ${\rm usp}(2N)$ is the Lie-algebra of the unitary symplectic group.

This group theoretical consideration implies a natural Riemannian metric on the  coset $\mathrm{Sp}(2N)/\mathrm{U}(N)$ represented as $\exp[i{\rm usp}(2N)/{\rm u}(N)]$ which is the induced Haar metric that is left-invariant under $\mathrm{Sp}(2N)$ and right-invariant under the real representation of $\mathrm{U}(N)$. The non-degenerate invariant length element is
\begin{equation}\label{def.inv.length}
{\rm d}s^2=-\Tr {\rm d}C{\rm d}C^{-1}=-\Tr ({\rm d}C\hat{\tau}_2)^2=-\Tr [S^{-1}{\rm d}S,\hat{\tau}_2]^2
\end{equation}
because of ${\rm d}C\hat{\tau}_2=S[S^{-1}{\rm d}S,\hat{\tau}_2]S^{-1}$ with the commutator $[X,Y]=XY-YX$. This metric corresponds to the volume element (given in terms of a wedge product) \begin{equation}
\begin{split}
{\rm d}\mu_{\mathrm{Sp}(2N)/\mathrm{U}(N)}(S)=&\left(\bigwedge_{a=1}^N\{S^{-1}{\rm d}S\}_{2a,2a}\wedge\{S^{-1}{\rm d}S\}_{2a,2a-1}\right)\wedge\left(\bigwedge_{1\leq a<b\leq N}\{S^{-1}{\rm d}S\}_{2a,2b}\wedge\{S^{-1}{\rm d}S\}_{2a-1,2b}\right),
\end{split}
\end{equation}
where $\{S^{-1}{\rm d}S\}_{a,b}$ represents the $1$-form entry in the $a$-th row and $b$-th column. This measure is not normalised as the coset $\mathrm{Sp}(2N)/\mathrm{U}(N)$ is non-compact. Moreover, it is rather complicated to deal with. 

To simplify it, we first of all multiply the measure with the Haar measure of the unitary group $\mathrm{U}(N)$, and consider $S\in\mathrm{Sp}(2N)$ instead of the coset. This is justified as we are only interested in properties of the covariance matrix $C=SS^\intercal$ (especially the final measure only depends on this combination) which is invariant under the real representation of $\mathrm{U}(N)$. Furthermore, the integration over $\mathrm{U}(N)$ yields a constant which cancels when properly normalising the probability measure. Hence, we consider the group invariant measure
\begin{equation}
\begin{split}
{\rm d}\mu_{\mathrm{Sp}(2N)}(S)=&\left(\bigwedge_{a=1}^N\{S^{-1}{\rm d}S\}_{2a,2a}\wedge\{S^{-1}{\rm d}S\}_{2a,2a-1}\wedge\{S^{-1}{\rm d}S\}_{2a-1,2a}\right)\\
&\hspace*{-2cm}\wedge\left(\bigwedge_{1\leq a<b\leq N}\{S^{-1}{\rm d}S\}_{2a,2b}\wedge\{S^{-1}{\rm d}S\}_{2b,2a}\wedge\{S^{-1}{\rm d}S\}_{2a-1,2b}\wedge\{S^{-1}{\rm d}S\}_{2b-1,2a}\right).
\end{split}
\end{equation}
which follows from the group invariant pseudo-Riemannian metric ${\rm d}s^2=\Tr (S^{-1}{\rm d}S)^2$ on $\mathrm{Sp}(2N)$. This metric is induced via restriction of the group invariant metric of the larger group $\mathrm{Gl}_{\mathbb{R}}(2N)$ which is given by ${\rm d}s^2=\Tr (G^{-1}{\rm d}G)^2$ with $G\in\mathrm{Gl}_{\mathbb{R}}(2N)$. The volume element for the latter one can be readily derived and simplified as follows
\begin{equation}
\begin{split}
{\rm d}\mu_{\mathrm{Gl}_{\mathbb{R}}(2N)}(G)=&\bigwedge_{a,b=1}^{2N}\{G^{-1}{\rm d}G\}_{a,b}=\frac{{\rm d}G}{|\det G|^N},
\end{split}
\end{equation}
where ${\rm d}G$ is the (oriented) product of all differentials. The restriction is then given by a Dirac delta function enforcing that $S\in\mathrm{Sp}(2N)\subset\mathrm{Gl}_{\mathbb{R}}(2N)$ is symplectic so that
\begin{equation}\label{measure}
\begin{split}
{\rm d}\mu_{\mathrm{Sp}(2N)}(S)=\delta(\hat{\tau}_2-S\hat{\tau}_2S^\intercal){\rm d}\mu_{\mathrm{Gl}_{\mathbb{R}}(2N)}(S)=\delta(\hat{\tau}_2-S\hat{\tau}_2S^\intercal){\rm d}S
\end{split}
\end{equation}
because $|\det S|=1$ when $S^{-1}=\hat{\tau}_2S^\intercal\hat{\tau}_2$. The measure~\eqref{measure} will be our reference measure for the ensemble we are going to construct.

\section{The Random Matrix Model with Fixed Marginals}\label{RMT}

As described in the main article and in App.~\ref{app:bh_calculation}, we fix each marginal covariance matrix\begin{equation}
C_{(i)}=\frac{1}{2}\left[\begin{array}{cc} 2\langle \Psi|\hat{q}_{i}\hat{q}_i|\Psi\rangle & \langle \Psi|\hat{q}_i\hat{p}_i+\hat{p}_i\hat{q}_i|\Psi\rangle \\ {} \langle \Psi|\hat{q}_i\hat{p}_i+\hat{p}_i\hat{q}_i|\psi\rangle & 2\langle \Psi|\hat{p}_i\hat{p}_i|\Psi\rangle \end{array}\right],
\end{equation}
for any $i=1,\ldots,N$. To simplify the computation and asymptotic analysis, we assume that they are all distinct. Furthermore, we are interested in the limiting probability distribution of the block $C_{(12)}$, we indeed can consider the correlation between the modes $i=1$ and $i=2$ without loss of generality as we have not chosen explicit form of those.

At this stage, we use a pair ordering of the matrix entries that describes individual degrees of freedom, so our canonical quadratures would be ordered as $(q_1,p_1,\dots,q_N,p_N)$. To define and compute what the conditional probability density of $C_{(12)}$ for given $C_{(i)}$ is, we introduce the $2N\times 2N$ matrices $\widehat{C}=\mathrm{diag}(C_{(1)},\ldots,C_{(N)})$,
\begin{equation}
\delta C=\left[\begin{array}{c|c|c} 0_{2\times 2} & C_{(12)} & 0_{2\times 2(N-2)} \\\hline  C_{(12)}^\intercal & 0_{2\times 2} & 0_{2\times 2(N-2)} \\\hline 0_{ 2(N-2)\times2} & 0_{ 2(N-2)\times2} & 0_{ 2(N-2)\times2(N-2)} \end{array}\right],
\end{equation}
\begin{equation}
\widehat{S}=\mathrm{diag}\left(\left[\begin{array}{cc} \{SS^\intercal\}_{1,1} & \{SS^\intercal\}_{1,2} \\ \{SS^\intercal\}_{2,1} & \{SS^\intercal\}_{2,2} \end{array}\right],\ldots,\left[\begin{array}{cc} \{SS^\intercal\}_{2N-1,2N-1} & \{SS^\intercal\}_{2N-1,2N} \\ \{SS^\intercal\}_{2N,2N-1} & \{SS^\intercal\}_{2N,2N} \end{array}\right]\right),
\end{equation}
and
\begin{equation}
\delta\widehat{S}=\left[\begin{array}{c|c|c} 0_{2\times 2} & \begin{array}{cc} \{SS^\intercal\}_{1,3} & \{SS^\intercal\}_{1,4} \\ \{SS^\intercal\}_{2,3} & \{SS^\intercal\}_{2,4} \end{array} & 0_{2\times 2(N-2)} \\\hline  \begin{array}{cc} \{SS^\intercal\}_{1,3} & \{SS^\intercal\}_{2,3} \\ \{SS^\intercal\}_{1,3} & \{SS^\intercal\}_{2,4} \end{array} & 0_{2\times 2} & 0_{2\times 2(N-2)} \\\hline 0_{ 2(N-2)\times2} & 0_{ 2(N-2)\times2} & 0_{ 2(N-2)\times2(N-2)} \end{array}\right].
\end{equation}
This allows us to write the conditional probability density of $C_{(12)}$ in the very compact form
\begin{equation}
\begin{split}
\rho(C_{(12)}|\widehat{C})=&\frac{\int_{\mathbb{R}^{2N\times 2N}}\delta(\delta C-\delta S)\delta(\widehat{C}-\widehat{S})\delta(\hat{\tau}_2-S\hat{\tau}_2S^\intercal){\rm d}S}{\int_{\mathbb{R}^{2N\times 2N}\times \mathbb{R}^{2\times 2}}\delta(\delta C-\delta S)\delta(\widehat{C}-\widehat{S})\delta(\hat{\tau}_2-S\hat{\tau}_2S^\intercal){\rm d}S\,{\rm d}\delta C}.
\end{split}
\end{equation}

\subsection{Pure two-mode Gaussian states}\label{sec:two-mode-equal-marginals}

As a reference for the ensuing discussion
we consider a pure two-mode bosonic Gaussian state, meaning $N=2$. In this case the two Dirac delta functions $\delta(\delta C-\delta S)\delta(\widehat{C}-\widehat{S})$ already fix all ten matrix  entries so that it is \begin{equation}
    SS^\intercal=\left[\begin{array}{c|c} C_{(1)} & C_{(12)}  \\\hline  C_{(12)}^\intercal & C_{(2)} \end{array}\right]=C.
\end{equation}
Since it also holds $S\hat{\tau}_2S^\intercal=\hat{\tau}_2$, this implies that the two symplectic eigenvalues $\nu_{+}\geq\nu_{-}\geq0$ of $C$ are equal to $1$. Since the von Neumann entropies of two parts of pure state are the same, and since for a Gaussian state of one oscillator the von Neumann entropy is a monotonous function of the symplectic eigenvalue, respectively $\mu_1$ and $\mu_2$, it follows that $\mu_1=\mu_2=\mu$.

Combining the lower bound~\eqref{lower.bound.nup} with $\nu_+=1$ with~\eqref{sev.2mod.b}, we arrive at the two equations
\begin{equation}
 1=\frac{\mu_1^2+\mu_2^2}{2}+\frac{d_+^2-d_-^2}{2} \qquad {\rm and} \qquad 0=\frac{(\mu_1^2-\mu_2^2)^2}{4}+(\mu_1^2+\mu_2^2)\frac{d_+^2-d_-^2}{2}+\mu_1\mu_2(d_+^2+d_-^2)
\end{equation}
under the condition $\mu_1=\mu_2$. Therefore, there is a unique solution
\begin{equation}
\begin{split}
    d_-=\sqrt{2(\mu_1^2-1)}\qquad{\rm and}\qquad d_+=0.
\end{split}
\end{equation}
Summarising the situation for the pure two-mode bosonic Gaussian state, the whole distribution of the matrix $C_{(12)}$ reduces to
\begin{equation}
\begin{split}
    \rho(C_{(12)}|\mu_1)
    =&\int_{[0,2\pi]}\frac{{\rm d}\varphi}{2\pi}\delta\left(C_{(12)}-\sqrt{\mu_1^2-1}\ S_{(1)}\left[\begin{array}{cc}
        \cos(\varphi) & \sin(\varphi) \\
        \sin(\varphi) & -\cos(\varphi)
    \end{array}\right]S_{(2)}^\intercal\right).
\end{split}
\end{equation}
Therefore, for the pure two-mode bosonic Gaussian state we obtain that the two-modes are entangled as expected since then  $d_-=\sqrt{2(\mu_1^2-1)}>\sqrt{2(\mu_1-1)^2}$.

\subsection{The Limit of a Large Number of Modes}

We would like to turn our attention to a large number of modes $N\gg1$. To deal with the Dirac delta functions we introduce Fourier transformations via the auxiliary matrices $\widehat{A}=\mathrm{diag}(A_{(1)},\ldots,A_{(N)})$ with $A_{(i)}\in{\rm Sym}(2)$, 
\begin{equation}
\delta A=\left[\begin{array}{c|c|c} 0_{2\times 2} & A_{(12)} & 0_{2\times 2(N-2)} \\\hline  (A_{(12)})^\intercal & 0_{2\times 2} & 0_{2\times 2(N-2)} \\\hline 0_{ 2(N-2)\times2} & 0_{ 2(N-2)\times2} & 0_{ 2(N-2)\times2(N-2)} \end{array}\right]\quad{\rm with}\ A_{(12)}\in\mathbb{R}^{2\times 2},
\end{equation}
$A=\widehat{A}+\delta A$ and $B=-B^\intercal=B^*\in{\rm Asym}(2N)$. Introducing two real increments $\epsilon,\tilde\epsilon>0$ to guarantee the convergence of the integrals, it is
\begin{equation}
\begin{split}
\rho(C_{(12)}|\widehat{C})=&\lim_{\tilde{\epsilon}\to0}\frac{\int \ee{\Tr(\epsilon\mathbf{1}_{2N}-\ii A)(\widehat{C}+\delta C-SS^\intercal)+\Tr(\hat{\tau}_2-S\hat{\tau}_2S^\intercal)B+\tilde{\epsilon}\Tr(\epsilon\mathbf{1}_{2N}-\ii A)^2+\tilde{\epsilon}\Tr B^2}{\rm d}S{\rm d}A{\rm d}B}{\int \ee{\Tr(\epsilon\mathbf{1}_{2N}-\ii A)(\widehat{C}+\delta C-SS^\intercal)+\Tr(\hat{\tau}_2-S\hat{\tau}_2S^\intercal)B+\tilde{\epsilon}\Tr(\epsilon\mathbf{1}_{2N}-\ii A)^2+\tilde{\epsilon}\Tr B^2}{\rm d}S {\rm d}A{\rm d}B{\rm d}\delta C}.
\end{split}
\end{equation}
We would like to point out that the imaginary unit in the $B$-dependent term is comprised in the second Pauli matrix $\tau_2$ and the limit  of $\epsilon\to0$ is not needed as it automatically drops out when integrating over $A$ and $B$ and taking the limit $\tilde{\epsilon}\to0$.
The integral over $S$ is Gaussian and can be readily carried out, leading to
\begin{equation}
\begin{split}
\rho(C_{(12)}|\widehat{C})
=\lim_{\tilde{\epsilon}\to0}\frac{\int e^{\Tr(\epsilon\mathbf{1}_{2N}-\ii A)(\widehat{C}+\delta C)+\Tr\hat{\tau}_2B+\tilde{\epsilon}\Tr(\epsilon\mathbf{1}_{2N}-\ii A)^2+\tilde{\epsilon}\Tr B^2}\det[(\epsilon\mathbf{1}_{2N}-\ii A)\otimes\mathbf{1}_{2}+B\otimes \tau_2]^{-N/2}\ {\rm d}A {\rm d}B}{\int e^{\Tr(\epsilon\mathbf{1}_{2N}-\ii A)(\widehat{C}+\delta C)+\Tr\hat{\tau}_2B+\tilde{\epsilon}\Tr(\epsilon\mathbf{1}_{2N}-\ii A)^2+\tilde{\epsilon}\Tr B^2}\det[(\epsilon\mathbf{1}_{2N}-\ii A)\otimes\mathbf{1}_{2}+B\otimes\tau_2]^{-N/2}\ {\rm d}A {\rm d}B {\rm d}\delta C}.
\end{split}
\end{equation}
When diagonalising $\tau_2$ and exploiting that $B$ is antisymmetric while $A$ is symmetric and a determinant keeps the same when transposing its argument we can further simplify the expression to
\begin{equation}
\begin{split}
\rho(C_{(12)}|\widehat{C})
=\lim_{\tilde{\epsilon}\to0}\frac{\int e^{\Tr(\epsilon\mathbf{1}_{2N}-\ii A)(\widehat{C}+\delta C)+\Tr\hat{\tau}_2B+\tilde{\epsilon}\Tr(\epsilon\mathbf{1}_{2N}-\ii A)^2+\tilde{\epsilon}\Tr B^2}\det[\epsilon\mathbf{1}_{2N}-\ii A+B]^{-N}\ {\rm d}A {\rm d}B}{\int e^{\Tr(\epsilon\mathbf{1}_{2N}-\ii A)(\widehat{C}+\delta C)+\Tr\hat{\tau}_2B+\tilde{\epsilon}\Tr(\epsilon\mathbf{1}_{2N}-\ii A)^2+\tilde{\epsilon}\Tr B^2}\det[\epsilon\mathbf{1}_{2N}-\ii A+B]^{-N}\ {\rm d}A {\rm d}B {\rm d}\delta C}.
\end{split}
\end{equation}

The interplay of the exponential function and the determinant yields a concentration of the integrand in the large $N$-limit. The asymptotic expansion can be performed by writing
\begin{equation}
\begin{split}
\det[\epsilon\mathbf{1}_{2N}-\ii A+B]^{-N}=\exp\left[-N\Tr\log(\epsilon\mathbf{1}_{2N}-\ii A+B)\right].
\end{split}
\end{equation}
The differentiation in $B$ and $A$ leads to the saddle point equations
\begin{eqnarray}
\hat{\tau}_2&=&\frac{N}{2}\left[(\epsilon\mathbf{1}_{2N}-\ii A+B)^{-1}-(\epsilon\mathbf{1}_{2N}-\ii A-B)^{-1}\right],\label{sad.1}\\
C_{(ab)}&=&\frac{N}{2}\left[\left\{(\epsilon\mathbf{1}_{2N}-\ii A+B)^{-1}\right\}_{(ab)}+\left\{(\epsilon\mathbf{1}_{2N}-\ii A-B)^{-1}\right\}_{(ab)}\right],\label{sad.2}
\end{eqnarray}
respectively, with $(a,b)=(1,1)\ldots,(N,N)$ as well as $(a,b)=(1,2)$. We note that when performing the differentiation one needs to respect the symmetries of $A$ and $B$ and that the term with $\tilde{\epsilon}$ can be neglected as we take the limit $\tilde{\epsilon}\to0$. The notation $\{X\}_{(ab)}$ stands for the $2\!\times\!2$ block of the matrix $X$ at position $(a,b)$.

The square root $\sqrt{\epsilon\mathbf{1}_{2N}-\ii A}$, for which we choose the branch cut along the negative real line, is well-defined for all $A\in{\rm Sym}(2N)$ via the spectral decomposition theorem because the Hermitian part of the matrix $\epsilon\mathbf{1}_{2N}-\ii A$ is positive definite due to $\epsilon>0$. In particular, we can even take the inverse of this matrix which allows us to define the matrices
\begin{equation}
\widetilde{B}=(\epsilon\mathbf{1}_{2N}-\ii A)^{-1/2}B(\epsilon\mathbf{1}_{2N}-\ii A)^{-1/2}\quad{\rm and}\quad D=(\epsilon\mathbf{1}_{2N}-\ii A)^{1/2}\hat{\tau}_2(\epsilon\mathbf{1}_{2N}-\ii A)^{1/2}.
\end{equation}
Then, equation~\eqref{sad.1} takes the simpler form
\begin{equation}
D=\frac{N}{2}\left[(\mathbf{1}_{2N}+\widetilde{B})^{-1}-(\mathbf{1}_{2N}-\widetilde{B})^{-1}\right]=-N\widetilde{B}(\mathbf{1}_{2N}-\widetilde{B}^2)^{-1}
\end{equation}
or equivalently
\begin{equation}
\mathbf{1}_{2N}-\widetilde{B}^2=-N\widetilde{B}D^{-1}.
\end{equation}
This equation shows that $\widetilde{B}$ and $D$ must commute. Thence, exploiting $\epsilon\approx0$, $B$ must share the same block structure as $A$, and the set of saddle point equations can be uniquely solved by linear combination and inversion yielding $B_{(ab)}=0$ for $a\neq b$ and $(a,b)\neq (1,2)$,
\begin{equation}
\begin{split}
B_{(i)}=&\frac{N}{2}\left[(\tau_2-C_{(i)})^{-1}+(\tau_2+C_{(i)})^{-1}\right] \qquad{\rm and}\qquad A_{(i)}=\frac{N}{2i}\left[(\tau_2-C_{(i)})^{-1}-(\tau_2+C_{(i)})^{-1}\right]
\end{split}
\end{equation}
for $i=3,\ldots,N$ and
\begin{equation}
\begin{split}
\left[\begin{array}{cc} B_{(1)} & B_{(12)} \\ -B_{(12)}^\intercal & B_{(2)} \end{array}\right]=&\frac{N}{2}\left(\left[\begin{array}{cc} \tau_2-C_{(1)} &- C_{(12)} \\ -C_{(12)}^\intercal & \tau_2-C_{(2)} \end{array}\right]^{-1}+\left[\begin{array}{cc} \tau_2+C_{(1)} & C_{(12)} \\ C_{(12)}^\intercal & \tau_2+C_{(2)} \end{array}\right]^{-1}\right),\\
\left[\begin{array}{cc} A_{(1)} & A_{(12)} \\ -A_{(12)}^\intercal & A_{(2)} \end{array}\right]=&\frac{N}{2\ii}\left(\left[\begin{array}{cc} \tau_2-C_{(1)} & -C_{(12)} \\ -C_{(12)}^\intercal & \tau_2-C_{(2)} \end{array}\right]^{-1}-\left[\begin{array}{cc} \tau_2+C_{(1)} & C_{(12)} \\ C_{(12)}^\intercal & \tau_2+C_{(2)} \end{array}\right]^{-1}\right).
\end{split}
\end{equation}
We underline that these saddle points can only be reached when $\widehat{C}+\delta C$ and  $\widehat{C}$ are positive definite and their symplectic eigenvalues are larger than $1$. Only then, the matrices $-\ii A\pm B=N(\widehat{C}+\delta C\pm\tau_2)^{-1}$ are positive definite Hermitian. The determinants prevent a deformation of the contour to the saddle point in the limit $N\to\infty$ when it does not satisfy this condition. One can implement the condition of symplectic eigenvalues larger than $1$ and positive definiteness by using the matrix Heaviside step function $\Theta(\widehat{C}+\delta C+\hat{\tau}_2)$ which is $1$ when the matrix $\widehat{C}+\delta C+\hat{\tau}_2$ is positive definite (and thus also $\widehat{C}+\delta C-\hat{\tau}_2=(\widehat{C}+\delta C+\hat{\tau}_2)^\intercal$) and zero else.

We denote the saddle point solution by $A_0$ and $B_0$ and expand in $\sqrt{N}\Delta A$ and $\sqrt{N}\Delta B$ up to the second order. For the determinant we get (up to a constant that will cancel in the ratio of the two integrals)
\begin{equation}
\begin{split}
&{\det}^{-1}\sqrt{(\epsilon\mathbf{1}_{2N}-\ii A)\otimes\mathbf{1}_{2N}+B\otimes\hat{\tau}_2}\\
\propto&\,{\det}^{N/2}(\hat{\tau}_2-\widehat{C}-\delta C){\det}^{N/2}(\hat{\tau}_2+\widehat{C}+\delta C)\exp[\ii\sqrt{N}\Tr (\widehat{C}+\delta C)\Delta A-\sqrt{N}\Tr\hat{\tau}_2\Delta B]\\
&\times\exp\left[\frac{1}{4}\bigl(\Tr[(\hat{\tau}_2+\widehat{C}+\delta C)(-\ii\Delta A+\Delta B)]^2+\Tr[(\hat{\tau}_2-\widehat{C}-\delta C)(-\ii\Delta A-\Delta B)]^2\bigl)\right].
\end{split}
\end{equation}
Remembering that $\hat{\tau}_2+\widehat{C}+\delta C$ and $X=\Delta A+\ii\Delta B$ are Hermitian, the exponential function can be evaluated,
\begin{equation}
\begin{split}
&\exp\left[-\frac{1}{2}\Tr[(\hat{\tau}_2+\widehat{C}+\delta C)(\Delta A+\ii\Delta B)]^2\right]\\
=&\exp\left[-\frac{1}{2}\Tr\left(\left[\begin{array}{cc} \tau_2+C_{(1)} & C_{(12)} \\ C_{(12)}^\intercal & \tau_2+C_{(2)} \end{array}\right]\left[\begin{array}{cc} X_{(1)} & X_{(12)} \\ X_{(12)}^\dagger & X_{(2)} \end{array}\right]\right)^2\right]\\
&\times\prod_{i=3}^N\exp\left[-\frac{1}{2}\Tr[(\tau_2+\widehat{C}_{(i)})X_{(i)}]^2-\Tr\left[\begin{array}{cc} \tau_2& 0 \\0 & \tau_2 \end{array}\right]\left[\begin{array}{c} \Delta B_{(1i)} \\ \Delta B_{(2i)}\end{array}\right]\tau_2\left[\begin{array}{cc} \Delta B_{(1i)}^\intercal, & \Delta B_{(2i)}^\intercal\end{array}\right]\right.\\
&\hspace*{1.5cm}\left.-\Tr\left[\begin{array}{cc} C_{(1)} & C_{(12)} \\ C_{(12)}^\intercal & C_{(2)} \end{array}\right]\left[\begin{array}{c} \Delta B_{(1i)} \\ \Delta B_{(2i)}\end{array}\right]C_{(j)}\left[\begin{array}{cc} \Delta B_{(1i)}^\intercal, & \Delta B_{(2i)}^\intercal\end{array}\right] \right]\\
&\times\prod_{3\leq a<b\leq N}\exp\left[-\Tr \tau_2\Delta B_{(ab)}\tau_2\Delta B_{(ab)}^\intercal-\Tr C_{(a)}\Delta B_{(ab)}C_{(b)}\Delta B_{(ab)}^\intercal\right]
\end{split}
\end{equation}
The integral over the off-diagonal $\Delta B_{(1j)}$, $\Delta B_{(2j)}$, and $\Delta B_{(ab)}$ can be readilly carried out as they are Gaussian integrals over eight-dimensional (for the combined integrals over $\Delta B_{(1j)}$ and $\Delta B_{(2j)}$) and four-dimensional (for $\Delta B_{(ab)}$) real vectors. The integrals over the full $2\!\times\!2$ Hermitian matrices $X_{(j)}$ can be readily performed by substituting $X_{(j)}=(\tau_2+C_{(j)})^{-1/2}Y_{(j)}(\tau_2+C_{(j)})^{-1/2}$ leading to the Jacobian $dX_{(j)}=\det^{-2}(\tau_2+C_{(j)})dY_{(j)}$. We do the same trick for the $4\times 4$ Hermitian matrix 
\begin{equation}
\begin{split}
\left[\begin{array}{cc} X_{(1)} & X_{(12)} \\ X_{(12)}^\dagger & X_{(2)} \end{array}\right]=\left[\begin{array}{cc} \tau_2+C_{(1)} & C_{(12)} \\ C_{(12)}^\intercal & \tau_2+C_{(2)} \end{array}\right]^{-1/2}\left[\begin{array}{cc} Y_{(1)} & Y_{(12)} \\ Y_{(12)}^\dagger & Y_{(2)} \end{array}\right]\left[\begin{array}{cc} \tau_2+C_{(1)} & C_{(12)} \\ C_{(12)}^\intercal & \tau_2+C_{(2)} \end{array}\right]^{-1/2}
\end{split}
\end{equation}
resulting in the Jacobian 
\begin{equation}
dX_{(11)}dX_{(12)}dX_{(22)}={\det}^{-4}\left[\begin{array}{cc} \tau_2+C_{(1)} & C_{(12)} \\ C_{(12)}^\intercal & \tau_2+C_{(2)} \end{array}\right]dY_{(11)}dY_{(12)}dY_{(22)}.
\end{equation}
Summarising everything we arrive at the result
\begin{equation}
\label{eq:famous-70}
\begin{split}
\rho(C_{(12)}|\widehat{C})=&\frac{
\Theta(\widehat{C}+\delta C+\hat{\tau}_2)
}
{Z_N}
\frac{
{\det}^{N-4}
\left[\begin{array}{cc} \tau_2+C_{(1)} & C_{(12)} \\ C_{(12)}^\intercal & \tau_2+C_{(2)} 
\end{array}\right]}{\prod_{j=3}^N
\det\sqrt{
\left[
\begin{array}{cc} \tau_2& 0 \\
0 & \tau_2 \end{array}\right]\otimes\tau_2+
\left[\begin{array}{cc} C_{(1)} & C_{(12)} \\
         C_{(12)}^\intercal & C_{(2)} \end{array}\right]
         \otimes C_{(j)}}}
\end{split}
\end{equation}
with $Z_N$ being the normalisation. We could drop the limit $\tilde{\epsilon}\to0$ as the Heaviside theta function restricts the support on a compact set and thus guaranteeing its integrability.

Next, we express everything in terms of the squared symplectic eigenvalues $\mu_i^2=\det C_{(i)}$ of $C_{(i)}$ and $\nu_\pm$, see~\eqref{sev.2mod.b}.

We may exploit the relations
\begin{equation}
\label{eq:matrix-equalities}
\det(\tau_2+C_{(i)})=\mu_i^2-1,\ \det\sqrt{\tau_2\otimes\tau_2+C_{(a)}\otimes C_{(b)}}=\mu_a^2\mu_b^2-1,\ {\rm and}\ \det\left[\begin{array}{cc} \tau_2+C_{(1)} & C_{(12)} \\ C_{(12)}^\intercal & \tau_2+C_{(2)} \end{array}\right]=(\nu_+^2-1)(\nu_-^2-1).
\end{equation}
The middle equation can be derived by noting that $C_{(a)}$ and $C_{(b)}$ can be separately diagonalised by two symplectic transformations without changing the direct product $\tau_2\otimes\tau_2$. Therefore, it is
\begin{equation}
\label{eq:clarification}
\det\left(\tau_2\otimes\tau_2+C_{(a)}\otimes C_{(b)}\right)
=\det\left[\left(
\begin{array}{cccc}
0 & 0 & 0 & 1 \\
0 & 0 & -1 & 0 \\
0 & -1 & 0 & 0 \\
1 & 0 & 0 & 0 
\end{array}\right)
+ 
\mu_a\mu_b\left(
\begin{array}{cccc}
1 & 0 & 0 & 0 \\
0 & 1 & 0 & 0 \\
0 & 0 & 1 & 0 \\
0 & 0 & 0 & 1 
\end{array}\right)\right]
=\left(\mu_a^2\mu_b^2-1\right)^2
\end{equation}
The Heaviside theta function becomes explicitly
\begin{equation}
    \Theta(\widehat{C}+\delta C+\hat{\tau}_2)=\Theta(\nu_+-1)\Theta(\nu_--1)\Theta(\sqrt{\mu_1\mu_2}\mathbf{1}_{2}-S_{(1)}^{-1}C_{(12)}(S_{(2)}^\intercal)^{-1})\prod_{j=1}^N\Theta(\mu_j-1)
\end{equation}
and thence reflects the accessible domain. When assuming already $\mu_j\geq1$ and $\nu_+\geq \nu_-$ and make use of the singular values $|c_\pm|=(d_+\pm d_-)/\sqrt{2}$, it further simplifies to
\begin{equation}
    \Theta(\widehat{C}+\delta C+\hat{\tau}_2)=\Theta(\nu_--1)\Theta(\sqrt{2\mu_1\mu_2}-d_+-d_-).
\end{equation}
Also the remaining determinants in the denominator can be simplified. For this purpose we denote the symplectic eigenvalues of the $4\times4$ matrix comprising the blocks $C_{(1)}$, $C_{(2)}$, and  $C_{(12)}$ by $\lambda_1$ and $\lambda_2$. Hence, it is
\begin{equation}
\begin{split}
&\det\sqrt{\left[\begin{array}{cc} \tau_2& 0 \\0 & \tau_2 \end{array}\right]\otimes\tau_2+\left[\begin{array}{cc} C_{(1)} & C_{(12)} \\ C_{(12)}^\intercal & C_{(2)} \end{array}\right]\otimes C_{(j)}}=(\nu_+^2\mu_j^2-1)(\nu_-^2\mu_j^2-1)
\end{split}
\end{equation}
The above can be derived by the same method as~\eqref{eq:clarification} above.
Collecting everything, we arrive at the following expression for the probability density
\begin{equation}\label{eq:rho_exact_Phi_factorized}
\begin{split}
&\rho(C_{(12)}|\widehat{C})\propto\frac{\Theta(\nu_--1)\Theta(\sqrt{2\mu_1\mu_2}-d_+-d_-)}{(\nu_+^2-1)^2(\nu_-^2-1)^2}\prod_{j=3}^N\underbrace{\frac{[1+\mu_1^2(\mu_j^2-1)/(\mu_1^2-1)][1+\mu_2^2(\mu_j^2-1)/(\mu_2^2-1)]}{[1+\nu_+^2(\mu_j^2-1)/(\nu_+^2-1)][1+\nu_-^2(\mu_j^2-1)/(\nu_-^2-1)]}}_{\displaystyle=:\Phi\left(\mu_1,\mu_2,C_{(12)},\mu_j\right)}\end{split}
\end{equation}
where we omit a normalization constant which certainly depends on $\mu_1,\dots,\mu_N$. 
The symplectic eigenvalues $\nu_{\pm}$ are seen as functions of the matrix $C_{(12)}$, see~\eqref{sev.2mod}. So far we only needed the assumption that $N\gg1$ to find this asymptotic approximation. The scaling of the diagonal blocks $C_{(i)}$ have been irrelevant in this discussion because we consider $\epsilon\approx0$ to be arbitrarily small.

It is also useful to give the probability distribution in terms of the variables $d_+$ and $d_-$ as the special orthogonal matrices involved naturally drop out in the distribution and therefore only contribute to the normalisation constant. The singular value decomposition of $ S_{(1)}^{-1}C_{(12)}(S_{(2)}^\intercal)^{-1}$ to the singular values $|c_\pm|$ yields a Jacobian which is proportional to $|c_+^2-c_-^2|$, see~\cite{mehta}. This term reads in terms of $d_\pm$ as follows $d_+d_-$. We recall that the Jacobian is the product of the absolute value of all roots in the restricted root system of the corresponding matrix space. For real  $2\!\times\!2$ matrices the root system is given by $|c_+-c_-|$ and $|c_++c_-|$.  

We arrive at the probability distribution
\begin{equation}\label{prob.dens.largeN.dpm}
\begin{split}
&\rho(d_+,d_-|\widehat{C})\propto\frac{\Theta(\nu_--1)\Theta(\sqrt{2\mu_1\mu_2}-d_+-d_-)\ d_+d_-}{ (\nu_+^2-1)^2(\nu_-^2-1)^2}\prod_{j=3}^N\frac{[1+\mu_1^2(\mu_j^2-1)/(\mu_1^2-1)][1+\mu_2^2(\mu_j^2-1)/(\mu_2^2-1)]}{[1+\nu_+^2(\mu_j^2-1)/(\nu_+^2-1)][1+\nu_-^2(\mu_j^2-1)/(\nu_-^2-1)]}.
\end{split}
\end{equation}
This time we understand the symplectic eigenvalues $\nu_\pm$ as a function of $d_\pm$, see~\eqref{sev.2mod.b}. We can actually express the probability density in the symplecctic eigenvalues $\nu_\pm$, only, by noting that
\begin{equation}
    \begin{split}\label{sev.dpm.rel}
        d_+^2-d_-^2=&\nu_+^2+\nu_-^2-\mu_1^2-\mu_2^2\qquad{\rm and}\qquad d_+^2+d_-^2=\frac{(\nu_+^2+\nu_-^2-\mu_1^2-\mu_2^2)^2}{4\mu_1\mu_2}+\frac{\mu_1^2\mu_2^2-\nu_+^2\nu_-^2}{\mu_1\mu_2}.
    \end{split}
\end{equation}
Assuming with restriction of generality that $\mu_1\geq\mu_2$, the eligible domain is then in terms of $\nu_\pm$
\begin{equation}
        \Sigma=\{(\nu_+,\nu_-)|\ 1\leq\nu_-\leq\mu_2 \quad{\rm and}\quad \mu_1-\mu_2+\nu_-\leq \nu_+\leq\mu_1+\mu_2-\nu_-\}.
\end{equation}
The Jacobian for the change of variables $(d_+,d_-)\to(\nu_+,\nu_-)$ is proportional to $\nu_+\nu_-(\nu_+^2-\nu_-^2)/(\mu_1\mu_2 d_+ d_-)$. This allows us to write 
\begin{equation}\label{prob.dens.largeN}
\begin{split}
\rho(\nu_+,\nu_-|\widehat{C})\propto&\frac{\nu_+\nu_-(\nu_+^2-\nu_-^2)}{\displaystyle  (\nu_+^2-1)^2(\nu_-^2-1)^2\prod_{j=3}^N\left[1+\frac{(\mu_j^2-1)(\mu_2^2-\nu_-^2)}{(\mu_j^2\mu_2^2-1)(\nu_-^2-1)}\right]\left[1+\frac{(\mu_j^2-1)(\mu_1^2-\nu_+^2)}{(\mu_j^2\mu_1^2-1)(\nu_+^2-1)}\right]}\chi_{\Sigma}(\nu_+,\nu_-) ,
\end{split}
\end{equation}
where $\chi_{\Sigma}(\nu_+,\nu_-)$ is the indicator function which restricts $\nu_\pm$ to the domain $\Sigma$. Additionally, we have massaged the product over the factors $\Phi(\mu_1,\mu_2,C_{(12)},\mu_j)$ a bit. The region of entanglement can be also expressed in terms of $\nu_\pm$ which becomes
\begin{equation}\label{ent.region.nupm}
\begin{split}
    \Sigma_{\rm ent}=&\left\{(\nu_+,\nu_-)\left| 1\leq \nu_-\leq \mathcal{B}_\nu(\mu_1,\mu_2) \ {\rm and}\  \mu_1-\mu_2+\nu_-\leq \nu_+\leq\sqrt{\frac{2(\mu_1^2+\mu_2^2)}{\nu_-^2+1}-1}\right.\right\}\end{split}
\end{equation}
with
\begin{equation}
    \mathcal{B}_\nu(\mu_1,\mu_2)=\sqrt{\frac{(\mu_1-\mu_2)^2}{4}+\mu_1+\mu_2-1}-\frac{\mu_1-\mu_2}{2},
\end{equation}
which can be computed by setting $\tilde{\nu}_-\geq1$, see~\eqref{sev.2mod.part.trans}, and plugging in~\eqref{sev.dpm.rel}, which leads to $(\nu_+^2+1)(\nu_-^2+1)\leq 2(\mu_1^2+\mu_2^2)$.

\begin{figure}
    \centering
    \includegraphics[height=7cm]{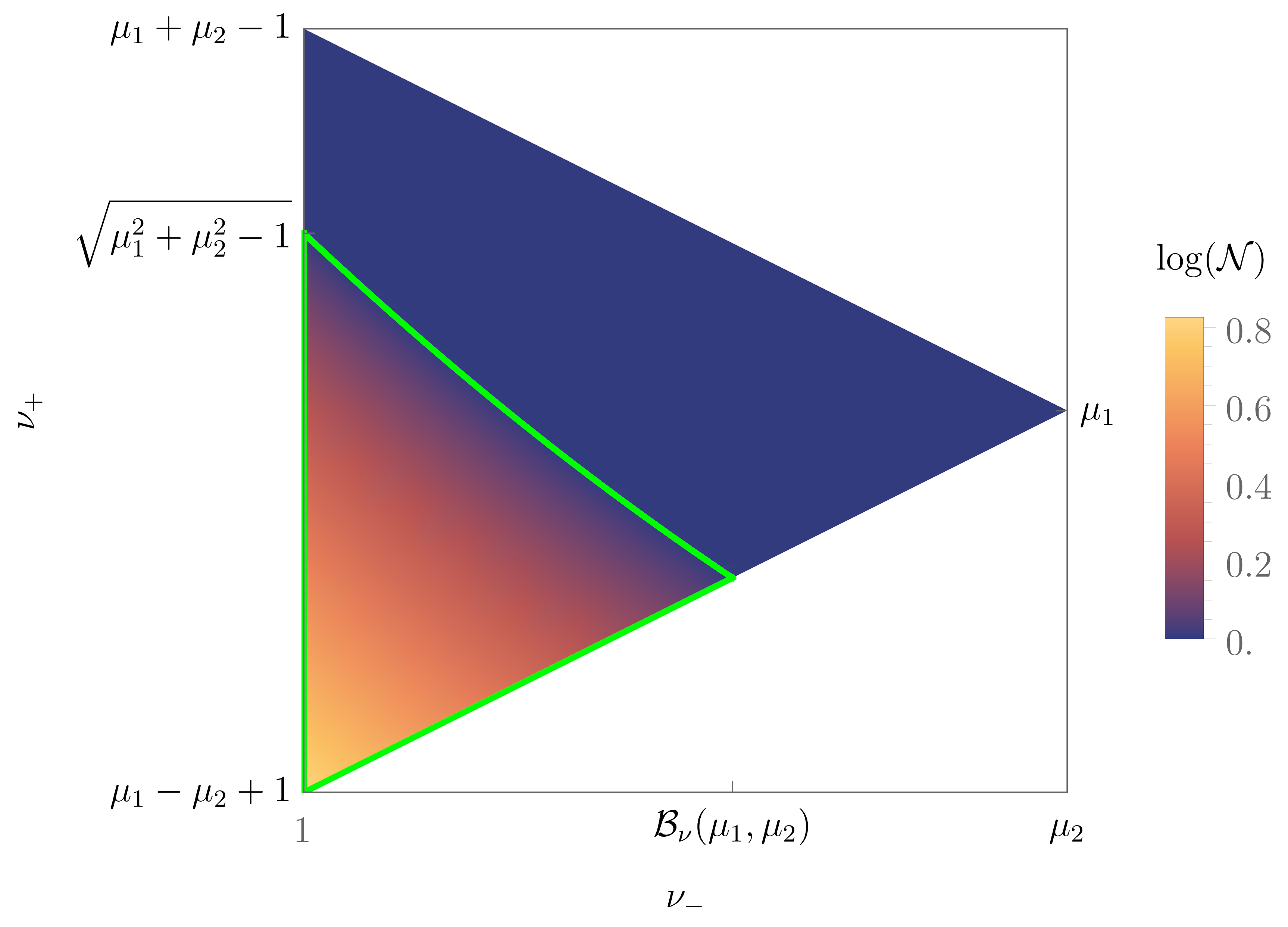}
    \caption{We show the equivalent of Fig.~1 from the main text in the coordinates $\nu_\pm$, where we plot the logarithmic negativity in the allowed region and indicate which part is entangled. Note that the transformation between $d_\pm$ and $\nu_\pm$ is non-linear, such that area sizes are not preserved (when comparing to Fig.~1 from the main text).}
    \label{fig:enter-label}
\end{figure}

Each mode (from $j=3,4,\dots$) contributes by a factor of $\Phi\left(\mu_1,\mu_2,C_{(12)},\mu_j\right)$ to the probability distribution, which have the following properties. Firstly, we have $\Phi\left(\mu_1,\mu_2,C_{(12)},\mu_j=1\right)=1$ which ensures that as expected additional pure modes in the total state do not affect $\rho$. 

Secondly, for $\mu_j>1$, we have $0\leq \Phi\left(\mu_1,\mu_2,C_{(12)},\mu_j\right)\leq 1$. The maximum value is uniquely attained at $C_{(12)}=0$ when $\{\nu_+,\nu_-\}=\{\mu_1,\mu_2\}$ where $\Phi\left(\mu_1,\mu_2,C_{(12)}=0,\mu_j\right)=1$. This can be shown by realising that we can also write
\begin{equation}
    \Phi\left(\mu_1,\mu_2,C_{(12)},\mu_j\right)=\frac{[1+(1-\mu_j^{-2})/(\mu_1^2-1)][1+(1-\mu_j^{-2})/(\mu_2^2-1)]}{[1+(1-\mu_j^{-2})/(\nu_+^2-1)][1+(1-\mu_j^{-2})/(\nu_-^2-1)]},
\end{equation}
which is evidently strictly increasing in $\nu_+$. Thus, it takes for any fixed $\nu_-$ its maximum value at $\nu_+=\mu_1+\mu_2-\nu_-$, i.e.
\begin{equation}
    \Phi\left(\mu_1,\mu_2,C_{(12)},\mu_j\right)|_{\nu_+=\mu_1+\mu_2-\nu_-}=\frac{[1+(1-\mu_j^{-2})/(\mu_1^2-1)][1+(1-\mu_j^{-2})/(\mu_2^2-1)]}{[1+(1-\mu_j^{-2})/((\mu_1+\mu_2-\nu_-)^2-1)][1+(1-\mu_j^{-2})/(\nu_-^2-1)]}.
\end{equation}
Taking the logarithmic derivative in $\nu_-$ leads to
\begin{equation}
\begin{split}
    &\partial_{\nu_-}[\Phi\left(\mu_1,\mu_2,C_{(12)},\mu_j\right)|_{\nu_+=\mu_1+\mu_2-\nu_-}]\\
    =&\frac{[\mu_1+\mu_2-2\nu_-][1-\mu_j^{-2}][(\mu_1+\mu_2-\nu_-)\nu_-([(\mu_1+\mu_2)/2-\nu_-]^2+3[\mu_1+\mu_2]^2/4-1-\mu_j^{-2})-\mu_j^{-2}]}{[1+(1-\mu_j^{-2})/((\mu_1+\mu_2-\nu_-)^2-1)][1+(1-\mu_j^{-2})/(\nu_-^2-1)]}.
\end{split}
\end{equation}
which is strictly positive whenever $\mu_j>1$ and $1\leq\nu_-\leq \mu_2\leq\mu_1$. Therefore, the maximum is only attained at $\nu_-=\mu_2$ and $\nu_+=\mu_1$ corresponding to $C_{(12)}=0$.

Thirdly, the factors of $1/(\nu_+^2-1)$ and $1/(\nu_-^2-1)$ ensure that $\Phi\left(\mu_1,\mu_2,C_{(12)},\mu_j\right)$ vanishes exactly at the boundary of admissible entries of $C_{(12)}$, \ie at the boundary of the support, see~\eqref{boundary}.

Fourthly, in the limit $\mu_j\to\infty$, there is a limiting shape of the factor
\begin{equation}
\lim_{\mu_j\to\infty}\Phi\left(\mu_1,\mu_2,C_{(12)},\mu_j\right)=
\frac{\mu_1^2 \mu_2^2 (\nu_-^2-1)(\nu_+^2-1) }{\nu_+^2\nu_-^2(\mu_1^2-1)(\mu_2^2-1)}
\end{equation}
which allows to derive lower bounds on the probability distribution from the total number of modes, even if the individual $\mu_j$ are not known.

\subsection{A first look at the Gaussian limit}\label{sec:Gaussian-Gaussian}

The full expression of
the conditional probability of a block 
$C_{(12)}$ is given
by
\eqref{eq:famous-70},
which we 
in \eqref{eq:rho_exact_Phi_factorized}
and following equations above have expressed 
in terms of symplectic eigenvalues of
the sub-matrices $C_{(1)}$ ($\mu_1$), $C_{(2)}$
($\mu_2$) and
$C_{(i)}$
($\mu_i$), and linear 
combinations of the singular values of off-diagonal sub-matrix $C_{(12)}$.
A detailed investigation of several limit cases will be carried out in Section~\ref{sec:asymptotic-limits}.
In this section we will take a step back and leave $C_{(12)}$ as a general $2\times 2$-matrix, and consider the important case when $C_{(12)}$ is small. We will specify in Sec.~\ref{sec:asymptotic-limits} what ``small" explicitly means. The outcome will be that the 
conditional probability $\rho(C_{(12)}|\widehat{C})$ is a Gaussian in the matrix $C_{(12)}$, hence 
the section title.

For each of the $N-4$ determinants in the numerator
\eqref{eq:famous-70} we take out the two diagonal blocks, use the first identity
in \eqref{eq:matrix-equalities},
which is
$\det(\tau_2+C_{(i)})=\mu_i^2-1$, and subtract a suitable multiple of the first two rows from the second two rows.
The $4\times 4$ determinant then reduces 
to
\begin{eqnarray}
{\det}\left[\begin{array}{cc} \tau_2+C_{(1)} & C_{(12)} \\ C_{(12)}^\intercal & \tau_2+C_{(2)} \end{array}\right]=\left(\mu_1^2-1\right)
\left(\mu_2^2-1\right)
\det\left[
\mathbf{1}_2 - 
(\tau_2+C_{(1)})^{-1}C_{(12)} (\tau_2+C_{(2)})^{-1}C_{(12)}^\intercal \right]&& 
\end{eqnarray}
involving  $G_{(i)}=(\tau_2+C_{(i)})^{-1}$
(for $i=1,2$) which is reminiscent to a  resolvent.
We use a relation valid for any $2\!\times\!2$ matrix $H$ that 
$\det\left[\mathbf{1}_2-H\right]=1-\Tr H+\det H$,
and the
same first equation in \eqref{eq:matrix-equalities} applied to
resolvents $G_{(1)}$
and $G_{(2)}$. 
Together they give
\begin{eqnarray}
{\det}\left[\begin{array}{cc} \tau_2+C_{(1)} & C_{(12)} \\ C_{(12)}^\intercal & \tau_2+C_{(2)} \end{array}\right]=\left(\mu_1^2-1\right)
\left(\mu_2^2-1\right)
\left(1-
\Tr\left[
(\tau_2+C_{(1)})^{-1}C_{(12)} (\tau_2+C_{(2)})^{-1}C_{(12)}^\intercal
\right]\right)
+ {\det}^2 C_{(12)}
\end{eqnarray}
The last term will drops out 
in the small $C_{(12)}$
limit as it is
of fourth order in the matrix elements of $C_{(12)}$.
Additionally, the trace can be simplified using 
$H^{-1}=\tau_2 H^\intercal\tau_2/\det(H)$, which is also true for every 
invertible $2\times 2$-matrix. This gives
$
(\tau_2+C_{(i)})^{-1}=
(\tau_2C_{(i)}\tau_2-\tau_2)/(\mu_i^2-1)$,
where we can recognize that $\tau_2C_{(i)}\tau_2$,
proportional to $C_{(i)}^{-1}$, is a symmetric matrix. Multiplying out the parenthesis we have
\begin{equation}
    \begin{split}
&\Tr\left[
(\tau_2+C_{(1)})^{-1}C_{(12)} (\tau_2+C_{(2)})^{-1}C_{(12)}^\intercal
\right]\\
=&\frac{1}{(\mu_1^2-1)(\mu_2^2-1)}\Big(\Tr\left[
\tau_2C_{(1)}\tau_2 C_{(12)} \tau_2C_{(2)}\tau_2C_{(12)}^\intercal
\right] 
- \Tr\left[
\tau_2C_{(12)} \tau_2C_{(2)}\tau_2C_{(12)}^\intercal
\right]  \\
&
- \Tr\left[
\tau_2C_{(1)}\tau_2C_{(12)} \tau_2C_{(12)}^\intercal \right]
+ 
\Tr\left[
\tau_2C_{(12)} \tau_2C_{(12)}^\intercal
\right]\Big)
    \end{split}
\end{equation}
The first trace in above is a quadratic function of the matrix elements of $C_{(12)}$ which we 
in this Section will leave unresolved. 
We effectively resolve it when we estimate the function $\Phi$ from
\eqref{eq:rho_exact_Phi_factorized} in Sections~\ref{sec:probability} 
and~\ref{sec:asymptotic-limits}.
The second and the third 
trace involve the product of a symmetric and an antisymmetric matrix and therefore vanish, while the last is
$\Tr \tau_2 C_{(12)} \tau_2 C_{(12)}^\intercal=2\det C_{(12)}$.
Each of the
$N-4$ determinants in the numerator
\eqref{eq:famous-70} is thus to quadratic order in $C_{(12)}$
\begin{eqnarray}
{\det}\left[\begin{array}{cc} \tau_2+C_{(1)} & C_{(12)} \\ C_{(12)}^\intercal & \tau_2+C_{(2)} \end{array}\right]\approx(\mu_1^2-1)(\mu_2^2-1)
-\Tr\left[
\tau_2C_{(1)}\tau_2 C_{(12)} \tau_2C_{(2)}\tau_2C_{(12)}^\intercal
\right] 
- 
\Tr\left[
\tau_2C_{(12)} \tau_2C_{(12)}^\intercal
\right] .
\end{eqnarray}

For each determinant in the denominator of
\eqref{eq:famous-70} we first note that it is the square root of the determinant of an 
$8\times 8$ matrix $\mathbf{M}$.
To resolve it we 
use the fact that $C_{(i)}$ can be diagonalized by a symplectic transformation without changing $\tau_2$. Hence the full matrix splits into four blocks, and its determinant can be written like
\begin{equation}
 \det \mathbf{M} = \det \left(\begin{array}{c|c}
\mu_i\left[\begin{array}{cc} C_{(1)} & C_{(12)} \\ C_{(12)}^\intercal & C_{(2)} \end{array}\right]   &
 (-i)\left[\begin{array}{cc} \tau_2& 0 \\ 0 & \tau_2 \end{array}\right] \\ \hline 
 i\left[\begin{array}{cc} \tau_2& 0 \\0 & \tau_2 \end{array}\right] &
  \mu_i\left[\begin{array}{cc} C_{(1)} & C_{(12)} \\ C_{(12)}^\intercal & C_{(2)} \end{array}\right]
 \end{array}\right)
= \det \left(\begin{array}{c|c}
\mathbf{1}
 &
i\left[\begin{array}{cc} \tau_2& 0 \\0 & \tau_2 \end{array}\right]\mu_i\left[\begin{array}{cc} C_{(1)} & C_{(12)} \\ C_{(12)}^\intercal & C_{(2)} \end{array}\right] 
  \\ \hline
  \mathbf{0}
  &
\mathbf{N}\mathbf{N}^\intercal
\end{array}\right)
\end{equation}
with a new 
$4\times 4$-matrix
\begin{equation}
\mathbf{N}=
\left[\begin{array}{cc} \tau_2& 0 \\0 & \tau_2 \end{array}\right]
+\mu_i
\left[\begin{array}{cc} C_{(1)} &  C_{(12)} \\  C_{(12)}^\intercal & C_{2)} \end{array}\right].
\end{equation}
To reach the above we have 
switched places of the first and last four columns of $\mathbf{M}$ (four permutations), used that $\tau_2^{-1}=\tau_2$
and that
$
\left[\begin{array}{cc} \tau_2& 0 \\0 & \tau_2 \end{array}\right]$
and
$
\left[\begin{array}{cc} C_{(1)} &  C_{(12)} \\  C_{(12)}^\intercal & C_{(2)} \end{array}\right]$
are respectively anti-symmetric
and symmetric,
and made suitable subtractions. 
The determinant of $\mathbf{N}$ is of the same type as the determinants in the numerator, only with all matrix elements in the 2-block scaled by $\mu_i$. The ratio of $N-2$ of these determinants from the numerator and all the determinants in the denominator is thus, up to quadratic order in $C_{(12)}$,
\begin{equation}
    \begin{split}
&\frac{{\det}^{N-2}\left[\begin{array}{cc} \tau_2+C_{(1)} & C_{(12)} \\ C_{(12)}^\intercal & \tau_2+C_{(2)} \end{array}\right]}{\prod_{j=3}^N\det\sqrt{\left[\begin{array}{cc} \tau_2& 0 \\0 & \tau_2 \end{array}\right]\otimes\tau_2+\left[\begin{array}{cc} C_{(1)} & C_{(12)} \\ C_{(12)}^\intercal & C_{(2)} \end{array}\right]\otimes C_{(j)}}}\\
\approx&\prod_{j=3}^N
\frac{(\mu_1^2-1)(\mu_2^2-1)
-\Tr\left[
\tau_2C_{(11)}\tau_2 C_{(12)} \tau_2C_{(22)}\tau_2C_{(12)}^\intercal
\right] 
- 
\Tr\left[
\tau_2C_{(12)} \tau_2C_{(12)}^\intercal
\right]}
{(\mu_j^2\mu_1^2-1)(\mu_j^2\mu_2^2-1)
-\mu_j^4\Tr\left[
\tau_2C_{(1)}\tau_2 C_{(12)} \tau_2C_{(2)}\tau_2C_{(12)}^\intercal
\right] 
- 
\mu_j^2\Tr\left[
\tau_2C_{(12)} \tau_2C_{(12)}^\intercal
\right]}
    \end{split}
\end{equation}
Absorbing combinations  that do not depend on $C_{(12)}$ in a normalization $Z_G$, and neglecting the positivity condition, which is automatically satisfied for small enough $C_{(12)}$, we find in the
the limit of small $C_{(12)}$
\begin{equation}
\begin{split}\label{eq:gauss-app}
\rho(C_{(12)}|\widehat{C})=&\frac{1}{Z_G}
\exp\biggl(\tilde\alpha\Tr \left[\tau_2C_{(12)}\tau_2C_{(12)}^\intercal\right]-\tilde\beta\Tr\left[\tau_2C_{(1)}\tau_2C_{(12)}\tau_2C_{(2)}\tau_2C_{(12)}^\intercal\right]\biggl),
\end{split}
\end{equation}
with
\begin{equation}
\begin{split}
\label{eq:tildealphabeta-SM-new}
\tilde\alpha=&\sum_{j=3}^N
\left[-\frac{1}{(\mu_1^2-1)(\mu_2^2-1)}
+ \frac{\mu_j^2}{(\mu_j^2\mu_1^2-1)(\mu_j^2\mu_2^2-1)}
\right]
=\sum_{j=3}^N
\frac{-(\mu_j^2-1)(\mu_j^2\mu_1^2\mu_2^2-1)}{(\mu_1^2-1)(\mu_2^2-1)(\mu_j^2\mu_1^2-1)(\mu_j^2\mu_2^2-1)}
,\\
\tilde\beta=&\sum_{j=3}^N
\left[\frac{1}{(\mu_1^2-1)(\mu_2^2-1)}
- \frac{\mu_j^4}{(\mu_j^2\mu_1^2-1)(\mu_j^2\mu_2^2-1)}\right]
=
\sum_{j=3}^N
\frac{(\mu_j^2-1)\left(\mu_j^2(\mu_1^2+\mu_2^2)-(\mu_j^2+1)\right)}{(\mu_1^2-1)(\mu_2^2-1)
(\mu_j^2\mu_1^2-1)(\mu_j^2\mu_2^2-1)}
\end{split}
\end{equation}
The inverse covariance matrix
\begin{equation}
\begin{split}
X=-\tilde\alpha\tau_2\otimes\tau_2 +\tilde{\beta} [\tau_2C_{(1)}\tau_2]\otimes[\tau_2C_{(2)}\tau_2]
\end{split}
\end{equation}
has two pairs of 
doubly degenerate eigenvalues equal to
\begin{equation}
\begin{split}
\Lambda_-=\mu_1\mu_2\tilde{\beta}+\tilde\alpha=&
\sum_{j=3}^N
\left[\frac{\mu_1\mu_2-1}{(\mu_1^2-1)(\mu_2^2-1)}
+ \frac{\mu_j^2-\mu_j^4\mu_1\mu_2}{(\mu_j^2\mu_1^2-1)(\mu_j^2\mu_2^2-1)} \right]\\
=&\sum_{j=3}^N
\frac{(\mu_j^2-1)
\left(\mu_j^2(-\mu_1^2\mu_2^2+\mu_1^3\mu_2+\mu_1\mu_2^3-\mu_1\mu_2)-\mu_1\mu_2+1\right)
}{(\mu_1^2-1)(\mu_2^2-1)(\mu_j^2\mu_1^2-1)(\mu_j^2\mu_2^2-1)},
\\
\Lambda_+=\mu_1\mu_2\tilde{\beta}-\tilde\alpha
=&
\sum_{j=3}^N
\left[\frac{\mu_1\mu_2+1}{(\mu_1^2-1)(\mu_2^2-1)}
- \frac{\mu_j^4\mu_1\mu_2+\mu_j^2}{(\mu_j^2\mu_1^2-1)(\mu_j^2\mu_2^2-1)}\right]\\
=&
\sum_{j=3}^N
\frac{(\mu_j^2-1)
\left(\mu_j^2(\mu_1^2\mu_2^2+\mu_1^3\mu_2+\mu_1\mu_2^3-\mu_1\mu_2)-\mu_1\mu_2-1\right)}{(\mu_1^2-1)(\mu_2^2-1)
(\mu_j^2\mu_1^2-1)(\mu_j^2\mu_2^2-1)}.
\end{split}
\end{equation}
This is the same  
as will be obtained below in
the representation using
the singular values of the
$C_{(12)}$
\textit{i.e.}
\eqref{eq:Lambda-159}, with the indentification
$\Lambda_{\pm}=
\mu_1\mu_2\tilde{\beta}\mp\tilde\alpha$. 
In these later more extensive investigation reported in Section~\ref{sec:asymptotic-limits}
we will also determine the eigenvectors
as well as more carefully outline the conditions for the various limits to apply.

\subsection{Asymptotic Formula for the Area of Entangled Two-Mode States (Case $\mu_1=\mu_2=\mu$)}\label{sec:area}

To get an asymptotic expansion for the area of the allowed region in the $d_+$-$d_-$-plane for the simplest case of two equal marginal symplectic eigenvalues $\mu_1=\mu_2=\mu$, we introduce polar coordinates with $d_-=d\cos{\varphi}$ and $d_+=d\sin{\varphi}$. We can translate the constraint~\eqref{region} to be in the eligible region
into a simple equation $d\leq d_{\max}(\varphi)$ with
\begin{align}
    d^2_{\max}(\varphi)=2\mu\sec^2(2\varphi)\left(\mu-\sqrt{2[\mu^2+(\mu^2-2)\cos(2\varphi)]\sin^2(\varphi)}\right)-2\sec(2\varphi)\,.
\end{align}
The entangled region corresponds to the area with $0\leq \varphi\leq\frac{\pi}{4}$ and
\begin{align}
    d_{\max}(\tfrac{\pi}{4}-\varphi)<d\leq d_{\max}(\varphi)\,.
\end{align}
We will compute its area $\mathcal{A}$ in the limit $\mu\to\infty$. This yields
\begin{align}
    \mathcal{A}(\mu)=\int^{\pi/4}_0 d\varphi \underbrace{\frac{d_{\max}^2(\varphi)- d_{\max}^2(\frac{\pi}{2}-\varphi)}{2}}_{\mathrm{Int}(\varphi,\mu)}\,.
\end{align}
We would like to find the asymptotics of $\mathcal{A}(\mu)$ as $\mu\to\infty$. For this, we can first expand the integrand,
\begin{align}
\mathrm{Int}(\varphi,\mu)=4\csc(4\varphi)(1-\sin(2\varphi)+O(1/\mu^2)\,,
\end{align}
whose leading order term is non-integrable around $\varphi=0$ which is only cured by the remaining terms. Therefore, integration and taking the series expansion in the limit $\mu\to\infty$ do not commute here, as we need to separate the integration on the scales $\varphi\gg \mu$ and $\varphi=O(\mu)$. We therefore evaluate the series expansion of $\mathcal{A}(\mu)$ by first computing the integral with the above expansion in the interval $[\frac{r}{\mu},\frac{\pi}{4}]$
\begin{align}
    I_1=\int^{\pi/4}_{r/\mu}\mathrm{Int}(\varphi,\mu)d\varphi=\mathrm{arctanh}\biggl[ \sin (\tfrac{2 r}{\mu })\biggl]+\log\biggl[ \cot (\tfrac{2 r}{\mu })\biggl]-\log (2)\,.
\end{align}
For the remaining interval $[0,\frac{r}{\mu}]$, we expand $\mathrm{Int}$ in terms of $\mu$ for $\varphi=\frac{\tilde{\varphi}}{\mu}$ yielding
\begin{align}
\mathrm{Int}(\tfrac{\tilde{\varphi}}{\mu},\mu)=2(\sqrt{1+\tilde{\varphi}^2}-\tilde{\varphi})\mu-2+O(1/\mu)\,,
\end{align}
whose integral is
\begin{align}
    I_2=\int^r_0\mathrm{Int}(\tfrac{\tilde{\varphi}}{\mu},\mu) \tfrac{d\tilde{\varphi}}{\mu}=\mathrm{arcsinh}(r)-r^2+r\sqrt{1+r^2}-\frac{2r}{\mu}+O(1/\mu^2)\,.
\end{align}
The two integrals combine to give the area $\mathcal{A}(\mu)=I_1+I_2$ and expand first for large $\mu$ and then large $r$ to find
\begin{align}
\begin{split}
    \mathcal{A}(\mu)=I_1+I_2&=\lim_{r\to\infty}\left(r\sqrt{1+r^2}-r^2+\mathrm{arcsinh}(r)-\log(4r)+\log(\mu)\right)+O(1/\mu^2)\\
    &=\log(\mu)+\tfrac{1}{2}-\log(2)+O(1/\mu^2)\,.
\end{split}
\end{align}
We therefore conclude that the region scales as $\mathcal{A}(\mu)\sim \log{\mu}$ in the limit $\mu\to\infty$. In contrast, the total area of the allowed region behaves as $\mu^2/2$, such that the relative fraction scales as $2\log\mu/\mu^2$, which establishes that the relative size of the entangled region vanishes rapidly for large $\mu$.

\subsection{Estimate for the Probability of Entangled Two-mode States}\label{sec:probability}

We would like to estimate the probability that the two modes $1$ and $2$ are entangled for a large number $N\gg1$. As we will see the function $\Phi\left(\mu_1,\mu_2,C_{(12)},\mu_j\right)$ plays a crucial role in this. 

When assuming that $\mu_3>1$ and $\mu_4>1$ are the third and fourth largest eigenvalue, the factor
\begin{equation}
\begin{split}
    &\frac{ d_+d_-}{(\nu_+^2-1)^2(\nu_-^2-1)^2}\Phi\left(\mu_1,\mu_2,C_{(12)},\mu_3\right)\Phi\left(\mu_1,\mu_2,C_{(12)},\mu_4\right)\\
    \leq&\max_{(d_-,d_+)\in\Sigma_{\rm ent}}\left\{ \frac{ d_+d_-(\mu_1^2\mu_3^2-1)(\mu_1^2\mu_4^2-1)(\mu_2^2\mu_3^2-1)(\mu_2^2\mu_4^2-1)}{(\mu_1^2-1)^2(\mu_2^2-1)^2(\nu_+^2\mu_3^2-1)(\nu_+^2\mu_4^2-1)(\nu_-^2\mu_3^2-1)(\nu_-^2\mu_4^2-1)}\right\}\\
    \leq& \frac{ 2\mu_1\mu_2(\mu_1^2\mu_3^2-1)(\mu_1^2\mu_4^2-1)(\mu_2^2\mu_3^2-1)(\mu_2^2\mu_4^2-1)}{(\mu_1^2-1)^2(\mu_2^2-1)^2(\mu_3^2-1)(\mu_4^2-1)(\mu_3^2-1)(\mu_4^2-1)},
\end{split}
\end{equation}
where $\Sigma_{\rm ent}$ was given in~\eqref{region.entangled}, and we exploited that $\nu_\pm\geq1$ and $d_\pm\leq \sqrt{2\mu_1\mu_2}$. Thus, the maximum of the probability density $\rho(d_+,d_-|\hat{C})$ is dictated by the remaining product $\prod_{j=5}^N\Phi\left(\mu_1,\mu_2,C_{(12)},\mu_j\right)$ when $N\gg1$.

As aforementioned, the symplectic eigenvalue $\nu_-$ is strictly decreasing in $d_-$. Since $\Phi\left(\mu_1,\mu_2,C_{(12)},\mu_j\right)$ is strictly increasing in $\nu_->1$, we can bound
\begin{equation}
    \Phi\left(\mu_1,\mu_2,C_{(12)},\mu_j\right)< \Phi\left(\mu_1,\mu_2,C_{(12)},\mu_j\right)|_{d_-=\mathcal{B}_+(d_+,\mu_1,\mu_2)}
\end{equation}
due to~\eqref{region.entangled.b}.  This means that the probability density $\rho(d_+,d_-|\hat{C})$ restricted to the region of entanglement takes its maximum on the boundary $d_-=\mathcal{B}_+(d_+,\mu_1,\mu_2)$ between the region of separability and entanglement  when $N$ is suitably large.

Next we, look for the the maximum of $\Phi\left(\mu_1,\mu_2,C_{(12)},\mu_j\right)$ along the line $d_-=\mathcal{B}_+(d_+,\mu_1,\mu_2)$. When taking the derivative of the denominator of
\begin{eqnarray}\label{Phi.expr.alt}
    \Phi\left(\mu_1,\mu_2,C_{(12)}|_{d_-=\mathcal{B}_+(d_+,\mu_1,\mu_2)},\mu_j\right)=\frac{[1+(1-\mu_j^{-2})/(\mu_1^2-1)][1+(1-\mu_j^{-2})/(\mu_2^2-1)]}{[1+(1-\mu_j^{-2})/(\nu_+^2-1)][1+(1-\mu_j^{-2})/(\nu_-^2-1)]}\biggl|_{d_-=\mathcal{B}_+(d_+,\mu_1,\mu_2)}
\end{eqnarray}
with respect to $d_+$, we obtain
\begin{equation}
\begin{split}
   &\partial_{d_+}\left[1+\frac{1-\mu_j^{-2}}{\nu_+^2|_{d_-=\mathcal{B}_+(d_+,\mu_1,\mu_2)}-1}\right]\left[1+\frac{1-\mu_j^{-2}}{\nu_-^2|_{d_-=\mathcal{B}_+(d_+,\mu_1,\mu_2)}-1}\right]\\
   =&\frac{d_+\mu_1\mu_2(1-\mu_j^{-2})(\mu_1^2+\mu_2^2-1-\mu_j^{-2})}{2\sqrt{(\mu_1+\mu_2)^2+2\mu_1\mu_2 d_+^2}(1+\mu_1\mu_2-\sqrt{(\mu_1+\mu_2)^2+2\mu_1\mu_2 d_+^2})^2}.
\end{split}
\end{equation}
Therefore, the minimum of the denominator and, thus, the maximum of $\Phi\left(\mu_1,\mu_2,C_{(12)}|_{d_-=\mathcal{B}_+(d_+,\mu_1,\mu_2)},\mu_j\right)$ is attained at $d_+=0$. Then, the factor takes the value
\begin{equation}
\begin{split}
    &\Phi\left(\mu_1,\mu_2,C_{(12)}|_{d_+=0,d_-=d_2},\mu_j\right)=\frac{4(\mu_j^2\mu_1^2-1)(\mu_j^2\mu_2^2-1)}{(\mu_1+1)(\mu_2+1)([\mu_j^2(\mu_1+\mu_2-1)-1]^2-\mu_j^2[\mu_1-\mu_2]^2)}
\end{split}
\end{equation}
because $\mathcal{B}_+(0,\mu_1,\mu_2)=d_2$, see~\eqref{axes.cross}.

Finally we estimate the area for the region of entanglement,
\begin{equation}
    \mathcal{A}_{\rm ent}=\int_{\Sigma_{\rm ent}}{\rm d}d_+{\rm d}d_-.
\end{equation}
It is actually contained in the triangle with the corners $(d_-,d_+)=(d_2,0),(d_3,0),(d_1,d_1)$. Thus, we have the simple upper bound
\begin{equation}
    \mathcal{A}_{\rm ent}<\frac{d_1}{2}(d_3-d_2).
\end{equation}
When $\mu_1\gg\mu_2>1$ it behaves like a constant in $\mu_1$
\begin{eqnarray}
  \mathcal{A}_{\rm ent}\overset{\mu_1\gg\mu_2>1}{<}  \frac{1}{2}\sqrt{\frac{(\mu_2+1)(\mu_1-1)^2}{\mu_2}},
\end{eqnarray}
while for the situation when both are large $\mu_1,\mu_2\gg1$ this bound becomes linear
\begin{eqnarray}
  \mathcal{A}_{\rm ent}\overset{\mu_1\gg\mu_2>1}{<}  \frac{1}{2}\min\{\mu_1,\mu_2\}.
\end{eqnarray}
This is actually rather loose and can be improved as we have done for the situation when $\mu_1=\mu_2=\mu$ in Sec.~\ref{sec:two-mode}.

Collecting everything, we find the following upper bound for the probability of entangled modes $1$ and $2$,
\begin{equation}\label{upper.ent}
\begin{split}
    P_{\rm ent}\leq& \frac{d_1(d_3-d_2)}{\hat{Z}_N}\frac{ \mu_1\mu_2(\mu_1^2\mu_3^2-1)(\mu_1^2\mu_4^2-1)(\mu_2^2\mu_3^2-1)(\mu_2^2\mu_4^2-1)}{(\mu_1^2-1)^2(\mu_2^2-1)^2(\mu_3^2-1)(\mu_4^2-1)(\mu_3^2-1)(\mu_4^2-1)}\\
    &\times\prod_{j=5}^N\frac{4(\mu_j^2\mu_1^2-1)(\mu_j^2\mu_2^2-1)}{(\mu_1+1)(\mu_2+1)([\mu_j^2(\mu_1+\mu_2-1)-1]^2-\mu_j^2[\mu_1-\mu_2]^2)}
\end{split}
\end{equation}
There is still the unknown normalisation constant $\hat{Z}_N$. To get rid of it we also need to derive a lower bound for the probability such that mode $1$ and $2$ are two separable modes.

For this purpose, we recall that $\Phi\left(\mu_1,\mu_2,C_{(12)}=0,\mu_j\right)=1$, meaning for $d_+=d_-=0$ which implies $\nu_+=\max\{\mu_1,\mu_2\}$ and $\nu_-=\min\{\mu_1,\mu_2\}$ which lies deeply in the interior of $\Sigma_{\rm sep}$. We construct a lower bound for the product $\prod_{j=3}^N\Phi\left(\mu_1,\mu_2,C_{(12)},\mu_j\right)$ for an area about this point.
To this aim, we consider the simple lower bound
\begin{equation}\label{lower.boud.prob.for.sep}
\begin{split}
    \prod_{j=3}^N\Phi\left(\mu_1,\mu_2,C_{(12)},\mu_j\right)=&\left[\prod_{j=3}^N\left(1+\frac{(\mu_j^2-1)(\mu_1^2-\nu_+^2)}{(\mu_1^2\mu_j^2-1)(\nu_+^2-1)}\right)\left(1+\frac{(\mu_j^2-1)(\mu_2^2-\nu_-^2)}{(\mu_2^2\mu_-^2-1)(\nu_-^2-1)}\right)\right]^{-1}\\
    \geq&\exp\left[-\sum_{j=3}^N\left(\frac{(\mu_j^2-1)(\mu_1^2-\nu_+^2)}{(\mu_1^2\mu_j^2-1)(\nu_+^2-1)}+\frac{(\mu_j^2-1)(\mu_2^2-\nu_-^2)}{(\mu_2^2\mu_-^2-1)(\nu_-^2-1)}\right)\right],
\end{split}
\end{equation}
where we have used the relation $1+x\leq e^x$ which holds true for an arbitrary real number $x$. Here, we have assumed that $\mu_1\geq\mu_2$ to keep the notation in its limit, meaning it is $\nu_+=\mu_1$ and $\nu_-=\mu_2$ for $d_+=d_-=0$. As we want to bound the probability of a separable two-mode state from below and the product over $\Phi\left(\mu_1,\mu_2,C_{(12)},\mu_j\right)$ exhibits a concentration phenomenon, we need to bound the product from below. A simple bound would be
\begin{equation}
    \prod_{j=3}^N\Phi\left(\mu_1,\mu_2,C_{(12)},\mu_j\right)\geq\ee{-2}
\end{equation}
which is realised by the inequalities
\begin{equation} \label{lower.boud.prob.for.sep.b}
\begin{split}
   \sum_{j=3}^N\frac{(\mu_j^2-1)(\mu_1^2-\nu_+^2)}{(\mu_1^2\mu_j^2-1)(\nu_+^2-1)}\leq&1\quad{\rm and}\quad\sum_{j=3}^N\frac{(\mu_j^2-1)(\mu_2^2-\nu_-^2)}{(\mu_2^2\mu_-^2-1)(\nu_-^2-1)}\leq1,
\end{split}
\end{equation}
for which need to find now suitable $\nu_\pm$.
These inequalities can be rewritten to
\begin{equation}
\begin{split}
    \nu_+^2\geq \mu_1^2-\frac{\mu_1^2-1}{1+\sum_{j=3}^N(\mu_j^2-1)/(\mu_1^2\mu_j^2-1)}\quad{\rm and}\quad \nu_-^2\geq \mu_2^2-\frac{\mu_2^2-1}{1+\sum_{j=3}^N(\mu_j^2-1)/(\mu_2^2\mu_j^2-1)}.
\end{split}
\end{equation}

The symplectic eigenvalues $\nu_\pm$ are continuous functions in $d_\pm$, see~\eqref{sev.2mod.b}. For small $d_\pm$ they behave like
\begin{equation}
\begin{split}
    \nu_+^2=& \mu_1^2+\left(1+\frac{1}{(\mu_1-\mu_2)^2}\right)\frac{d_+^2}{2}-\left(1+\frac{1}{(\mu_1+\mu_2)^2}\right)\frac{d_-^2}{2}+\mathcal{O}(d_+^4+d_-^4),\\
    \nu_-^2=& \mu_2^2+\left(1-\frac{1}{(\mu_1-\mu_2)^2}\right)\frac{d_+^2}{2}-\left(1-\frac{1}{(\mu_1+\mu_2)^2}\right)\frac{d_-^2}{2}+\mathcal{O}(d_+^4+d_-^4)
\end{split}
\end{equation}
when $\mu_1>\mu_2>1$.
Therefore, a part of the range over the eligible region will be
\begin{equation}
\begin{split}
    d_+\in&\left[0,D_+\min\left\{\frac{1}{\mu_1+\mu_2},\sqrt{\frac{(\mu_1-\mu_2)^2(\mu_1^2-1)}{1+\sum_{j=3}^N(\mu_j^2-1)/(\mu_1^2\mu_j^2-1)}}\right\}\right],\\
    d_-\in&\biggl[0,\underbrace{D_-\min\left\{\frac{1}{\mu_1+\mu_2},\sqrt{\frac{\mu_2^2-1}{1+\sum_{j=3}^N(\mu_j^2-1)/(\mu_2^2\mu_j^2-1)}}\right\}}_{=:d_-^{\rm(low)}}\biggl]
\end{split}
\end{equation}
with $D_\pm>0$ two functions of $\mu_j$ that are of order $\mathcal{O}(1)$. The minimum with the point $\frac{1}{\mu_1+\mu_2}$, see~\eqref{d1.def}, guarantees that the rectangle spanned by these intervals lies fully in $\Sigma_{\rm sep}$ and have a finite distance to the boundary $d_+=\mathcal{B}_+(d_-,\mu_1,\mu_2)$ and it justifies the Taylor expansion of $\nu_\pm$ in $d_\pm$ in the case that $\mu_1$ and $\mu_2$ are growing in the limit $N\to\infty$.

We notice that the two scales of $d_+$ and $d_-$
 differ when $\mu_1\approx\mu_2$. 
When $\mu_1=\mu_2=\mu$, the two symplectic eigenvalues become
 \begin{equation}
     \nu_\pm=\mu^2+\frac{d_+^2-d_-^2}{2}\pm\sqrt{2}\mu d_+
 \end{equation}
 which shows the proper adjustment for the scale of $d_+$ as follows
\begin{equation}
    d_+\in\biggl[0,\underbrace{D_+\min\left\{\frac{1}{\mu_1+\mu_2},\sqrt{\frac{(\mu_1-\mu_2)^2(\mu_1^2-1)}{1+\sum_{j=3}^N(\mu_j^2-1)/(\mu_1^2\mu_j^2-1)}+\frac{(\mu_1^2-1)^2}{(\mu_1+\mu_2)^2[1+\sum_{j=3}^N(\mu_j^2-1)/(\mu_1^2\mu_j^2-1)]^2}}\right\}}_{=:d_+^{\rm(low)}}\biggl].
\end{equation}

Surely, the rectangle we have constructed is not very explicit due to $D_\pm>0$, though they are of order $1$ which is appropriate for our purposes, nor does it cover the whole region given by~\eqref{lower.boud.prob.for.sep} under intersecting it with $\Sigma_{\rm sep}$. An explicit calculation of the full region defined by this is too involved and unnecessary for our purposes, as we can readily give a lower bound for having mode $1$ and $2$ to be separable which is
\begin{equation}\label{lower.sep}
\begin{split}
    P_{\rm sep}\geq& \frac{\ee{-1}(d_-^{\rm(low)}d_+^{\rm(low)})^2}{4\hat{Z}_N}\min_{(d_-,d_+)\in[0,d_-^{\rm(low)}]\times[0,d_+^{\rm(low)}]}\left\{ \frac{ 1}{(\nu_+^2-1)^2(\nu_-^2-1)^2}\right\},
\end{split}
\end{equation}
where we used
\begin{equation}
    \int_{[0,d_-^{\rm(low)}]\times[0,d_+^{\rm(low)}]}\frac{ d_+d_-{\rm d}d_+{\rm d}d_-}{(\nu_+^2-1)^2(\nu_-^2-1)^2}\geq\min_{(d_-,d_+)\in[0,d_-^{\rm(low)}]\times[0,d_+^{\rm(low)}]}\left\{ \frac{ 1}{(\nu_+^2-1)^2(\nu_-^2-1)^2}\right\}\ \int_{[0,d_-^{\rm(low)}]\times[0,d_+^{\rm(low)}]} d_+d_-{\rm d}d_+{\rm d}d_-.
\end{equation}
The minimum is bounded from below by
\begin{equation}
    \min_{(d_-,d_+)\in[0,d_-^{\rm(low)}]\times[0,d_+^{\rm(low)}]}\left\{ \frac{ 1}{(\nu_+^2-1)^2(\nu_-^2-1)^2}\right\}\geq \frac{1}{(\mu_1+\mu_2)^2(\mu_1+\mu_2-2)^2(\min\{\mu_1^2,\mu_2^2\}-1)^2}
\end{equation}
because of~\eqref{upper.bound.nup} and~\eqref{upper.bound.num.b}. Evidently, it is particularly useful as it is independent of the number $N$ of all modes and and the other symplectic eigenvalues $\mu_{j\geq 3}$.

We combine~\eqref{upper.ent} and~\eqref{lower.sep} to find the following upper bound for the ratio of the probability to have an entangled or a separable two-mode state in the reduced system which is given by
\begin{equation}\label{ratio.prob}
\begin{split}
    \frac{P_{\rm ent}}{P_{\rm sep}}\leq& \frac{4\ee{}\,d_1(d_3-d_2)}{(d_-^{\rm(low)}d_+^{\rm(low)})^2}\frac{ \mu_1\mu_2(\mu_1+\mu_2)^2(\mu_1+\mu_2-2)^2(\min\{\mu_1^2,\mu_2^2\}-1)^2(\mu_1^2\mu_3^2-1)(\mu_1^2\mu_4^2-1)(\mu_2^2\mu_3^2-1)(\mu_2^2\mu_4^2-1)}{(\mu_1^2-1)^2(\mu_2^2-1)^2(\mu_3^2-1)(\mu_4^2-1)(\mu_3^2-1)(\mu_4^2-1)}\\
    &\times\prod_{j=5}^N\frac{4(\mu_j^2\mu_1^2-1)(\mu_j^2\mu_2^2-1)}{(\mu_1+1)(\mu_2+1)([\mu_j^2(\mu_1+\mu_2-1)-1]^2-\mu_j^2[\mu_1-\mu_2]^2)}.
\end{split}
\end{equation}
We recall that $\mu_3$ and $\mu_4$ are the largest and second largest symplectic eigenvalue of $\mu_3,\ldots,\mu_N$. The product leads to an exponential suppression of entangled states when
\begin{equation}\label{condition}
    \lim_{N\to\infty}\sum_{j=5}^N\left[\frac{4(\mu_j^2\mu_1^2-1)(\mu_j^2\mu_2^2-1)}{(\mu_1+1)(\mu_2+1)([\mu_j^2(\mu_1+\mu_2-1)-1]^2-\mu_j^2[\mu_1-\mu_2]^2)}-1\right]=-\infty.
\end{equation}
We note that each single summand is certainly negative. The question is whether it converges to a finite number or not. The latter always happens whenever $\mu_1$ or $\mu_2$ remain larger than $1$ when $N\to\infty$, meaning in a general situation. 
Actually, when $\mu_1$ and $\mu_2$ are of order $\mathcal{O}(1)$ including that they are very close to $1$ a Gaussian approximation of the probability density is possible, see Sec.~\ref{sec:asymptotic-limits} which concentrates about $d_+=d_-=0$ so that even in this case, which we have to exclude here, the probability vanishes to find the two modes entangled.

In the limit $N\to\infty$, we do not distinguish in the kind of double scaling limit where actually also the marginal symplectic eigenvalues $\mu_j=\mu_j^{(N)}$ can be $N$-dependent. Indeed, for each $N\in\mathbb{N}$ one can consider a distribution of marginal symplectic eigenvalues $\diff P_N(\mu)=\sum_{j=1}^N\delta(\mu-\mu_j^{(N)})\diff \mu$. This distribution can be viewed as measures on $[1,\infty)$, with $\int_{[1,\infty)} \diff P_N=N$.
In the limit $N\to\infty$ the sequence of distributions are required to converge to a limiting measure in the measure theoretic sense.
In physical applications, like our treatment of black holes in Sec.~\ref{app:bh_calculation}, it would be typical to have that the support of preceding distributions is contained in the following ones, \ie $\mathrm{supp}\, \diff P_{N-1}\subset  \mathrm{supp}\,\diff P_N$ 
for all $N$. Yet this is not really required, nor do we require that the measure converges to a continuous density or a sum of Dirac delta functions or any other kind of measure.  The limit~\eqref{condition} as well as the discussion in the subsection~\ref{sec:asymptotic-limits} only considers the convergence or divergence of some series which are simply sequences of positive numbers.
For better readability we drop the superscript $(N)$ henceforth and write $\mu_j$ instead of $\mu^{(N)}_j$.

One final comment to this problem, we believe that the bound~\eqref{ratio.prob} we found is actually very loose and one can surely find a tighter one. The question is whether it is possible to find the proper scaling, though one cannot expect something simple as there are two many parameters involved.

\subsection{Asymptotic Limits}
\label{sec:asymptotic-limits}

In this section we give various asymptotic limits of the probability density~\eqref{prob.dens.largeN} in the limit $N\to\infty$ of pure states comprising an infinite number of modes.
As mentioned in the previous subsection, we consider double scaling limits where the number of modes $N$ grows while the marginal symplectic eigenvalues $\mu_j=\mu_j^{(N)}$ might be $N$-dependent. Interestingly, the classification of the asymptotic analysis is independent of the limiting distribution of $\mu_j$. It is only impacted by whether certain sums are finite or diverge and in case of the latter what their order is in terms of $N$ and/or $\mu_1$ and $\mu_2$.

When considering the probability density of $C_{(12)}$ in terms of $\nu_\pm$, see~\eqref{prob.dens.largeN}, we notice that there is a competition between the the asymptotic behaviour of the sum
\begin{equation}\label{sum}
    \alpha_N=\sum_{j=3}^{N}\frac{\mu_j^2-1}{\mu_j^2}
\end{equation}
and the maximal boundary values $\mu_1\pm(\mu_2-\nu_-)$ and $\mu_2$ reached by $\nu_\pm$. First and foremost, we assume that infinitely many of the remaining marginal symplectic eigenvalues do not converge too fast to $\mu_j\to1$, if they do at all. Too fast means that the limit of the above sum is vanishing $\lim_{N\to\infty}\sum_{j=3}^{N}(1-\mu_j^{-2})=0$. 
This condition only means that the whole quantum state in fact becomes separable states in each of the modes corresponding to $\mu_j\to1$ in this limit. Certainly, when this is the case even the discussion in the two subsections~\ref{sec:Gaussian-Gaussian} and~\ref{sec:probability} will fail. 

We first go through the cases when the limit of the sum~\eqref{sum} stays finite. This case is likely to be unphysical in the black hole application as we expect a quasi-continuum of the marginal symplectic eigenvalues in the limit of large black hole masses and, thus, long black hole life times. We note that for a tiny black hole, the modes would always be discrete. A quasi-continuum only arises in the limit of large life times of the black hole.  Mathematically one can indeed choose $\mu_j=1+1/j^\gamma$ with $\gamma>1$ which leads to a finite series.

\begin{enumerate}
    \item 
In the case that
\begin{equation}
    \lim_{N\to\infty}\alpha_N\in(0,\infty),\quad\lim_{N\to\infty}(\mu_1-\mu_2)\in[0,\infty), \quad{\rm and}\quad \lim_{N\to\infty}(\mu_2-1)\in(0,\infty]
\end{equation}
the product cannot be approximated and converges to an infinite product, i.e.,
\begin{equation}\label{asymp.case1}
\begin{split}
\rho(\nu_+,\nu_-|\widehat{C})\propto&\frac{\nu_+\nu_-(\nu_+^2-\nu_-^2)}{\displaystyle  (\nu_+^2-1)^2(\nu_-^2-1)^2\prod_{j=3}^\infty\left[1+\frac{\mu_j^2-1}{\mu_j^2(\nu_-^2-1)}\right]\left[1+\frac{\mu_j^2-1}{\mu_j^2(\nu_+^2-1)}\right]}\chi_{\Sigma}(\nu_+,\nu_-).
\end{split}
\end{equation}
This result holds true regardless how large $\mu_1$ and $\mu_2$ may become as long as they are close enough, meaning their difference is at most of order one.  In particular, the symplectic eigenvalues $\nu_\pm$ will be of order one. When the marginal symplectic eigenvalues $\mu_1$ and $\mu_2$ diverge to infinity while $\mu_1-\mu_2=\mathcal{O}(1)$ (to guarantee that $\nu_+$ is allowed to be of order one) the distribution~\eqref{asymp.case1} becomes independent of them as they only enter via the boundary.

The only simplification in the case $\mu_1=\mu_2=\mu$ is in the domain $\Sigma$ which becomes $\nu_-\leq\nu_+\leq 2\mu-\nu_-$ and $1\leq\nu_-\leq \mu$.

\item
When 
\begin{equation}
    \lim_{N\to\infty}\alpha_N\in(0,\infty),\quad\lim_{N\to\infty}(\mu_1-\mu_2)=\infty, \quad{\rm and}\quad \lim_{N\to\infty}(\mu_2-1)\in(0,\infty)
\end{equation}
the symplectic eigenvalue $\nu_+$ is in leading order equal to $\mu_1-\mu_2\gg1$ while its domain $[\mu_1-\mu_2+\nu_-,\mu_1+\mu_2+\nu_-]$ has the finite length $2(\mu_2-\nu_-)$. Therefore, we can expand in $\nu_+=\mu_1-\mu_2+\nu_-+2\delta\nu_+$ with $\delta\nu_+\in[0,\mu_2-\nu_-]$ is
\begin{equation}\label{asymp.case2}
\begin{split}
\rho(\mu_1-\mu_2+\nu_-+2\delta\nu_+,\nu_-|\widehat{C})\propto&\frac{\nu_-}{\displaystyle  (\nu_-^2-1)^2\prod_{j=3}^\infty\left[1+\frac{\mu_j^2-1}{\mu_j^2(\nu_-^2-1)}\right]}\Theta([\mu_2-\nu_-][\nu_--1])\Theta(\delta\nu_+-\mu_2+\nu_-).
\end{split}
\end{equation}

\item
For 
\begin{equation}
    \lim_{N\to\infty}\alpha_N\in(0,\infty),\quad\lim_{N\to\infty}(\mu_1-\mu_2)=\infty, \quad{\rm and}\quad 1> \lim_{N\to\infty}\frac{\mu_1-\mu_2}{\mu_1+\mu_2}>0.
\end{equation}
the symplectic eigenvalue $\nu_+$ must be now expanded like $\nu_+=(\mu_1+\mu_2)\delta\nu_+$ with $\delta\nu_+\in[(\mu_1-\mu_2)/(\mu_1+\mu_2),1]$ is
\begin{equation}\label{asymp.case3}
\begin{split}
\rho((\mu_1+\mu_2)\delta\nu_+,\nu_-|\widehat{C})\propto&\frac{\nu_-}{\displaystyle \delta\nu_+ (\nu_-^2-1)^2\prod_{j=3}^\infty\left[1+\frac{\mu_j^2-1}{\mu_j^2(\nu_-^2-1)}\right]}\Theta(\nu_--1)\Theta\left([1-\delta\nu_+]\left[\delta\nu_+-\frac{\mu_1-\mu_2}{\mu_1+\mu_2}\right]\right).
\end{split}
\end{equation}
We notice that the $\mu_2$-dependence vanishes in the boundary condition of $\nu_-$. Furthermore, the two symplectic eigenvalues become statistically independent as the distributions factorise.

The previous case can be regained by taking the limit $[\mu_1-\mu_2]/[\mu_1+\mu_2]\to1$ which implies that $1-\delta\nu_+$ will be of order $\mu_2/(\mu_1+\mu_2)\ll1$.

\item 
Finally it can be
\begin{equation}
    \lim_{N\to\infty}\alpha_N\in(0,\infty),\quad\lim_{N\to\infty}(\mu_1-\mu_2)=\infty, \quad{\rm and}\quad \lim_{N\to\infty}\frac{\mu_1-\mu_2}{\mu_1+\mu_2}=0,
\end{equation}
which can be obtained from the previous case by taking the limit $[\mu_1-\mu_2]/[\mu_1+\mu_2]\to0$. As we encounter a logarithmic singularity, the proper expansion is $\nu_+=(\mu_1-\mu_2)^{\delta\nu_+}(\mu_1+\mu_2)^{1-\delta\nu_+}$ which results in
\begin{equation}\label{asymp.case4}
\begin{split}
\rho\left((\mu_1-\mu_2)^{\delta\nu_+}(\mu_1+\mu_2)^{1-\delta\nu_+},\nu_-|\widehat{C}\right)\propto&\frac{\nu_-}{\displaystyle  (\nu_-^2-1)^2\prod_{j=3}^\infty\left[1+\frac{\mu_j^2-1}{\mu_j^2(\nu_-^2-1)}\right]}\Theta(\nu_--1)\Theta\left(1-\delta\nu_+^2\right).
\end{split}
\end{equation}
Since this limit is can be seen as a asymptotic result of the previous case it is not surprisig that also now the two symplectic eigenvalues are statistically independent.

\end{enumerate}

The other cases are given by a diverging limit of the sum~\eqref{sum} which is more interesting for our purposes. For this analysis we introduce the function
\begin{eqnarray}\label{betaN}
    \beta_N(\mu)=\sum_{j=3}^{N}\frac{\mu_j^2-1}{\mu_j^2\mu^2-1}
\end{eqnarray}
to simplify the notation. This function is strictly decreasing in $
\mu$ so that $\beta_N(\mu_1)\leq\beta_N(\mu_2)$ when $\mu_1\geq\mu_2$. It is also more suitable to start from the expression~\eqref{prob.dens.largeN} for the level density in those cases when $\nu_\pm$ will be close to $\mu_1$ and $\mu_2$. Additionally, we would like to point out that 
\begin{equation}\label{inequ.cons1}
    \lim_{N\to\infty}\alpha_N=\lim_{N\to\infty}\sum_{j=3}^N\frac{\mu_j^2-1}{\mu_j^2}=\infty \quad{\rm and}\quad \lim_{N\to\infty}\mu\in[0,\infty) \quad \Rightarrow\quad \lim_{N\to\infty}\beta_N(\mu)=\infty
\end{equation}
which can be reversed to
\begin{equation}\label{inequ.cons2}
    \lim_{N\to\infty}\alpha_N=\lim_{N\to\infty}\sum_{j=3}^N\frac{\mu_j^2-1}{\mu_j^2}=\infty \quad {\rm and}\quad  \lim_{N\to\infty}\beta_N(\mu)\in[0,\infty) \quad \Rightarrow\quad \lim_{N\to\infty}\mu=\infty.
\end{equation}
This is due to the bound of the following difference
\begin{equation}
    \left|\alpha_N-\beta_N(\mu)\right|\leq\max\{|\mu^2-2|,\mu^2-1\}\beta_N(\mu).
\end{equation}

These implications will be helpful in discussing the following cases where we start with $\lim_{N\to\infty}\beta_N(\mu_2)\in[0,\infty)$. Employing the knowledge~\eqref{inequ.cons1} we have learned, we know that $\mu_1,\mu_2\to\infty$ when $N\to\infty$. The limit for $\mu_1$ follows from the fact $\beta_N(\mu_1)\leq\beta_N(\mu_2)$.

\begin{enumerate}
\setcounter{enumi}{4}
    
\item   We consider \label{item:high_excitation_limit}
\begin{equation}
\begin{split}
    \lim_{N\to\infty}\alpha_N=\infty \quad{\rm and}\quad  \lim_{N\to\infty}\beta_N(\mu_2)\in(0,\infty).
\end{split}
\end{equation}
    As $\beta(\mu_2)$ is finite we know from~\eqref{inequ.cons1} that $\mu_2\to\infty$ and, hence, $\mu_1\to\infty$. This, however, implies that $\beta_N(\mu_1)\approx \alpha_N/\mu_1^2$ as well as $\beta_N(\mu_2)\approx \alpha_N/\mu_2^2$. Thus we expand in the variables
\begin{equation}
    \nu_+=\mu_1+\mu_2\delta\nu_+ \qquad{\rm and}\qquad \nu_-=\mu_2\delta\nu_-
\end{equation}
    with
\begin{equation}
    |\delta\nu_+|\leq 1-\delta\nu_-\qquad{\rm and}\qquad 1\geq\delta\nu_-\geq \frac{1}{\mu_2}\overset{ N\to\infty}{\longrightarrow}0.
\end{equation}

    This time we make use of the uniform bound in $\mu_j$
    \begin{equation}
        \frac{\mu_j^2-1}{\mu_j^2\mu_2^2-1}\leq\frac{1}{\mu_2^2}+\frac{1}{\mu_2^2(\mu_2^2-1)}+\frac{1}{\mu_2^2-1}=\mathcal{O}\left(\frac{1}{\mu_2^2}\right)
    \end{equation}    
     to find
    \begin{equation}
        \sum_{j=3}^N\frac{(\mu_j^2-1)^{n+1}(\mu_2^2-\nu_-^2)^{n+1}}{(\mu_j^2\mu_2^2-1)^{n+1}(\nu_-^2-1)^{n+1}}\leq\beta_N(\mu_2)\frac{(\mu_2^2-\nu_-^2)^{n+1}}{(\nu_-^2-1)^{n+1}}\mathcal{O}\left(\frac{1}{\mu_2^{2n}}\right)=\mathcal{O}\left(\frac{1}{\mu_2^{2n}}\right)\overset{n\geq1,\ N\to\infty}{\longrightarrow}0.
    \end{equation}
    Similarly we find such a limit for the term with $\mu_1$ and $\nu_+$. Therefore, the probability density can be approximated like
\begin{equation}\label{asymp.case5}
\begin{split}
\rho\left(\mu_1+\mu_2\delta\nu_+,\mu_2\delta\nu_-|\widehat{C}\right)=&\frac{(\mu_1/\mu_2+\delta\nu_+)^2-\delta\nu_-^2}{Z(\mu_1,\mu_2)(\mu_1/\mu_2+\delta\nu_+)^3\delta\nu_-^3}\exp\left[-\frac{\beta_N(\mu_2)}{\delta\nu_-^2}-\frac{\beta_N(\mu_1)}{(1+\mu_2\delta\nu_+/\mu_1)^2}\right]\Theta(1-\delta\nu_--|\delta\nu_+|)
\end{split}
\end{equation}
with $Z(\mu_1,\mu_2)$ the normalisation. As long as $\beta_N(\mu_2)$ is finite while $\alpha_N\to\infty$, $Z(\mu_1,\mu_2)$ remains finite regardless which relation $\mu_1$ and $\mu_2$ may satisfy.
The integrability is guaranteed by the very same singularities of the prefactors with those in the exponent. In terms of $\nu_\pm$ the probability density is
\begin{equation}\label{asymp.case5.nupm}
\begin{split}
\rho\left(\nu_+,\nu_-|\widehat{C}\right)=&\frac{\alpha_N(\nu_+^2-\nu_-^2)}{Z(\mu_1,\mu_2)\nu_+^3\nu_-^3}\exp\left[-\frac{\alpha_N}{\nu_-^2}-\frac{\alpha_N}{\nu_+^2}\right]\chi_\Sigma(\nu_+,\nu_-).
\end{split}
\end{equation}
We recall that $\mu_1,\mu_2\gg1$ in this case which allows us to replace $\beta_N(\mu_1)=\alpha_N/\mu_1^2$ and  $\beta_N(\mu_2)=\alpha_N/\mu_2^2$ and to omit the term $-1$ in $\nu_\pm^2-1$.

    The expression~\eqref{asymp.case5} concentrates in the corner where the two mode system can be entangled. As we know from the previous section we know that, nonetheless, the probability of having mode $1$ and $2$ to be entangled is exponentially small. We can make a simpler estimate for the current specific case. Combining the bounds~\eqref{entangled.bound.nup} and~\eqref{entangled.bound.num} of $\nu_\pm$ with the fact that $\mu_1\geq\mu_2\gg1$, we know that the area in the $\nu_-$--$\nu_+$ plane where the two modes are entangled must lie in the triangle with the corners $(\nu_-,\nu_+)=(1,\mu_1-\mu_2+1),(\sqrt{\mu_1+\mu_2},\mu_1-\mu_2+\sqrt{\mu_1+\mu_2}),(1,\sqrt{\mu_1^2+\mu_2^2+1})$.  The function
    \begin{equation}
        F(\nu)=\frac{1}{\nu^3}\ee{-\alpha_N/\nu^2}
    \end{equation}
    takes the maximal value $F(\sqrt{2\alpha_N/3})=\ee{-3/2}(2\alpha_N/3)^{3/2}$ and is strictly increasing when $\alpha_N>3\nu^2/2$. Moreover, it is
    \begin{equation}
        \int (\nu_+^2-\nu_-^2)\chi_{\Sigma_{\rm ent}}(\nu_+,\nu_-)d\nu_-d\nu_+=\mathcal{O}\left(\mu_1^{7/2}\right)
    \end{equation}
    because $\nu_-\leq\sqrt{\mu_1+\mu_2}=\mathcal{O}(\sqrt{\mu_1})$ and $\nu_+\leq\sqrt{\mu_1^2+\mu_2^2+1}=\mathcal{O}(\mu_1)$. The bound for $\nu_-$ also implies $\alpha_N\gg3\nu_-^2/2$ and that we can bound $F(\nu_-)$ by its value at $\nu_-=\sqrt{\mu_1+\mu_2}$, while for $F(\nu_+)$ we take the bound $F(\sqrt{2\alpha_N/3})$. Collecting everything the probability to have an entangled two-mode state for the modes $1$ and $2$ is at most of order
    \begin{equation}\label{eq:cse5.Pent.order}
        P_{\rm ent}=\mathcal{O}\left(\mu_1^{2}\alpha_N^{5/2}\exp\left[-\frac{\alpha_N}{\mu_1+\mu_2}\right]\right)=\mathcal{O}\left(\beta_N^{5/2}(\mu_1)\mu_1^{7}\exp\left[-\beta_N(\mu_1)\frac{\mu_1^2}{\mu_1+\mu_2}\right]\right).
    \end{equation}
    The exponent diverges to $-\infty$ because $\beta_N(\mu_1)=\alpha_N/\mu_1^2$ and $\beta_N(\mu_2)=\alpha_N/\mu_2^2$ remain finite.

    Anew, there might be some simplifications of~\eqref{asymp.case5} in specific sub-limits. For instance when $\beta(\mu_1)\to0$, which  is possible as we only restrict $\beta_N(\mu_2)>0$, we find 
\begin{equation}\label{asymp.case5a}
\begin{split}
\rho\left(\mu_1+\mu_2\delta\nu_+,\mu_2\delta\nu_-|\widehat{C}\right)\propto&\frac{(\mu_1/\mu_2+\delta\nu_+)^2-\delta\nu_-^2}{(\mu_1/\mu_2+\delta\nu_+)^3\delta\nu_-^3}\exp\left[-\frac{\beta_N(\mu_2)}{\delta\nu_-^2}\right]\Theta(1-\delta\nu_--|\delta\nu_+|),
\end{split}
\end{equation}
which is still integrable. When $\mu_1\gg \mu_2$ it becomes
\begin{equation}\label{asymp.case5b}
\begin{split}
\rho\left(\mu_1+\mu_2\delta\nu_+,\mu_2\delta\nu_-|\widehat{C}\right)\propto&\frac{1}{\delta\nu_-^3}\exp\left[-\frac{\beta_N(\mu_2)}{\delta\nu_-^2}\right]\Theta(1-\delta\nu_--|\delta\nu_+|),
\end{split}
\end{equation}
which is particularly simple despite that it is evidently non-Gaussian in $d_\pm$. Also the case $\mu_1=\mu_2=\mu$ is possible with the probability density
\begin{equation}\label{asymp.case5c}
\begin{split}
\rho\left(\mu+\mu\delta\nu_+,\mu\delta\nu_-|\widehat{C}\right)\propto&\frac{(1+\delta\nu_+)^2-\delta\nu_-^2}{(1+\delta\nu_+)^3\delta\nu_-^3}\exp\left[-\frac{\beta_N(\mu)}{\delta\nu_-^2}-\frac{\beta_N(\mu)}{(1+\delta\nu_+)^2}\right]\Theta(1-\delta\nu_--|\delta\nu_+|).
\end{split}
\end{equation}
    
    \item   The results for the limit
\begin{equation}
\begin{split}
    \lim_{N\to\infty}\alpha_N=\lim_{N\to\infty}\frac{\mu_1-\mu_2}{\sqrt{\alpha_N}}=\infty, \quad \lim_{N\to\infty}\frac{\mu_1-\mu_2}{\mu_2}\in(0,\infty], \quad{\rm and}\quad  \lim_{N\to\infty}\beta_N(\mu_2)=0.
\end{split}
\end{equation}
    can be derived along the lines of the previous case. We modify the double scaling of the symplectic eigenvalues $\nu_\pm$ as follows
\begin{equation}
    \nu_+=\mu_1+\mu_2\delta\nu_+ \qquad{\rm and}\qquad \nu_-=\mu_2\sqrt{\beta_N(\mu_2)}\delta\nu_-
\end{equation}
    with
\begin{equation}
    |\delta\nu_+|\leq 1-\sqrt{\beta_N(\mu_2)}\delta\nu_-\overset{ N\to\infty}{\longrightarrow}1 \qquad{\rm and}\qquad \frac{1}{\sqrt{\beta_N(\mu_2)}}\geq\delta\nu_-\geq \frac{1}{\mu_2\sqrt{\beta_N(\mu_2)}}.
\end{equation}
    We note that it is once again $\mu_1,\mu_2\to\infty$ due to~\eqref{inequ.cons1} so that $\beta_N(\mu_2)\approx\alpha_N/\mu_2^2$ but it is still $\mu_2\sqrt{\beta_N(\mu_2)}\approx\sqrt{\alpha_N}\to\infty$. Therefore, the domain simplifies in the limit $N\to\infty$ to
\begin{equation}
    |\delta\nu_+|\leq 1 \qquad{\rm and}\qquad \delta\nu_-\geq0.
\end{equation}
    
    The above parametrisation works out as $\beta_N(\mu_2)\to0$ also enforces $0\leq\beta_N(\mu_1)\leq\beta_N(\mu_2)\to0$. 
    Thence, the upper bound of the terms in the Taylor series of $\sum_{j=3}^N\log(\Phi(\mu_1,\mu_2,C_{(12)},\mu_j)$ look very similar to the previous case. In particular the probability density becomes
\begin{equation}\label{asymp.case6}
\begin{split}
\rho\left(\mu_1+\mu_2\delta\nu_+,\mu_2\delta\nu_-|\widehat{C}\right)=&\frac{2}{\log[(\mu_1+\mu_2)/(\mu_1-\mu_2)]}\frac{1}{(\mu_1/\mu_2+\delta\nu_+)\delta\nu_-^3}\exp\left[-\frac{1}{\delta\nu_-^2}\right]\Theta(1-|\delta\nu_+|).
\end{split}
\end{equation}
    This is a case where the two symplectic eigenvalues $\nu_\pm$ are asymptotically statistically independent.

    This limit comprises an obvious simplifications when $\mu_1\gg\mu_2$ when we find
\begin{equation}\label{asymp.case6.a}
\begin{split}
\rho\left(\mu_1+\mu_2\delta\nu_+,\mu_2\delta\nu_-|\widehat{C}\right)=&\frac{2}{\delta\nu_-^3}\exp\left[-\frac{1}{\delta\nu_-^2}\right]\Theta(1-|\delta\nu_+|).
\end{split}
\end{equation}
    
    \item   One would expect that the scaling
\begin{equation}
\begin{split}
    \lim_{N\to\infty}\alpha_N=\infty,\quad \lim_{N\to\infty}\frac{\mu_1-\mu_2}{\sqrt{\alpha_N}}\in(0,\infty], \quad{\rm and}\quad  \lim_{N\to\infty}\beta_N(\mu_2)=\lim_{N\to\infty}\frac{\mu_1-\mu_2}{\mu_2}=0
\end{split}
\end{equation}
looks almost similar, however, the probability distribution~\eqref{asymp.case7} develops a logarithmic singularity. Therefore, the proper scaling is
\begin{equation}
    \nu_+=(\mu_1-\mu_2)^{1-\delta\nu_+}(\mu_1+\mu_2)^{\delta\nu_+} \qquad{\rm and}\qquad \nu_-=\mu_2\sqrt{\beta_N(\mu_2)}\delta\nu_-
\end{equation}
    with
\begin{equation}
\begin{split}
    &0\leq\frac{\log\bigl[\bigl(\mu_1-\mu_2+\mu_2\sqrt{\beta_N(\mu_2)}\delta\nu_-\bigl)/(\mu_1-\mu_2)\bigl]}{\log[(\mu_1+\mu_2)/(\mu_1-\mu_2)]}\leq\delta\nu_+\leq \frac{\log\bigl[\bigl(\mu_1+\mu_2-\mu_2\sqrt{\beta_N(\mu_2)}\delta\nu_-\bigl)/(\mu_1-\mu_2)\bigl]}{\log[(\mu_1+\mu_2)/(\mu_1-\mu_2)]}\\
    &{\rm and}\qquad \frac{1}{\sqrt{\beta_N(\mu_2)}}\geq\delta\nu_-\geq \frac{1}{\mu_2\sqrt{\beta_N(\mu_2)}}.
\end{split}
\end{equation}
Since it is $\beta_N(\mu_2)\approx\alpha_N/\mu_2^2$, $\mu_1+\mu_2\gg\mu_1-\mu_2$, and $1\ll\alpha_N=\mathcal{O}([\mu_1-\mu_2]^2)$ this domain simplifies to
\begin{equation}
    \delta\nu_+\in(0,1) \qquad{\rm and}\qquad \delta\nu_-\geq 0.
\end{equation}

We have still the ordering of the scales $\nu_+\gg\nu_-\gg1$ like in the previous case. Thence, the Taylor expansion works in the very same way, especially we need
\begin{equation}
\begin{split}
    0\leq&\frac{\sum_{j=3}^N(1-\mu_j^{-2})^{n+1}}{\nu_+^{2n+2}}\leq\frac{\alpha_N}{(\mu_1-\mu_2)^2}\left(\frac{\mu_1-\mu_2}{\mu_1+\mu_2}\right)^{2(n+1)\delta\nu_+}\overset{\delta\nu_+>0,\ n\geq0,\ N\to\infty}{\longrightarrow}0,\\
    0\leq&\frac{\sum_{j=3}^N(1-\mu_j^{-2})^{n+1}}{\nu_-^{2n+2}}\leq\mathcal{O}\left(\frac{1}{\alpha_N^{n}}\right)\overset{n\geq1,\ N\to\infty}{\longrightarrow}0.
\end{split}
\end{equation}
This allows to approximate the probability distribution as follows
\begin{equation}\label{asymp.case7}
\begin{split}
\rho\left((\mu_1-\mu_2)^{1-\delta\nu_+}(\mu_1+\mu_2)^{\delta\nu_+} ,\mu_2\sqrt{\beta_N(\mu_2)}\delta\nu_-|\widehat{C}\right)=&\frac{2}{\delta\nu_-^3}\exp\left[-\frac{1}{\delta\nu_-^2}\right]\Theta(\delta\nu_+-\delta\nu_+^2).
\end{split}
\end{equation}
It looks almost like the one of~\eqref{asymp.case6.a} with the subtle difference that the scaling of $\nu_+$ is of a completely different nature.

\item   When studying the scaling
\begin{equation}
\begin{split}
    \lim_{N\to\infty}\alpha_N=\infty, \quad \lim_{N\to\infty}\frac{\mu_1-\mu_2}{\sqrt{\alpha_N}}=0, \quad{\rm and}\quad  \lim_{N\to\infty}\beta_N(\mu_2)=\lim_{N\to\infty}\frac{\mu_1-\mu_2}{\mu_2}=0,
\end{split}
\end{equation}
    we need to change the scaling to
\begin{equation}
    \nu_+=(\mu_2\sqrt{\beta_N(\mu_2)})^{1-\delta\nu_+}(\mu_1+\mu_2)^{\delta\nu_+} \qquad{\rm and}\qquad \nu_-=\mu_2\sqrt{\beta_N(\mu_2)}\delta\nu_-
\end{equation}
    with
\begin{equation}
\begin{split}
    &0\leq\frac{\log\bigl[\bigl(\mu_1-\mu_2+\mu_2\sqrt{\beta_N(\mu_2)}\delta\nu_-\bigl)/(\mu_2\sqrt{\beta_N(\mu_2)})\bigl]}{\log[(\mu_1+\mu_2)/(\mu_2\sqrt{\beta_N(\mu_2)})]}\leq\delta\nu_+\leq \frac{\log\bigl[\bigl(\mu_1+\mu_2-\mu_2\sqrt{\beta_N(\mu_2)}\delta\nu_-\bigl)/(\mu_2\sqrt{\beta_N(\mu_2)})\bigl]}{\log[(\mu_1+\mu_2)/(\mu_2\sqrt{\beta_N(\mu_2)})]}\\
    &{\rm and}\qquad \frac{1}{\sqrt{\beta_N(\mu_2)}}\geq\delta\nu_-\geq \frac{1}{\mu_2\sqrt{\beta_N(\mu_2)}}.
\end{split}
\end{equation}
Once again it can be simplified because of $\beta_N(\mu_2)\approx\alpha_N/\mu_2^2$ and $\mu_1+\mu_2\gg\sqrt{\alpha_N}$ so that
\begin{equation}
    \delta\nu_+\in(0,1) \qquad{\rm and}\qquad \delta\nu_-\geq 0.
\end{equation}
    Then we find the very same result as in the previous case
\begin{equation}\label{asymp.case8}
\begin{split}
\rho\left((\mu_2\sqrt{\beta_N(\mu_2)})^{1-\delta\nu_+}(\mu_1+\mu_2)^{\delta\nu_+},\mu_2\sqrt{\beta_N(\mu_2)}\delta\nu_- |\widehat{C}\right)=&\frac{2}{\delta\nu_-^3}\exp\left[-\frac{1}{\delta\nu_-^2}\right]\Theta(\delta\nu_+-\delta\nu_+^2)
\end{split}
\end{equation}
    with the subtlety that the scaling has changed. Evidently, the result above does not change when setting $\mu_1=\mu_2=\mu$ which is now allowed in this case but has been forbidden in the previous cases.

\end{enumerate}

There are two limits left when $\beta_N(\mu_2)\to\infty$. To achieve this, we define another function given by the quotient
    \begin{equation}\label{gammaN.def}
        \gamma_N(\mu_1,\mu_2)=\frac{\beta_N(\mu_1)(\mu_2-\mu_2^{-1})}{\beta_N(\mu_2)(\mu_1-\mu_1^{-1})}\in[0,1]
    \end{equation}
 to shorten the expressions.
    The limits are given by the following asymptotic regimes.

\begin{enumerate}
\setcounter{enumi}{8}

\item   We start with
\begin{equation}\label{asymp.case9.lim}
\begin{split}
    \lim_{N\to\infty}\alpha_N=\lim_{N\to\infty}\beta_N(\mu_2)=\infty \quad{\rm and}\quad \lim_{N\to\infty}\gamma_N(\mu_1,\mu_2)=0.
\end{split}
\end{equation}
    This double scaling allows a finite $\lim_{N\to\infty}\mu_2<\infty$ as well as the limit $\lim_{N\to\infty}\mu_2=\infty$. Regardless, which case it is, it certainly holds
    \begin{equation}
        0\leq\frac{1}{\mu_1}\frac{\mu_2-\mu_2^{-1}}{\beta_N(\mu_2)}\leq\frac{1}{\mu_2}\frac{\mu_2-\mu_2^{-1}}{\beta_N(\mu_2)}\leq\frac{1}{\beta_N(\mu_2)}\overset{N\to\infty}{\longrightarrow}0.
    \end{equation}
    Therefore, we may choose the parametrisation
    \begin{equation}
        \nu_+=\mu_1-\frac{\mu_2-\mu_2^{-1}}{\beta_N(\mu_2)}\delta\nu_+ \qquad{\rm and}\qquad \nu_-=\mu_2-\frac{\mu_2-\mu_2^{-1}}{\beta_N(\mu_2)}\delta\nu_-
    \end{equation}
    which are drawn from the domains
    \begin{equation}
        0\leq|\delta\nu_+|\leq\delta\nu_-\leq\frac{2\mu_2\beta_N(\mu_2)}{\mu_2+1}\overset{N\to\infty}{\longrightarrow}\infty
    \end{equation}
    
    We need to expand $\sum_{j=3}^N\log(\Phi(\mu_1,\mu_2,C_{(12)},\mu_j)$. For this reason, it is important to notice that
    \begin{equation}
        \sum_{j=3}^N\frac{(\mu_j^2-1)^{n+1}(\mu_2^2-\nu_-^2)^{n+1}}{(\mu_j^2\mu_2^2-1)^{n+1}(\nu_-^2-1)^{n+1}}\leq\beta_N(\mu_2)\frac{(\mu_2^2-\nu_-^2)^{n+1}}{(\nu_-^2-1)^{n+1}}=\mathcal{O}\left(\frac{1}{\beta_N^n(\mu_2)}\right)\overset{n\geq1,\ N\to\infty}{\longrightarrow}0
    \end{equation}
    For the term with $\nu_+$ it is 
    \begin{equation}
        \sum_{j=3}^N\frac{(\mu_j^2-1)^{n+1}(\mu_1^2-\nu_+^2)^{n+1}}{(\mu_j^2\mu_1^2-1)^{n+1}(\nu_+^2-1)^{n+1}}\leq\beta_N(\mu_1)\frac{(\mu_1^2-\nu_+^2)^{n+1}}{(\nu_+^2-1)^{n+1}}=\mathcal{O}\left(\frac{\beta_N(\mu_1)(\mu_2-\mu_2^{-1})^{n+1}}{\beta_N^{n+1}(\mu_2)(\mu_1-\mu_1^{-1})^{n+1}}\right)\overset{n\geq0,\ N\to\infty}{\longrightarrow}0.
    \end{equation}
    This allows a truncated Taylor series which becomes exact in the limit $N\to\infty$. The probability density drastically simplifies to
\begin{equation}\label{asymp.case9}
\begin{split}
\rho\left(\mu_1-\frac{\mu_2-\mu_2^{-1}}{\beta_N(\mu_2)}\delta\nu_+,\mu_2-\frac{\mu_2-\mu_2^{-1}}{\beta_N(\mu_2)}\delta\nu_-|\widehat{C}\right)=&\frac{\mu_1-\mu_2-\frac{\mu_2-\mu_2^{-1}}{\beta_N(\mu_2)}(\delta\nu_+-\delta\nu_-)}{\mu_1-\mu_2-4\frac{\mu_2-\mu_2^{-1}}{\beta_N(\mu_2)}}\ee{-\delta\nu_-}\Theta(\delta\nu_--|\delta\nu_+|).
\end{split}
\end{equation}
The polynomial prefactor may simplify further depending on whether $\mu_1-\mu_2$ is much smaller or larger than $(\mu_2-\mu_2^{-1})/\beta_N(\mu_2)$.

        This limit is actually a Gaussian limit when going over to the variables $d_\pm$. Then we need to expand $\nu_\pm$ in small $d_\pm$, i.e.
\begin{equation}\label{nupm.expand.dpm}
    \nu_+\approx\mu_1+\frac{d_+^2}{2(\mu_1-\mu_2)}-\frac{d_-^2}{2(\mu_1+\mu_2)} \quad{\rm and}\quad \nu_-\approx\mu_2-\frac{ d_+^2}{2(\mu_1-\mu_2)}-\frac{d_-^2}{2(\mu_1+\mu_2)}
\end{equation}
in particularly we have assumed here that $(\mu_1-\mu_2)^2\gg(\mu_2-\mu_2^{-1})(\mu_1-\mu_2)/\beta_N(\mu_2)$. Certainly when $(\mu_1-\mu_2)^2$ and $(\mu_2-\mu_2^{-1})(\mu_1-\mu_2)/\beta_N(\mu_2)$ are of the same order one or $(\mu_1-\mu_2)^2$ is even smaller, the approximation~\eqref{nupm.expand.dpm} fails and  one needs to take the exact expression~\eqref{sev.2mod.b}.
    Hence, it is
\begin{equation}\label{asymp.case9.gauss}
\begin{split}
\rho\left(d_+,d_-|\widehat{C}\right)=&\frac{\beta_N^2(\mu_2)d_+d_-}{2(\mu_2-\mu_2^{-1})^2(\mu_1^2-\mu_2^2)}\exp\left[-\frac{\beta_N(\mu_2)d_+^2}{2(\mu_2-\mu_2^{-1})(\mu_1-\mu_2)}-\frac{\beta_N(\mu_2)d_-^2}{2(\mu_2-\mu_2^{-1})(\mu_1+\mu_2)}\right],
\end{split}
\end{equation}
where the Heaviside theta functions can be suppressed as they are naturally satisfied when assuming $d_\pm\geq0$. 

We would like to point out two subtleties. Firstly, this approximation does not cover the case when $\mu_1=\mu_2=\mu$ since then the third limit in~\eqref{asymp.case9.lim} is violated. Secondly, this present case comprises the cases when $\beta_N(\mu_1)/\beta_N(\mu_2)\to0$ and/or $(\mu_2-1)/(\mu_1-1)\to0$ because of the third limit in~\eqref{asymp.case9.lim} and that $\mu_2\leq\mu_1$ and $\beta_N(\mu_1)<\beta_N(\mu_2)$. Therefore, we can safely conclude for the remaining cases that $\beta_N(\mu_1)$ and $\beta_N(\mu_2)$ are of the same order as well as $\mu_2-1$ and $\mu_1-1$. What remains open is how large the difference $\mu_1-\mu_2$ is.

\item  There is another Gaussian limit in $d_\pm$ given by the scaling \label{item:gaussian_limit}
\begin{equation}
\begin{split}
    \lim_{N\to\infty}\alpha_N=\lim_{N\to\infty}\beta_N(\mu_2)=\infty \quad{\rm and}\quad \lim_{N\to\infty}\gamma_N(\mu_1,\mu_2)\in(0,1].
\end{split}
\end{equation}
    As we have learned from the previous case, the last two conditions imply that $\beta_N(\mu_1)\to\infty$ such that $\beta_N(\mu_1)/\beta_N(\mu_2)$ converges to a non-zero value and that $(\mu_1-1)/(\mu_2-1)$ are also converging to a non-zero number.

    As before we want to expand in a neighbourhood of $\nu_+=\mu_1$ and $\nu_-=\mu_2$ for which we choose this time
    \begin{equation}
        \nu_+=\mu_1-f_+\delta\nu_++f_-\delta\nu_- \qquad{\rm and}\qquad \nu_-=\mu_2-f_-\delta\nu_-
    \end{equation}
    with some scaling variables $f_\pm\ll\mu_{1}-1,\mu_2-1$ small enough and to be fixed. We will see that this choice is very convenient.
    The problem in the present case is that the expansion of $\sum_{j=3}^N\log[\Phi(\mu_1,\mu_2,C_{(12)},\mu_j]$ up to a linear order in the expansion variables $\delta\nu_\pm$ may not always yield an integrable probability density. Actually, we will see that $f_+$ is at most of order $f_-$ and that the choice 
    \begin{equation}
        f_+(\mu_1)=\frac{\mu_1-\mu_1^{-1}}{2\beta_N(\mu_1)}
    \end{equation}
    works well. Indeed it is $f_+/(\mu_{1}-1),f_+/(\mu_2-1)\to0$ when $\beta_N(\mu_1)\to\infty$. The expansion only in $f_+\delta\nu_+$ can be performed by noticing
\begin{equation}
\begin{split}
    \sum_{j=3}^N\log[\Phi(\mu_1,\mu_2,C_{(12)},\mu_j]=&\sum_{j=3}^N\log[\Phi(\hat\nu_+,\mu_2,C_{(12)},\mu_j]-\sum_{j=3}^N\log\left[1+\frac{(\mu_j^2-1)(\hat\nu_+^2-\nu_+^2)}{(\mu_j^2\hat\nu_+^2-1)(\nu_+^2-1)}\right]
\end{split}
\end{equation}
where we defined $\hat\nu_+=\nu_++f_+\delta\nu_+$. Only the second term depends on $f_+\delta\nu_+$, so that we can use
\begin{equation}
\begin{split}
    \sum_{j=3}^N\left(\frac{(\mu_j^2-1)(\hat\nu_+^2-\nu_+^2)}{(\mu_j^2\hat\nu_+^2-1)(\nu_+^2-1)}\right)^{n+1}=\mathcal{O}\left(\frac{1}{\beta_N^n(\mu_1)}\right)\overset{\geq1,\ N\to\infty}{\longrightarrow}0
\end{split}
\end{equation}
to find the approximation
\begin{equation}
\begin{split}
    \sum_{j=3}^N\log[\Phi(\mu_1,\mu_2,C_{(12)},\mu_j]\approx&\sum_{j=3}^N\log[\Phi(\hat\nu_+,\mu_2,C_{(12)},\mu_j]-\delta\nu_+.
\end{split}
\end{equation}
The expansion $f_-\delta\nu_-$ must be performed up the second because of the mentioned reason, which leads to
\begin{equation}
\begin{split}
    &\sum_{j=3}^N\log[\Phi(\hat\nu_+,\mu_2,C_{(12)},\mu_j]=\mathcal{O}(f_-^3)-\left[\frac{\beta_N(\mu_2)}{\mu_2-\mu_2^{-1}}-\frac{\beta_N(\mu_1)}{\mu_1-\mu_1^{-1}}\right]f_-\delta\nu_-\\
    -&\sum_{j=3}^N\left[\frac{(3\mu_1^2+1)(\mu_j^2-1)(\mu_j^2\mu_1^2-1)-2\mu_1^2(\mu_j^2-1)^2}{(\mu_j^2\mu_1^2-1)^2(\mu_1^2-1)^2}+\frac{(3\mu_2^2+1)(\mu_j^2-1)(\mu_j^2\mu_2^2-1)-2\mu_2^2(\mu_j^2-1)^2}{(\mu_j^2\mu_2^2-1)^2(\mu_2^2-1)^2}\right]f_-^2\delta\nu_-^2.
\end{split}
\end{equation}
While the linear term times the maximal value of $f_-\delta\nu_-=\mu_2-1$ vanishes when $\mu_1\to\mu_2$, the quadratic term times $f_-^2\delta\nu_-^2=(\mu_2-1)^2$ diverges to infinity and is positive, i.e.,
\begin{equation}
  (\mu_2-1)^2\sum_{j=3}^N\frac{(3\mu_1^2+1)(\mu_j^2-1)(\mu_j^2\mu_1^2-1)-2\mu_1^2(\mu_j^2-1)^2}{(\mu_j^2\mu_1^2-1)^2(\mu_1^2-1)^2}  \geq\frac{(\mu_1^2+1)(\mu_2-1)^2}{(\mu_1^2-1)^2}\beta_N(\mu_1)\to\infty
\end{equation}
an similarly for the $\mu_2$-dependent term.
For a limiting ratio $\lim_{N\to\infty}\gamma_N(\mu_1,\mu_2)\in(0,1)$, the linear term is diverging as it is $\beta_N(\mu_2)[1-\gamma_N(\mu_1,\mu_2)]/(\mu_2-\mu_2^{-1})\to\infty$ while the square term will be of lower order once we choose $f_-=(\mu_2-\mu_2^{-1})/\beta_N(\mu_2)[1-\gamma_N(\mu_1,\mu_2)]$.
Therefore, there is a transition going on when $\gamma_N(\mu_1,\mu_2)\to1$. 

To cover the above described behaviour we choose
\begin{equation}
\begin{split}
    &f_-(\mu_1,\mu_2)=\left[\frac{2\beta_N(\mu_2)[1-\gamma_N(\mu_1,\mu_2)]}{\mu_2-\mu_2^{-1}}\right.\\
    +&\left.\sqrt{\sum_{j=3}^N\left[\frac{(3\mu_1^2+1)(\mu_j^2-1)(\mu_j^2\mu_1^2-1)-2\mu_1^2(\mu_j^2-1)^2}{(\mu_j^2\mu_1^2-1)^2(\mu_1^2-1)^2}+\frac{(3\mu_2^2+1)(\mu_j^2-1)(\mu_j^2\mu_2^2-1)-2\mu_2^2(\mu_j^2-1)^2}{(\mu_j^2\mu_2^2-1)^2(\mu_2^2-1)^2}\right]}\right]^{-1}.
\end{split}
\end{equation}
One can now check that the domain becomes
\begin{equation}
    \begin{split}
        0\leq \delta\nu_+\leq2\frac{f_-(\mu_1,\mu_2)}{f_+(\mu_1)}\delta\nu_- \qquad{\rm and}\qquad 0\leq \delta\nu_-\leq\frac{\mu_2-1}{f_-(\mu_1,\mu_2)}\overset{N\to\infty}{\longrightarrow}\infty.
    \end{split}
\end{equation}
For the latter we notice that $f_-(\mu_1,\mu_2)$ is bound from below by a constant times $(\mu_2-1)/\sqrt{\beta_N(\mu_2)}$.
The probability density is then
\begin{equation}\label{asymp.case10}
\begin{split}
&\rho\left(\mu_1-f_+(\mu_1)\delta\nu_++f_-(\mu_1,\mu_2)\delta\nu_-,\mu_2-f_-(\mu_1,\mu_2)\delta\nu_-|\widehat{C}\right)\\
\propto&\Theta\left(2\frac{f_-(\mu_1,\mu_2)}{f_+(\mu_1)}\delta\nu_--\delta\nu_+\right)\ \left[\mu_1-\mu_2-f_+(\mu_1)\delta\nu_++2f_-(\mu_1,\mu_2)\delta\nu_-\right]\\
&\times\exp\left[-\delta\nu_+-[\gamma_N^{-1}(\mu_1,\mu_2)-1]\frac{f_-(\mu_1,\mu_2)}{f_+(\mu_1)}\delta\nu_--\left(1-\frac{f_-(\mu_1,\mu_2)[1-\gamma_N(\mu_1,\mu_2)]}{f_+(\mu_2)}\right)^2\delta\nu_-^2\right].
\end{split}
\end{equation}

When $\lim_{N\to\infty}\gamma_N(\mu_1,\mu_2)\in(0,1)$, meaning $\mu_1$ and $\mu_2$ are not too close to each other the linear term in $\delta\nu_-$ will dominate and $f_-(\mu_1,\mu_2)/f_+(\mu_1)\approx\gamma_N(\mu_1,\mu_2)/[1-\gamma_N(\mu_1,\mu_2)]$ as well as $f_-(\mu_1,\mu_2)/f_+(\mu_2)\approx1/[1-\gamma_N(\mu_1,\mu_2)]$, we find the simplification
\begin{equation}\label{asymp.case10a}
\begin{split}
&\rho\left(\mu_1-f_+(\mu_1)\delta\nu_++f_-(\mu_1,\mu_2)\delta\nu_-,\mu_2-f_-(\mu_1,\mu_2)\delta\nu_-|\widehat{C}\right)\\
\propto& \left(\mu_1-\mu_2-f_+(\mu_1)[\delta\nu_+-2\gamma_N(\mu_1,\mu_2)\delta\nu_-]\right)\ee{-\delta\nu_+-\delta\nu_-}\Theta\left(\frac{2\gamma_N(\mu_1,\mu_2)}{1-\gamma_N(\mu_1,\mu_2)}\delta\nu_--\delta\nu_+\right).
\end{split}
\end{equation}
    When we want to go back to the variables $d_\pm$, we can use~\eqref{nupm.expand.dpm} for which we need to assume that $\frac{\beta_N(\mu_2)}{\mu_2-\mu_2^{-1}}-\frac{\beta_N(\mu_1)}{\mu_1-\mu_1^{-1}}\gg1$. What we 
    find is
\begin{equation}\label{asymp.case10.Gauss}
\begin{split}
\rho(d_+,d_-|\widehat{C})=&4\Lambda_+\Lambda_- d_+d_-\ee{-\Lambda_+ d_+^2-\Lambda_- d_-^2},
\end{split}
\end{equation}
where
\begin{equation}
\label{eq:Lambda-159}
\begin{split}
\Lambda_\pm=&\frac{1}{\mu_1\mp\mu_2}\left[\frac{\beta_N(\mu_2)}{\mu_2-\mu_2^{-1}}\mp\frac{\beta_N(\mu_1)}{\mu_1-\mu_1^{-1}}\right]\\
=&\sum_{j=3}^N\frac{(\mu_j^2-1)[\mu_j^2\mu_1\mu_2(\mu_1\pm\mu_2)^2\mp(\mu_1\mu_2\mu_j^2\pm1)(\mu_1\mu_2\pm1)]}{(\mu_1^2-1)(\mu_2^2-1)(\mu_1^2\mu_j^2-1)(\mu_2^2\mu_j^2-1)}.
\end{split}
\end{equation}  
The result of the previous case can be simply regain by taking the limit $\gamma_N(\mu_1,\mu_2)\to0$ as it is then $(\mu_1-\mu_2)\Lambda_+\approx(\mu_1+\mu_2)\Lambda_-\approx\beta_N(\mu_2)/(\mu_2-\mu_2^{-1})$.

When we want to set $\mu_1=\mu_2=\mu$, it is $\gamma_N(\mu,\mu)=1$ and $f_-(\mu,\mu)/f_+(\mu)\to\infty$. Therefore, we find
\begin{equation}\label{asymp.case10b}
\begin{split}
&\rho\left(\mu-f_+(\mu)\delta\nu_++f_-(\mu,\mu)\delta\nu_-,\mu-f_-(\mu,\mu)\delta\nu_-|\widehat{C}\right)=2\delta\nu_-\ee{-\delta\nu_+-\delta\nu_-^2}.
\end{split}
\end{equation}
When going over to $d_\pm$ for $\mu_1=\mu_2=\mu$ we must use the following simplified relation to the symplectic eigenvalues $\nu_\pm$,
\begin{equation}
    \nu_\pm^2=\mu^2+\frac{d_+^2-d_-^2}{2}\pm\sqrt{2}\mu d_+,
\end{equation}
which is actually exact and no approximation, cf., Eq.~\eqref{sev.2mod.b}. The corresponding probability is still Gaussian and of the form
\begin{equation}\label{asymp.case10b.Gauss}
\begin{split}
\rho\left(d_+,d_-|\widehat{C}\right)=&\left[\frac{1}{\mu}\partial_\mu\frac{\beta_N^2(\mu)}{(\mu-\mu^{-1})^2}\right] d_+d_- \exp\left[-\partial_\mu\frac{\beta_N(\mu)}{\mu-\mu^{-1}} d_+^2-\frac{\beta_N(\mu)}{\mu^2-1}d_-^2\right].
\end{split}
\end{equation}
As can be readily checked, this formula nicely fits together with~\eqref{asymp.case10.Gauss} when taking the limit $\mu_1,\mu_2\to\mu$. We underline that the scale of the variances of $d_+$ and $d_-$ are slightly different when $\mu\to1$ or $\mu\to\infty$.

\end{enumerate}

\section{Application to black hole}\label{app:bh_calculation}

\subsection{Discrete mode set}
We consider the set of discrete modes which is commonly used for the Hawking effect, since the original article~\cite{Hawking75}, see also~\cite{fabbri_modeling_2005}.
Consider a continuous family of mode operators $ a_\omega$ with $\omega>0$, which obey the commutation relations $\left[ a_\omega, a^\dagger_{\omega'}\right]=\delta(\omega-\omega')$.
Here, we will associate these with spherically symmetric positive frequency modes of outgoing radiation at asymptotic infinity $\mathcal{I}^+$ (but they could equally well correspond, say, to the right-moving sector of a massless field in 1+1D).
A discrete and complete set of mode operators can be defined as
\begin{equation}
a_{nj}=\frac1{\sqrt{\Delta \omega}}\integral\omega{j \Delta \omega}{(j+1) \Delta \omega} \ee{\ii 2\pi n \omega/\Delta \omega} a_\omega
\end{equation}
for integers  $n\in\mathbb{Z}$ and $j\in\mathbb{N}_0$, and
where we  refer to $\Delta \omega>0$ as the modes' band width. It is easily checked, that these operators obey the properly normalized commutation relations $\left[a_{nj},a_{n'j'}^\dagger\right]=\delta_{nn'}\delta_{jj'}$.

The shape of the mode functions associated with these modes shares characteristics with the sinc function, as one would expect based on its Fourier transform.
To illustrate this, let us briefly consider the construction (for the right-moving sector of a massless field) in 1+1D Minkowski spacetime where the mode operators $a_\omega$ is associated with the plane wave mode
If the mode operator $a_\omega$ is associated to a plane wave mode
\begin{equation}
u_\omega(x) =\frac1{\sqrt{4\pi \omega}} \ee{\ii \omega x},
\end{equation}
then the mode operator $a_{nj}$ is associated to the mode function
\begin{equation}
\begin{split}
&u_{nj}(x)=\frac1{\sqrt{\Delta \omega}}\integral{\omega}{j \Delta \omega}{(j+1)\Delta \omega}\ee{-\ii2\pi n \omega /\Delta \omega}u_\omega(x)
=\frac1{\sqrt{4\pi \Delta \omega}}\integral{\omega}{j \Delta \omega}{(j+1)\Delta \omega}\frac{\ee{\ii \omega\left(x-\frac{2\pi n}{\Delta \omega}\right)}}{\sqrt{\omega}}.
\end{split}
\end{equation}
This integral has a solution in terms of Fresnel functions. 
The resulting functions $u_{nj}(x)$ are oscillatory in their real and imaginary part, with a wave length determined by $j$. 
Their absolute value $|u_{nj}(x)|$  yields a hull function whose shape is not dependent on $j$, except for an overall factor determining the amplitude.
The absolute value  $|u_{nj}(x)|$ is symmetric around a peak at $x_n=2\pi n /\Delta \omega$. To the left and the right of this peak, $|u_{nj}(x)|$  has zeros at all multiples of $2\pi / \Delta \omega$.
In particular, all functions $u_{nj}$ share their zeros (except for their respective peaks).
Overall, the amplitude decays as  $|u_{nj}(x)|\sim 1/|x-x_n|$ away from the  peak.

\subsection{Model of evaporating black hole radiation}

To model the evaporating black hole, we divide up its lifetime into a large number of epochs.
These epochs are to be chosen such that the mass of the black hole during one epoch is approximately constant, and that its mass decreases by one Planck mass $m_P$ from one epoch to the next.
That is, in the $n$-th epoch the black hole mass is
\begin{equation}\label{eq:Mn}
    M_n=M_1-(n-1) m_P 
\end{equation}
where $M_1$ is the black hole's initial mass. Denoting by 
$M_R$ its final mass (which may be vanishing or some comparatively small remaining mass), the lifetime of the black hole is divided into $(M_1-M_R)/m_P$ epochs.

The length of the $n$-th epoch $\Delta t_n$, with respect to the time coordinate at asymptotic infinity $\mathcal{I}^+$, 
corresponds to the time it takes the black hole to loose one Planck mass.
Following~\cite{Page2005,SekinoSusskind}, we assume that this time is determined by the relation
\begin{equation}
\Delta t_n= \kappa  \,t_P(M_n/m_P)^2 ,
\end{equation}
with a constant $\kappa$.\footnote{
The constant $\kappa$  is determined by balancing the emitted radiation energy with the mass loss of the black hole and depends on the type and number of fields considered in the model (see~\cite{Page2005}). Page estimated it to be on the order of up to $10^4$ or less~\cite{Page2005}. In the present model a value of $\kappa\approx 1.2\times 10^3$ maintains this energy-mass balance.}

We model the radiation emanated from the black hole by discrete modes as those introduced in the previous subsection. Therein it was shown that these modes have a characteristic width of $2\pi/\Delta \omega$, and that increasing the index $n\to n+1$ shifts the center peak of the mode function by the same amount.
Here, to model the evaporating hole, we use varying band widths 
\begin{equation}
\Delta \omega_n=2\pi/\Delta t_n= 2\pi m_P^2/(\kappa t_P M_n^2),
\end{equation}
and assume that the radiation emanating from the black hole in the $n$-th epoch is captured by the modes $a_{nj}$ for $j=0,1,2,\dots$ with bin width $\Delta \omega_n$.\footnote{Varying band widths $\Delta \omega_n$ cause the commutator $\left[a_{nj},a_{n'j'}^\dagger\right]$ of modes from different epochs  to be non-zero, which we neglect here.}

We assume that the marginals of the bin modes correspond to a global thermal field state with the Hawking temperature corresponding to the black hole mass during the epoch, 
\begin{equation}
        T_n=(m_P^2c^2)/( 8\pi k_B M_n).
\end{equation}
Hence we take the number excitation value and symplectic eigenvalue of the individual modes $a_{nj}$ to be
    \begin{equation}\label{eq:anjdagger_anj_exptval}
    \begin{split}
        \exptval{a_{nj}^\dagger a_{nj}}&= \frac1{\ee{\hbar (j+1)\Delta\omega_n /(k_B T_n)}-1},
        \qquad
        \mu_{nj}=\coth\left(\frac{(j+1)\hbar\Delta\omega_{n}}{2k_BT_n}\right)
=\coth\left(\frac{(j+1) 8\pi^2  m_P}{\kappa   M_n} \right).
    \end{split}
    \end{equation}
This approximation is motivated by the fact that it converges quickly to the precise value for large frequency, \ie as $j\beta\Delta\omega\to\infty$.
For lower frequencies the approximation underestimates the exact expectation value.
However, since the limiting shape of $\Phi(\mu_1,\mu_2,C_{(12)},\mu_{nj})$ as $\mu_{nj}\to\infty$, limits the impact of this deviation as far as these modes enter the expressions for the probability distribution as  background modes (corresponding to $i\geq3$ in~\eqref{eq:rho_exact_Phi_factorized}), we apply the above simple shape of marginal values for all modes.

\subsection{Exact distribution}
We now proceed with the evaluation of the probability distribution~\eqref{prob.dens.largeN} applied to our assumptions on the black hole radiation above. For the evaluation, it is convenient to approximate the product over all modes by an integral.
To this end, we apply $\exp(\log(\dots))$ to both sides of~\eqref{prob.dens.largeN} and obtain
\begin{equation}
\begin{split}
&\rho(d_+,d_-|\widehat{C})\propto \frac{\Theta(\nu_--1)\Theta(\sqrt{2\mu_1\mu_2}-d_+-d_-)\ d_+d_-}{ (\nu_+^2-1)^2(\nu_-^2-1)^2 \Phi\left(\mu_1,\mu_2,C_{(12)},\mu_1\right)\Phi\left(\mu_1,\mu_2,C_{(12)},\mu_2\right)} 
\exp\left(\sum_n\sum_j \log \Phi\left(\mu_1,\mu_2,C_{(12)},\mu_{nj}\right)\right)
\end{split}
\end{equation}
Here we replaced the index $i=(n,j)$ by the double index which enumerates all modes from the different black hole epochs. Only $\mu_1=\mu_{n_1j_1}$ and $\mu_2=\mu_{n_2j_2}$ are kept as short hand notation for the symplectic eigenvalue of the two specific modes for which we consider the probability distribution over their correlations.

The sum over $j$ can be approximated by an integration as follows. We write $\mu_{nj} =\coth((j+1) \eta_n)$ and use 
\begin{equation}
    \eta_n=\frac{\hbar \Delta\omega_n}{2T_n k_B}=\frac{8\pi^2 m_P}{\kappa M_n}\ll1
\end{equation}
to approximate 
\begin{equation}
\sum_{j=0}^\infty f( \mu_1,\mu_2, d_+,d_-,\mu_j)\approx\frac1{\eta_n}\integral{x}0\infty f( \mu_1,\mu_2, d_+,d_-,\coth(x))
=:\frac{\kappa M_n}{8 \pi^2m_P}F(\mu_1,\mu_2,c_+,c_-),
\end{equation}
where
\begin{equation}
    f( \mu_1,\mu_2, d_+,d_-,\mu_j) = \log\Phi= \log \frac{[1+\mu_1^2(\mu_j^2-1)/(\mu_1^2-1)][1+\mu_2^2(\mu_j^2-1)/(\mu_2^2-1)]}{[1+\nu_+^2(\mu_j^2-1)/(\nu_+^2-1)][1+\nu_-^2(\mu_j^2-1)/(\nu_-^2-1)]}.
\end{equation}
Using
\begin{equation}
\begin{split}
   &\integral{x}0\infty \log \left(1+\frac{\mu^2(\coth(x)^2-1)}{\mu^2-1}\right)
=\frac14 \left(\pi^2 +\left(\log\frac{\mu-1}{\mu+1}\right)^2 \right)
   =\frac{\pi^2}4+\mathrm{arccoth}^2(\mu)
\end{split}
\end{equation}
we obtain
\begin{equation}
    \begin{split}
        \integral{x}0\infty f( \mu_1,\mu_2, d_+,d_-,\coth(x)) 
        =&\frac14\left( \left(\log\frac{\mu_1-1}{\mu_1+1}\right)^2+\left(\log\frac{\mu_2-1}{\mu_2+1}\right)^2-\left(\log\frac{\nu_+-1}{\nu_++1}\right)^2-\left(\log\frac{\nu_--1}{\nu_-+1}\right)^2 \right)\\
        =& \mathrm{arccoth}^2(\mu_1)+\mathrm{arccoth}^2(\mu_2)-\mathrm{arccoth}^2(\nu_+)-\mathrm{arccoth}^2(\nu_-)=:F(\mu_1,\mu_2,\nu_+,\nu_-).
    \end{split}
\end{equation}
Due to the mass of the black hole  during each epoch to correspond to a multiple of the Planck mass $m_P$, see~\eqref{eq:Mn}, the remaining sum over $n$ gives
\begin{equation}\label{eq:eta_n_sum}
    \begin{split}
        &\sum_n\frac1{\eta_n}
        = \frac{\kappa }{8\pi^2}  \sum_{n=1}^{R} n 
        = \frac{\kappa}{16\pi^2} \frac{M_1+M_R}{m_P}\left(1+\frac{M_1-M_R}{m_P}\right)
   \approx  \frac{\kappa }{16\pi^2}  \frac{M_1^2-M_R^2}{m_P^2}.
    \end{split}
\end{equation}
This results in the following expression for the probability distribution
\begin{equation}
\begin{split}
&\rho(d_+,d_-|\widehat{C})\propto \frac{\Theta(\nu_--1)\Theta(\sqrt{2\mu_1\mu_2}-d_+-d_-)\ d_+d_-}{ (\nu_+^2-1)^2(\nu_-^2-1)^2 \Phi\left(\mu_1,\mu_2,C_{(12)},\mu_1\right)\Phi\left(\mu_1,\mu_2,C_{(12)},\mu_2\right)} 
\exp\left( \frac{\kappa }{16\pi^2}  \frac{M_1^2-M_R^2}{m_P^2} F(\mu_1,\mu_2,\nu_+,\nu_-)\right)
\end{split}
\end{equation}
where the symplectic eigenvalues $\nu_\pm$ are given as functions of $\mu_1,\mu_2,d_\pm$ in~\eqref{sev.2mod.b}.

\subsection{Asymptotic limits}
The numerical evaluation of the expression for the probability density derived in the previous section quickly becomes challenging when trying to consider black holes with initial masses of the order of astrophysical black holes.
Hence, to understand the case of large black hole initial masses it is useful to employ the asymptotic limits derived in Sec~\ref{sec:asymptotic-limits}.
To this end, we need to construct a sequence of distributions, as discussed in the beginning of Sec.~\ref{sec:asymptotic-limits}.
For our purpose, we construct the sequence of distributions $\diff P_K(\mu)=\sum_{i=0}^{N_K}\delta(\mu-\mu_i^{(K)})d\mu$ such that the distributions correspond to black holes of larger and larger initial mass $M_1$ as $K\to\infty$.
The idea here is that for black holes of astrophysical scales the exact distribution is well captured by the appropriate asymptotic limits.

Concretely such a sequence can be constructed as follows:
In our model each epoch (indexed by $n$) contains an infinite number of modes (labelled by arbitrary large $j$). However, effectively only a finite number of these modes influence the distribution because as $j\to\infty$ we have that $\mu_{nj}-1\to0$ decays exponentially fast.
In fact to have $\mu_{nj}=\coth(j\eta_n) = 1+\epsilon$ we need only logarithmically many modes in $\epsilon$.
\begin{equation}
    \begin{split}
        1+\epsilon=\coth(j\eta_n) \Leftrightarrow j= \mathrm{arccoth}(1+\epsilon)\frac1{\eta_n}= \mathrm{arccoth}(1+\epsilon)\frac{\kappa M_n}{8\pi^2 m_P}\approx \frac{\log2-\log\epsilon}2 \frac{\kappa M_n}{8\pi^2 m_P}.
    \end{split}
\end{equation}
Hence we may define the distribution $\diff P_K$ such that it
comprises $K$ subsequent epochs, \ie it corresponds to a black hole evolving from an at initial mass $M_1=M_R+K\,m_P$ to a final mass $M_R$,
and where from each epoch it contains the first $\lceil K/\eta_n\rceil$ modes.
Thus the total number of modes contained in $\diff P_K$ is $N_K=\sum_{n=1}^{K}\lceil K/\eta_n\rceil$, for which $N_K\propto K^3$ as $K\to\infty$. The number $N_K$ will be set equal to the matrix dimension $N$ in the random matrix model proposed by us.

To identify the relevant asymptotic limits we need to calculate $\alpha_N$ in~\eqref{sum} and $\beta_N(\mu)$ in~\eqref{betaN}.
Both can be approximated by integrals similar as above through $\sum_j f(\eta_n j)\approx \eta_n^{-1}\integral{x}0\infty$ where $\mu_{nj}=\coth(j\eta_n)$ is replaced by $\coth(x)$.
For  $\alpha_N$ we obtain
\begin{equation}
\label{eq:alpha-N-all-epochs}
\begin{split}
    &\alpha_N+\frac{\mu_{1}^2-1}{\mu_{1}^2}+\frac{\mu_{2}^2-1}{\mu_{2}^2}=\sum_n\sum_j\frac{\mu_{nj}^2-1}{\mu_{nj}^2}
    \approx \sum_n\frac1{\eta_n}\underbrace{\integral{x}0\infty [1-\tanh(x)^2]}_{=1}\approx  \frac{\kappa }{16\pi^2}  \frac{M_1^2-M_R^2}{m_P^2}
    \\ & \Rightarrow \alpha_N\approx \frac{\kappa }{16\pi^2}  \frac{M_1^2-M_R^2}{m_P^2}+\frac1{\mu_1^2}+\frac1{\mu_2^2}-2.
\end{split}
\end{equation}
Here it is clear that in the limit $N\to\infty$, we have $M_1\to\infty$ and hence $\alpha_N\to\infty$.

For $\beta_N(\mu)$ (assuming $\mu>1$) we obtain
\begin{eqnarray}\label{eq:beta-N-all-epochs}
\begin{split}
&    \beta_N(\mu)+\frac{\mu_1^2-1}{\mu_1^2\mu^2-1}+\frac{\mu_2^2-1}{\mu_2^2\mu^2-1} =\sum_n\sum_{j}\frac{\mu_{nj}^2-1}{\mu_{nj}^2\mu^2-1}
    \approx\sum_n\frac1{\eta_n}\integral{x}0\infty \frac{\coth(x)^2-1}{\coth(x)^2\mu^2-1}
    =\sum_n\frac{\mathrm{arccoth}(\mu)}{\eta_n \mu} 
    \\
    & \Rightarrow   \beta_N(\mu)\approx \frac{\kappa }{16\pi^2}  \frac{M_1^2-M_R^2}{m_P^2} \frac{ \mathrm{arccoth}(\mu)}\mu -\frac{\mu_1^2-1}{\mu_1^2\mu^2-1}-\frac{\mu_2^2-1}{\mu_2^2\mu^2-1} 
    .
    \end{split}
\end{eqnarray}
If $\mu$ is kept constant as $N\to\infty$, then $\beta_N(\mu)$ diverges since the first term grows with $M_1\to\infty$.
However, when considering the most highly excited modes in our black hole model, which are given by the lowest lying modes of the two earliest epochs, $\beta_N(\mu)$ does not diverge to infinity. This becomes evident when considering 
that for $\mu=\coth(j \eta_n)$ we have
$\mathrm{arccoth}(\mu)/\mu=j\eta_n\tanh(j\eta_n)$ which has the expansion
\begin{equation}
    \frac{\kappa }{16\pi^2}  \frac{M_1^2-M_R^2}{m_P^2} j\eta_n \tanh(j\eta_n) \stackrel{j \eta_n \ll1}\sim \frac{\kappa }{16\pi^2}  \frac{M_1^2-M_R^2}{m_P^2} j^2 \eta_n^2 
= \frac{ 4 j^2 }{\kappa} \frac{M_1^2-M_R^2}{M_n^2}\stackrel{M_n=M_1,M_2}\approx \frac{4j^2}\kappa.
\end{equation}
Also, if $\mu=\mu_{1},\mu_{2}$ the last two terms in~\eqref{eq:beta-N-all-epochs} vanish in the limit $N\to\infty$, hence, $\lim \beta_N(\mu_{11})\approx \frac4\kappa$.

We apply two of the asymptotic limits derived in Sec.~\ref{sec:asymptotic-limits} for our discussion of the black hole.
On the one hand, this is the Gaussian approximation in Case~\ref{item:gaussian_limit} which yields the probability distribution over correlations between two modes that are moderately excited (relative to the highest excited modes in our model).
On the other hand, Case~\ref{item:high_excitation_limit} yields an expression for the highest excited modes in our model. In particular, this limit allows us to check that for these modes, while a considerable size of correlations is expected, still the probability to find them entangled rapidly decays to zero with the black holes initial mass.

\subsubsection{Gaussian approximation for lowly excited modes}
If we consider two fixed values for $\mu_1\geq\mu_2$ and leave them fixed as $N\to\infty$, then it follows from~\eqref{eq:beta-N-all-epochs} that the quotient $\gamma_N$, defined in~\eqref{gammaN.def}, takes the finite limiting value
\begin{equation}
        \gamma_N(\mu_1,\mu_2)=\frac{\beta_N(\mu_1)(\mu_2-\mu_2^{-1})}{\beta_N(\mu_2)(\mu_1-\mu_1^{-1})}\stackrel{N\to\infty}\longrightarrow
        \frac{ \mathrm{arccoth}(\mu_1)(\mu_2^2-1)}{ \mathrm{arccoth}(\mu_2)(\mu_1^2-1)}.
\end{equation}
Hence, we find this case to corresponds to Case~\ref{item:gaussian_limit} of Sec.~\ref{sec:asymptotic-limits} yielding a Gaussian approximation to the probability distribution.
The coefficients of the  Gaussian approximation are given in~\eqref{eq:Lambda-159}.
Employing an integral approximation to the sum as above (where it is convenient to change to the integration variable $\mu=\coth(x)$ for which $\diff \mu=(1-\coth(x)^2)\diff x$), we obtain
\begin{equation}
\begin{split}
&\Lambda_\pm+L_\pm = \sum_n \sum_j\frac{(\mu_{nj}^2-1)[\mu_{nj}^2\mu_1\mu_2(\mu_1\pm\mu_2)^2\mp(\mu_1\mu_2\mu_{nj}^2\pm1)(\mu_1\mu_2\pm1)]}{(\mu_1^2-1)(\mu_2^2-1)(\mu_1^2\mu_{nj}^2-1)(\mu_2^2\mu_{nj}^2-1)}
\\& \approx
\sum_n \frac1{\eta_n}\integral{\mu}1\infty \frac{ [\mu^2\mu_1\mu_2(\mu_1\pm\mu_2)^2\mp(\mu_1\mu_2\mu^2\pm1)(\mu_1\mu_2\pm1)]}{(\mu_1^2-1)(\mu_2^2-1)(\mu_1^2 \mu^2-1)(\mu_2^2\mu^2-1)}
\\&=
 \frac{\kappa }{16\pi^2}  \frac{M_1^2-M_R^2}{m_P^2}
 \frac{ (\mu_1\pm\mu_2)\left((\mu_1^2-1 )  \mathrm{arccoth}(\mu_2)\mp\mathrm{arccoth}(\mu_1) (\mu_2^2-1)  \right)}{(\mu_1^2-\mu_2^2)  \left(\mu_1^2-1\right) \left(\mu_2^2-1\right) }
,
\end{split}
\end{equation}  
and where $L_\pm$ are the summands corresponding to $\mu_1$ and $\mu_2$, \ie
\begin{equation}
    \begin{split}
        L_\pm &=
        \frac{1}{4} \left(\pm\frac{(2\mp 2\mu_1 \mu_2)}{\left(\mu_1^2+1\right) \left(\mu_2^2+1\right)} +\frac{3}{(\mu_1\mp1) (\mu_2-1)}+\frac{3}{(\mu_1\pm1) (\mu_2+1)} -\frac{2}{\mu_1 \mu_2-1}-\frac{2}{\mu_1 \mu_2+1}\right).
    \end{split}
\end{equation}

\subsubsection{Approximation for highly excited modes}
The two most highly excited modes in our model are the two lowest lying modes of the two earliest epochs.
In the limit of large black hole masses, the asymptotic limit of Case~\ref{item:high_excitation_limit} applies and 
yields a limit whose shape (relative to the marginal symplectic values) is independent of the black hole initial mass.

As $N\to\infty$ the marginal symplectic eigenvalues of the two most excited modes scale as
\begin{equation}
    \mu \sim \frac{\kappa M_1}{8\pi^2m_P},
\end{equation}
and from above we have that
\begin{equation}
    \alpha_n\sim \frac{\kappa (M_1^2-M_R^2)}{16\pi^2m_P^2},\quad \beta_N(\mu)\sim\frac4\kappa.
\end{equation}
If we express the size of symplectic eigenvalues $\mu_\pm$ relative to the dominating term in $\mu$, \ie
\begin{equation}
    \nu_\pm=\frac{\kappa M}{8\pi^2m_P} y_\pm,
\end{equation}
then we have $\alpha_N/\nu_\pm^2\sim \frac{4\pi^2}{\kappa}$ and the asymptotic expression for the probability distribution~\eqref{asymp.case5.nupm}
takes the shape
\begin{equation}
    \rho\left(\nu_+,\nu_-|\hat C\right)=\frac{4\pi^2 (y_+^2-y_-^2)}{Z(\mu_1,\mu_2) \kappa y_+^3 y_-^3}\exp\left(-\frac{4\pi^2}\kappa \left(\frac1{y_+^2}+\frac1{y_-^2}\right)\right) \chi_\Sigma(\nu_+,\nu_-). 
\end{equation}
This distribution has its maximum at the values
$
y_\pm^{\mathrm{max}}= {2\pi  \sqrt{2\pm\sqrt{2}} }/{\sqrt{\kappa}},
$
which for large values of $\kappa$ may seem to lie close to a pure and hence entangled state of these two modes.
However, the estimate in~\eqref{eq:cse5.Pent.order} which in our case reads
\begin{equation}
    P_{\rm ent} = \mathcal{O}\left( \kappa^2 M_1^7 \exp\left(-\frac{M_1}{4\pi^2}\right) \right)
\end{equation}
shows that the probability to have any entanglement between the two modes is heavily suppressed for large initial black hole masses.

\bibliography{prl-supplement}